\documentclass[traditabstract]{aa}
\usepackage{graphicx}
\usepackage{txfonts}
\usepackage{amsmath}
\usepackage{longtable}
\usepackage{xcolor}
\usepackage{natbib}
\usepackage{threeparttable}
\usepackage{pdfpages}
\usepackage{verbatim}
\usepackage{morefloats} 
\usepackage[hyperfootnotes=true, hidelinks, colorlinks=true, citecolor=blue, linkcolor=blue, linktocpage, bookmarks=true, urlcolor=blue]{hyperref}
\usepackage{longtable, multirow, pdflscape}
\usepackage{soul}
\usepackage{placeins}

\bibpunct{(}{)}{;}{a}{}{,}

\begin{document}

\newcommand{\hi}{\mbox{H\,{\sc i}}}
\newcommand{\zabs}{$z_{\rm abs}$}
\newcommand{\zmin}{$z_{\rm min}$}
\newcommand{\zmax}{$z_{\rm max}$}
\newcommand{\zq}{$z_{\rm q}$}
\newcommand{\zg}{$z_{\rm g}$}
\newcommand{\mgii}{Mg~{\sc ii}}
\newcommand{\lya}{Ly$\alpha$}
\newcommand{\kms}{km\,s$^{-1}$}
\newcommand{\cmsq}{cm$^{-2}$}
\newcommand{\degree}{\ensuremath{^\circ}}
\newcommand{\Msun}{$M_{\odot}$} 


\definecolor{green}{rgb}{0,0.4,0}
\newcommand{\SB}[1]{{\color{red} [SB:~ #1]}}
\newcommand{\NRJ}[1]{{\color{magenta} [NG:~ #1]}}
\newcommand{\JK}[1]{{\color{green} [JK:~ #1]}}
\newcommand{\UPDT}[1]{{\color{blue} [#1]}}
\newcommand{\TODO}[1]{{\color{orange} [TODO:~ #1]}}

\newcommand{\lapp}{\mbox{\raisebox{-0.3em}{$\stackrel{\textstyle <}{\sim}$}}}
\newcommand{\gapp}{\mbox{\raisebox{-0.3em}{$\stackrel{\textstyle >}{\sim}$}}}
\newcommand{\be}{\begin{equation}}
\newcommand{\en}{\end{equation}}
\newcommand{\di}{\displaystyle}
\def\tworule{\noalign{\medskip\hrule\smallskip\hrule\medskip}} 
\def\onerule{\noalign{\medskip\hrule\medskip}} 
\def\bl{\par\vskip 12pt\noindent}
\def\bll{\par\vskip 24pt\noindent}
\def\blll{\par\vskip 36pt\noindent}
\def\rot{\mathop{\rm rot}\nolimits}

\titlerunning{MALS Galactic \hi}
\authorrunning{Gupta et al.}
\title{The MeerKAT Absorption Line Survey (MALS) data release 3: Cold atomic gas associated with the Milky Way\thanks{The MALS images and spectra are publicly available at \href{https://mals.iucaa.in}{https://mals.iucaa.in}.}}

\author{
N. Gupta\href{https://orcid.org/0000-0001-9174-1186}{\includegraphics[width=8pt]{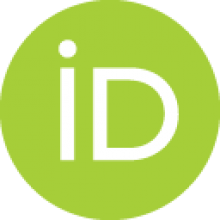}}  \inst{1}
\and
J. Kerp\inst{2}
\and
S. A. Balashev\inst{3}
\and 
A. P. M. Morelli\inst{2}
\and
F. Combes\inst{4}
\and
J-.K. Krogager\inst{5,6}
\and
E. Momjian\inst{7}
\and
D. Borgaonkar\inst{1}
\and
P. P. Deka\inst{1}
\and
K. L. Emig\inst{8}
\and
J. Jose\inst{9}
\and
G. I. G. J\'ozsa\inst{10, 11}
\and
H.-R. Kl\"ockner\inst{10}
\and
K. Moodley\inst{12,13}
\and
S. Muller\inst{14}
\and
P. Noterdaeme\inst{15}
\and
P. Petitjean\inst{15}
\and
J. D. Wagenveld\inst{10}
}

\institute{
Inter-University Centre for Astronomy and Astrophysics, Post Bag 4, Ganeshkhind, Pune 411 007, India \\\email{\href{ngupta@iucaa.in}{ngupta@iucaa.in}}
\and
Argelander-Institut für Astronomie, Universit\"at Bonn, Auf dem Hügel 71, D-53121 Bonn, Germany 
\and
Ioffe Institute, 26 Politeknicheskaya st., St. Petersburg, 194021, Russia
\and
Observatoire de Paris, Coll\`ege de France, PSL University, Sorbonne University, CNRS, LUX, Paris, France
\and
Universit\'e Lyon, ENS de Lyon, CNRS, Centre de Recherche Astrophysique de Lyon UMR5574, F-69230 Saint-Genis-Laval, France
\and
French-Chilean Laboratory for Astronomy, IRL 3386, CNRS and Universidad de Chile, Santiago, Chile
\and
National Radio Astronomy Observatory, P. O. Box O, Socorro, NM 87801, USA
\and
National Radio Astronomy Observatory, 520 Edgemont Road, Charlottesville, VA 22903, USA
\and
ThoughtWorks Technologies India Private Limited, Yerawada, Pune 411 006, India
\and
Max-Planck-Institut f\"ur Radioastronomie, Auf dem H\"ugel 69, D-53121 Bonn, Germany
\and
Department of Physics and Electronics, Rhodes University, P.O. Box 94 Makhanda 6140, South Africa
\and
School of Mathematics, Statistics \& Computer Science, University of KwaZulu-Natal, Westville Campus, Durban 4041, South Africa
\and
Astrophysics Research Centre, University of KwaZulu-Natal, Durban 4041, South Africa
\and
Department of Space, Earth and Environment, Chalmers University of Technology, Onsala Space Observatory, Sweden
\and
Institut d'Astrophysique de Paris, UMR 7095, CNRS-SU, 98bis bd Arago, 75014  Paris, France
}

\date{Received \today; accepted 2025}

\abstract
  {}
{
We present results of a blind search for Galactic \hi\ 21-cm absorption lines toward 19130 radio sources brighter than 1\,mJy at 1.4\,GHz, using 390 pointings of the MeerKAT Absorption Line Survey (MALS), each pointing centered on a source brighter than 200\,mJy.  The spectral resolution, the median spatial resolution, and the median 3$\sigma$ optical depth sensitivity ($\tau_{3\sigma}$) are 5.5\,\kms, $\sim9^{\prime\prime}$, and 0.381, respectively.  We used the spectra of the central sources and the other  off-axis radio sources within the telescope pointings to constrain the properties of \hi\ gas in the local interstellar medium (LISM) of the Galaxy.  
 }
{
Through an automated procedure, we detected 3640 \hi\ absorption features over $\sim$800\,deg$^2$.  This represents the largest Galactic \hi\ absorption line catalog to date.
We used \hi\ 21-cm emission line measurements from HI4PI, an all sky single-dish survey, and far-infrared maps from COBE/DIRBE and IRAS/ISSA in addition to the Gaussian decomposition of the HI4PI  into cold (CNM), lukewarm (LNM), and warm (WNM) neutral medium phases for our analyses.
}
{
We find a strong linear correlation with a coefficient of 0.84 between the \hi\ 21-cm emission line column densities ($N_\mathrm{HI}$) and the visual extinction ($A_\mathrm{V}$) measured toward the pointing center, along with the confinement of the absorption features to a narrow range in radial velocities (-25$<v_{\rm LSR}$[\kms]$<$+25). This implies that the detected absorption lines form a homogeneous sample of \hi\ clouds in the LISM.  
For central sight lines (median $\tau_{3\sigma}$=0.008), the detection rate is 82$\pm$5\%.  All the central MALS sight lines with \hi\ absorption have $N_\mathrm{HI}({\rm CNM})\,+\,N_\mathrm{HI}({\rm LNM}) \geq N_\mathrm{HI}({\rm WNM})$. 
The \hi\ 21-cm absorption optical depth is linearly correlated to $N_\mathrm{HI}$ and $A_\mathrm{V}$, with a correlation coefficient in excess of 0.8 up to $N_\mathrm{HI} \simeq 2\cdot 10^{21}\,\mathrm{cm^{-2}}$ or, equivalently, $A_\mathrm{V} \simeq 1$\,mag. 
Above this threshold, $A_\mathrm{V}$ traces the total hydrogen content, and consequently, $A_\mathrm{V}$ and the single-dish $N_\mathrm{HI}$ scale, differently.
The slopes of $N_\mathrm{HI}$ distributions of central sight lines with \hi\ 21-cm absorption detections and non-detection differ at $>2\sigma$.  A similar difference is observed for H$_2$ detections and non-detections in damped Lyman-alpha systems at $z\gapp1.8$, implying that turbulence-driven WNM-to-CNM conversion is the common governing factor for the presence of \hi\ 21-cm and H$_2$ absorption.
Through a comparison of central and off-axis absorption features, we find the optical depth variations ($\Delta\tau$) to be higher for pointings centered on regions with a higher $N_\mathrm{HI}$ and CNM fraction.  However, no such dependence is observed for the covering fraction of the absorbing structures over 0.1 - 10\,pc.  The slope (2.327 $\pm$ 0.153) of root mean square (rms) fluctuations in optical depth variations  in the quiescent gas associated with LISM is shallower than the earlier measurements in the disk.  The densities (20-30\,cm$^{-3}$) inferred from |$\Delta\tau$| at the median separation (1.5\,pc) of the sample are typical of the CNM values.  The negligible (median $\sim$ 0\,\kms) velocity shifts between central and off-axis absorbers are in line with the hypothesis that the CNM/LNM clouds freeze out of the extended WNM phase.
}
{}

\maketitle

\section{Introduction} 
\label{sec:intro}  

Our basic understanding of physical conditions in the neutral interstellar medium (ISM) is primarily based on observations of the Milky Way \citep[e.g.,][]{Kulkarni88, Dickey1990, KalberlaKerp2009, Mcclure-Griffiths2023}.  The framework is key to understanding the formation of stars and the evolution of distant galaxies for which detailed studies are not possible.
In the Milky Way's ISM, \hi\ gas exhibits a wide range of temperatures ($T_{\rm K} \sim$ 20 - 8000\,K) and volume densities ($\rho$ $\sim$ 0.1 - 100\,cm$^{-3}$).  It is primarily observed by the hyperfine transition of the \hi\ atom at a rest-frame frequency of 1420.405752\,MHz (21.106\,cm).  The relative population of hydrogen atoms in the two hyperfine states is represented by the spin temperature ($T_{\rm Spin}$), which is affected by both radiative and collisional processes \citep[e.g.,][]{Field1959}.  The energy gap of only $\Delta E = 5.87\cdot 10^{-6}$\,eV required to excite the transition and the tiny Lorentzian line width imply that the shape of the 21-cm line is a pure measure of the gas dynamics and pressure.  Therefore, it can be used as a probe for the physical conditions of all neutral gas phases.

The observed spectrum at frequency $\nu$, $T_b(\nu)$ of a 21-cm line from an isothermal \hi\ cloud with optical depth $\tau(\nu)$, in front of a radio source with the brightness temperature $T_{\rm rad}$, is given by
\begin{equation}
    T_b(\nu) = T_{\rm rad}(\nu)e^{-\tau(\nu)} + T_{\rm Spin}(\nu)(1 - e^{-\tau(\nu)}).
\label{eq:radtrans}
\end{equation}
Thus, the line may be observed in emission or absorption depending on whether $T_{\rm Spin}$ $>$ $T_{\rm rad}$ or not.
Since \hi\ 21-cm optical depth is inversely proportional to the spin temperature, it is thought that \hi\ absorption lines arise from the cold ($\sim$100\,K) regions of the interstellar \hi\ gas.  This was indeed found in past, less sensitive absorption line observations \citep[e.g.,][]{Radhakrishnan1972, Crovisier1978}.  In the literature, this neutral gas component is called the cold-neutral medium (CNM; with gas temperatures of $T_{\rm K} <$200\,K), while the diffuse warmer phase, thought to be detected only in emission, is denoted as the warm-neutral medium (WNM; with $T_{\rm K} \sim$ 4000 - 8000\,K). Both these phases coexist in the ISM over a certain kinetic gas pressure range set by the static balance of heating and cooling processes \citep[e.g.,][]{Wolfire03}. 

More sensitive absorption line surveys and sophisticated magnetohydrodynamic modeling at the beginning of the century revealed the presence of a third gas phase with an intermediate temperature, denoted as the unstable or lukewarm medium (LNM). \citet[][]{Heiles03} from the Arecibo Millenium survey of 79 sources, \citet[][]{Roy2013} from the \hi\ 21-cm absorption line spectroscopy of 33 sources, and \cite{KalberlaHaud2018} using the all sky HI4PI survey attributed $>30$\%, $>28$\%, and about $41$\% of the \hi\ to the LNM phase, respectively. However, the most sensitive ($\tau_{1\sigma}<$0.001 per 0.42\,\kms)\footnote{1$\sigma$ optical depth sensitivity} and the largest Galactic absorption line survey comprising 57 lines of sight recently reported a much lower LNM fraction of only about 20\% \citep[21-SPONGE survey;][]{Murray2018-sponge}. 

Numerical simulations \citep[e.g.,][]{Audit2005, Maclow2005, Dobbs2012} enable the modeling of the ISM structure from astronomical units (AU)  to the scale of tens of parsecs. In this work, we focus on the density and phase transitions of neutral atomic hydrogen, comprising the CNM, LNM, and WNM.
The cold atomic phase is much denser than the warm phase, and therefore much likelier to form cloudlets embedded in the diffuse warm neutral gas.
The CNM is sufficiently dense to enable the formation of H$_2$ on the surfaces of the dust grains. The enhanced fraction of H$_2$ in the CNM in comparison with the WNM then allows H$_2$ to shield itself against destructive UV photons forming -- the \hi-to-H$_2$ transition layer \cite[e.g.,][]{Abgrall1999, Sternberg2014}. 
While the WNM may exist as an individual phase, the LNM likely resides in regions with both CNM and WNM \citep[e.g.,][]{Audit2005, Maclow2005, Dobbs2012}.

Observationally, the relative contributions of the three atomic gas phases to the \hi\ 21-cm line signal depend on their density and temperature (Eq.\,\ref{eq:radtrans}). In the higher-density CNM, the collisional processes drive $T_\mathrm{Spin}$ toward the kinetic temperature of the gas, that is, $T_{\rm Spin}$ $\sim$ $T_{\rm K}$, whereas in the warmer atomic gas, $T_{\rm Spin}$ $<$ $T_{\rm K}$ \citep[][]{Liszt01}. The information from these different gas components may superpose on the radial velocity space in such a way as to prevent a unique decomposition of these individual effects. In fact, depending on the quality of the data and the methodology used for the decomposition of the various phases in emission and absorption line signal, the observed $T_{\rm Spin}$($\nu$) may actually be the column-density weighted mean of the spin temperatures along the sight line \citep[e.g.,][]{Field1959, KulkarniHeiles1988}.
Further improvements in our understanding of the different atomic gas phases will come from the large samples of the \hi\ absorption line spectrum that have an adequate sensitivity to individually detect all three \hi\ gas phases.  
The spatial and temporal optical depth variations ($\Delta\tau\gapp$0.05) at AU to parsec scales observed toward extended radio sources \citep[e.g.,][]{Faison2001, Brogan2005, Roy2012, Rybarczyk2020} and pulsars \citep[e.g.,][]{Frail1994, Johnston2003} are interpreted as being caused by superposed \hi\ filaments and sheets \citep[][]{Heiles97}, or a turbulent cascade over \hi\ structures on all scales \citep[][]{Deshpande00}. Spatially close-by absorption line measurements toward radio loud AGNs can provide crucial constraints on the turbulent structures of these clouds \citep[see][for a review]{Stanimivoric2018}.

In this paper, we present the Galactic \hi\ absorption line catalog of 3640 absorption features detected along 19130 sight lines from 390 out of 391 pointings\footnote{One pointing with poor quality was excluded.} of the MeerKAT Absorption Line Survey \citep[MALS;][]{Gupta17mals}, an ongoing large program at the MeerKAT telescope \citep[][]{Jonas16}. Of these, 3158 absorbers are visually confirmed and 1870 also have a peak signal-to-noise ratio (S/N) of $>6$.    This dramatically increases the sample of publicly available \hi\ 21-cm absorption line spectra.
Each MALS pointing in the L band covering 900 - 1670 MHz is centered on a radio source brighter than 200\,mJy at $\sim$1\,GHz. 
The radiation from the bright central AGN as well as numerous others, hereafter referred to as off-axis sources, within MALS pointings with a field of view corresponding to full width at half maximum (FWHM) of 88$^\prime$ at 1\,GHz, inevitably pass through the Milky Way environment via the Galactic halo. 

In contrast to the searches of absorption lines associated with distant galaxies \citep[e.g.,][]{Gupta18j1243}, the MALS pointings are selected to avoid the Galactic plane.
Consequently, MALS sight lines mostly pass through the local interstellar medium (LISM) at high Galactic latitudes. This, as is described later in the paper, is due to the optimization of the survey footprint for extragalactic science objectives. The LISM comprises the low-volume density local cavity \citep[][]{Frisch1983}. The Sun is localized within this irregularly shaped cavity \citep[][]{Lallement2019}, which may have been created by a series of nearby supernovae  \citep[e.g.,][]{Cox1987, Zucker2022}.  The focus of MALS is toward the high-Galactic-latitude sky. Here, high-velocity clouds and tidal gas stripped off from dwarf galaxies may also lie along the lines of sight \citep[][]{Wakker1997, Putman2021}.  The large number of high-latitude lines of sight also provide an opportunity to constrain the vertical distribution of \hi\ clouds above the Galactic disk \citep[e.g.,][]{Crovisier1978, Dickey2022, Wenger2024}.

This paper is structured as follows.  In Sect.~\ref{sec:obscat}, we present the details of the observations, calibration, and imaging of the \hi\ 21-cm line.  The details of the automated \hi\ 21-cm absorption line search from 391 pointings, the characterization of absorption features, and the resultant catalog are also presented in this section.
The data products and results of the OH 18-cm main lines at 1665 and 1667\,MHz, and the OH satellite line at 1612\,MHz, will be presented in a future paper.    In Sect.~\ref{sec:res}, we present the \hi\ 21-cm absorption line detection rates.  Multifrequency full-sky surveys showing the structure of the LISM in \hi\ emission and far-infrared (FIR) radiation in great detail are available. We cross-correlate this multifrequency information and decompose the \hi\ gas toward MALS pointings into CNM, LNM, and WNM. From these, we derive the integrated properties of the absorbing gas toward the pointing centers. The properties of the absorption lines toward the central and off-axis lines of sight are then used to investigate the parsec scale structure in the absorbing gas. The results and future prospects are summarized in Sect.~\ref{sec:summ}.

\section{Observations, data analysis, and catalog}      
\label{sec:obscat}   

\subsection{Observations and data analysis}      
\label{sec:obs}   

%
\citet[][]{Gupta21salt} described the target selection process optimized for the requirements of the extragalctic \hi\ 21-cm and OH 18-cm absorption line search, and ensuring reasonable observability with the MeerKAT telescope in the southern hemisphere \citep[see also][]{Krogager19}. 
The sky coverage of the Galactic component of MALS presented here is based on 391 pointings observed at L-band between April 1, 2020, and January 18, 2021 \citep[][]{Deka24dr1}.  Each MALS pointing was centered on a radio source brighter than $\sim$200\,mJy at 1\,GHz and declination, $\delta \lesssim$ $+20^{\circ}$ in the NRAO Very Large Array (VLA) Sky Survey  \citep[NVSS; ][]{Condon98} or the Sydney University Molonglo Sky Survey \citep[SUMSS;][]{Mauch03}, and observed for 56\,minutes, split into three scans of 1120\,s duration at different hour angles to improve the uv-coverage.  The total 856\,MHz bandwidth for these observations was centered on 1283.9869\,MHz and split into 32768 frequency channels.   This mode of the SKA Reconfigurable Application Board (SKARAB) correlator corresponds to a channel spacing of 26.123\,kHz, resulting in a spectral resolution of  5.5\,\kms\ at the rest frequency of the \hi\ 21-cm line. 
The correlator dump time was 8\,s. For dual, linearly polarized L-band feeds with orthogonal polarizations labeled X and Y, the data were acquired for all four polarization products: XX, XY, YX, and YY. On average 59 antennas of the MeerKAT-64 array participated in these observations.
%

A typical $\sim$3.5\,hr long L-band observing run comprised three targets. For flux density scale, delay, and bandpass calibrations, the observations included scans of 5-10 minutes duration on PKS\,1939–638 and / or PKS\,0408–658 at the start, middle and end.  Each scan on a target source was bracketed by a 60\,s long scan on a complex gain calibrator. 
For Galactic \hi\ analysis, we generated a measurement set comprising only XX and YY polarization products over 400 frequency channels centered on 1420.2709\,MHz in the topocentric frame.  This is a subset of L-band {\tt SPW\_10} defined in \citet[][]{Gupta21} and is hereafter labeled as {\tt LSPW\_10G}.  These data were processed using the Automated Radio Telescope Imaging Pipeline (ARTIP) based on NRAO's Common Astronomy Software Applications (CASA) package \citep[][]{Casa22}.  The details of ARTIP are provided in \citet[][]{Gupta21}.  

For the flux density calibration, we used the model based on {\tt Stevens-Reynolds 2016} for PKS\,1939–638 \citep{Partridge16} whereas for PKS\,0408–658 a model with $S_{\rm 1284\,MHz}= 17.066$\,Jy  and $\alpha=-1.179$ was used.  Next, the pipeline proceeded with delay, bandpass, and temporal complex gain calibration steps.  
The presence of Galactic \hi\ emission or absorption in the spectrum of the bandpass calibrator can limit the spectral line sensitivity and fidelity of the target source spectra. Through the examination of \hi\ emission and dust properties we rule out the presence of significant \hi\ absorption toward both  bandpass calibrators as follows. 
Specifically, we inspected the whole primary beam area for both bandpass calibrators in FIR, optical extinction and single-dish \hi\ 21-cm emission line datasets. To search for small-scale CNM gas within the field of view of MeerKAT, we first evaluated the \hi\ emission spectra from the Leiden/Argentine/Bonn \citep[LAB;][]{LAB2005} and HI4PI \citep{HI4PI} surveys. A small scale cold gas structure would show up as a narrow bright emission line in HI4PI. Since the beam ratio of the 64-m Parkes (HI4PI) and the 30-m Villa Elisa (LAB) radio telescope is $(\Omega_{64m}/\Omega_{30m})^2 = 4.6$, an unresolved small angular scale CNM structure observed with Villa Elisa would be identified by an up to 4.6 times higher peak-brightness temperature with Parkes. However, for the two fields of interest, the \hi\ spectra from both telescopes appear to be indistinguishable. This implies that toward both  fields,  the \hi\ 21-cm emission is smoothly distributed at the angular scales of the Villa Elisa beam. To complement these investigations we inspected the Planck dust temperature map \citep[][]{Planck2014} and the optical extinction maps of \citet[][]{SFD1998}. Neither FIR excess emission associated with cold dust nor high optical extinction were detected within the entire field of interest and, in particular, at the lines of sight toward both bandpass calibrators (see Sect.~\ref{sec:phase} for further details).  
%

Encouraged by the absence of any obvious \hi\ absorption toward PKS\,0408–658 and  PKS\,1939–638, we tested two bandpass calibration strategies.  In the first approach, the data from all the frequency channels in the dataset were used to determine the bandpass solutions.  In the second approach, we masked the frequency channels in the range, 1419.4 $-$ 1421.4\,MHz, on the calibrators.  The bandpass solutions were then interpolated across the masked frequency range and applied to the target source.  The target source spectra from both approaches are indistinguishable, confirming the conclusions based on \hi\ emission, FIR and visual extinction.  The spectra presented here are based on the second cautionary approach, ensuring uncompromising sensitivity to detect Galactic \hi\ emission and absorption in the MALS spectra. 

After calibration, the target source visibilities were processed for continuum and cube imaging.  The continuum dataset was generated by averaging 50 line-free channels.  It was imaged using {\tt robust=0} weighting as defined in CASA and {\tt w-projection} algorithm with 128 planes as the gridding algorithm in combination with {\tt multi-scale multi-term multifrequency synthesis} ({\tt MTMFS}) for deconvolution, with nterms=1 and four pixel scales (0, 2, 3, and 5) to model the extended emission \citep[][]{Rau11}. 
Imaging masks were appropriately adjusted using the {\tt Python Blob Detection and Source Finder} \citep[{\tt PyBDSF\footnote{PyBDSF version 1.10.1};}][]{mohan2015} between major cycles during imaging and self-calibration runs.  
This ensured that at any stage the artifacts in the vicinity of bright sources are excluded from the CLEANing process and the source model.   The relevant details of how this is achieved through {\tt PyBDSF} are presented in \citet[][]{Deka24dr1}. 
Overall, the pipeline performed three rounds of phase-only and a single round of amplitude and phase self-calibration.  The final 6k$\times$6k continuum images with a pixel size of $2^{\prime\prime}$ have a span of $3\fdg3$.

For cube imaging, the self-calibration solutions obtained from the continuum imaging were applied to the line dataset and continuum subtraction was performed using the model; that is, CLEAN components obtained from the last round of self-calibration. The continuum subtracted visibilities were then Fourier inverted to obtain spectral line cubes, which may then be deconvolved using CLEAN for line emission \citep[for example, see][]{Boettcher21, Maina22}. 
For Galactic science, we generated two sets of spectral-line cubes in the kinematic local standard of rest (LSRK\footnote{Hereafter, referred as LSR.}) system with {\tt robust=0}, which is suited for detecting absorption lines, and {\tt robust=0.4}, for detecting diffuse \hi\ emission. The {\tt robust $>$ 0.4} values yield synthesized beams with pronounced wings, and therefore were not preferred.   In this paper, we focus on absorption lines from {\tt robust=0} cubes detected toward compact radio sources. We note that these cubes have not been deconvolved for line signal.  The details of deconvolving the line signal will be presented in future papers focusing on extended absorption and emission lines.  We note that the radio continuum and spectral-line properties presented here are corrected for the attenuation of the primary beam pattern using the {\tt katbeam} (version\,0.1) model\footnote{https://github.com/ska-sa/katbeam}.

\subsection{Absorption line catalog}      
\label{sec:speccat}   

We extracted spectra from the {\tt robust = 0} cubes toward pixels corresponding to the peak flux density of 19274 radio sources brighter than 1\,mJy\,beam$^{-1}$ at 1.4\,GHz within $48\farcm5$ of the pointing center.  The continuum and spectral-line properties of individual sources are made available as a catalog  described in Table~\ref{tab:cat_cols}. The source name ({\tt Source\_name}) based on its position in the continuum image is given in column\,1.  Columns 2-9 provide details of the pointing in which the source is detected.  This includes {\tt Pointing\_id} based on the position of the central source in the NVSS or SUMSS, the dates of the UHF- and L-band observations ({\tt Obs\_date\_U} and {\tt Obs\_date\_L}),  the spectral window ID ({\tt SPW\_id}), and the details of the synthesized beam; that is, the major axis ({\tt Maj\_cbeam}), the minor axis ({\tt Min\_cbeam}), and the position angle ({\tt PA\_cbeam}). 
It is worth noting that the catalog columns have been defined to support all Galactic and extragalactic absorption line releases.  For the Galactic \hi\ absorption line catalog presented here, {\tt SPW\_id} = {\tt LSPW\_10G} consists of 400 frequency channels centered on 1420.2709\,MHz (see Sect.~\ref{sec:obs}). 

The position ({\tt Peak\_pos}) -- that is, the right ascension and declination in J2000 of the pixel at which the spectrum is extracted -- is provided in column\,10.  Columns\,11 and 12 provide the Galactic coordinates, longitude ({\tt Peak\_pos\_l}) and latitude ({\tt Peak\_pos\_b}) corresponding to {\tt Peak\_pos}.  The angular distance of the source from the pointing center ({\tt Distance\_pointing}) and the {\tt Scale\_factor} to correct for the primary beam attenuation are given in columns\,13 and 14, respectively. The peak flux density ({\tt Peak\_flux}) is noted in column\,15.
Columns\,16 and onward are concerned with the properties of the absorption lines from the automated search and Gaussian decomposition.  These include spectral root mean square (rms) noise ({\tt Spec\_rms}) estimated from line-free frequency channels (column\,16).  The process of absorption line detection and characterization is described in Sect.~\ref{sec:auto}.  The details of how {\tt Vis\_flag} (column\,17), indicating whether an absorption system is statistically reliable and passed the visual inspection, are presented in Sect.~\ref{sec:purity}.

\subsection{Automated line search and Gaussian decomposition}
\label{sec:auto}

\begin{figure*}
    \centering
    \vbox{
         \includegraphics[trim=0 1.0cm 0 0, width=\hsize]{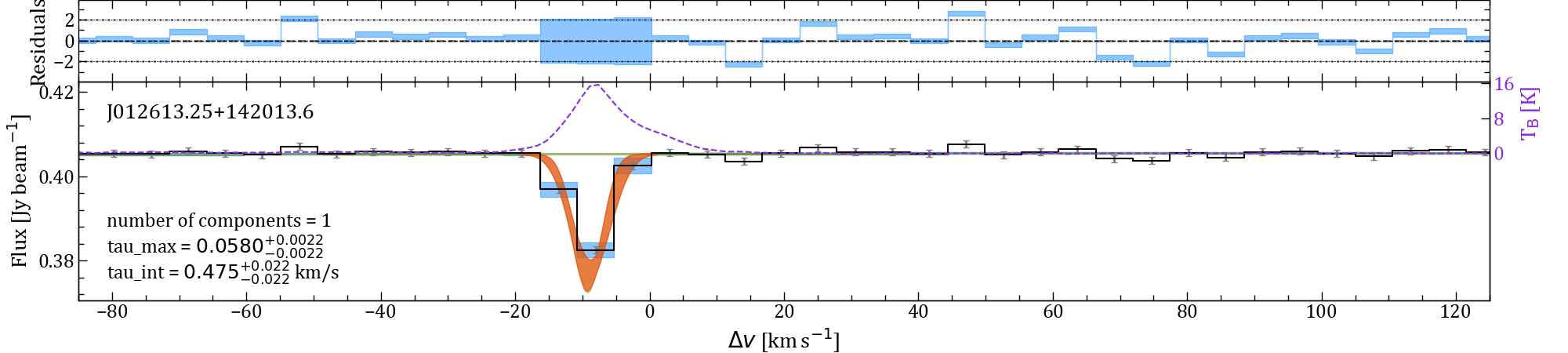} 
         \includegraphics[trim=0 1.0cm 0 0, width=\hsize]{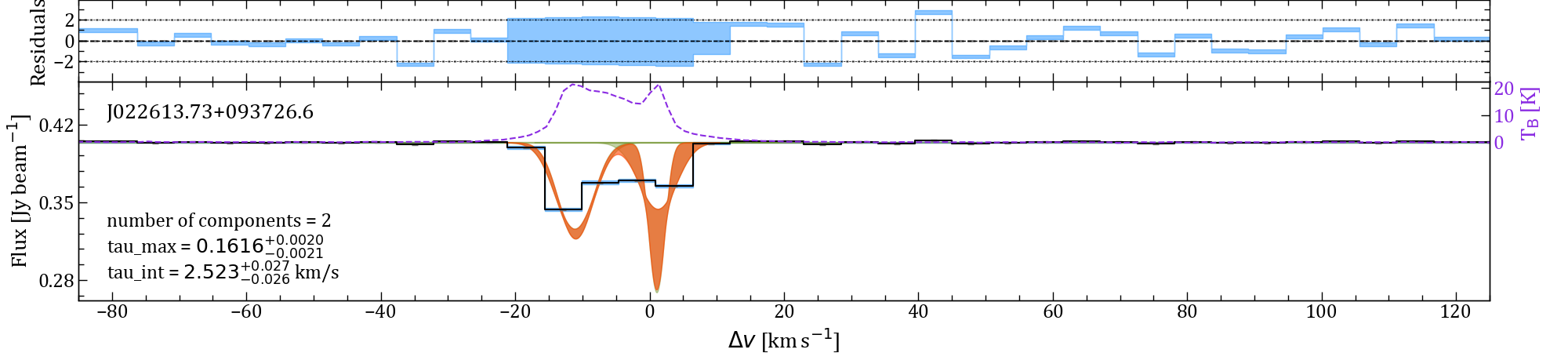} 
         \includegraphics[trim=0 0cm 0 0, width=\hsize]{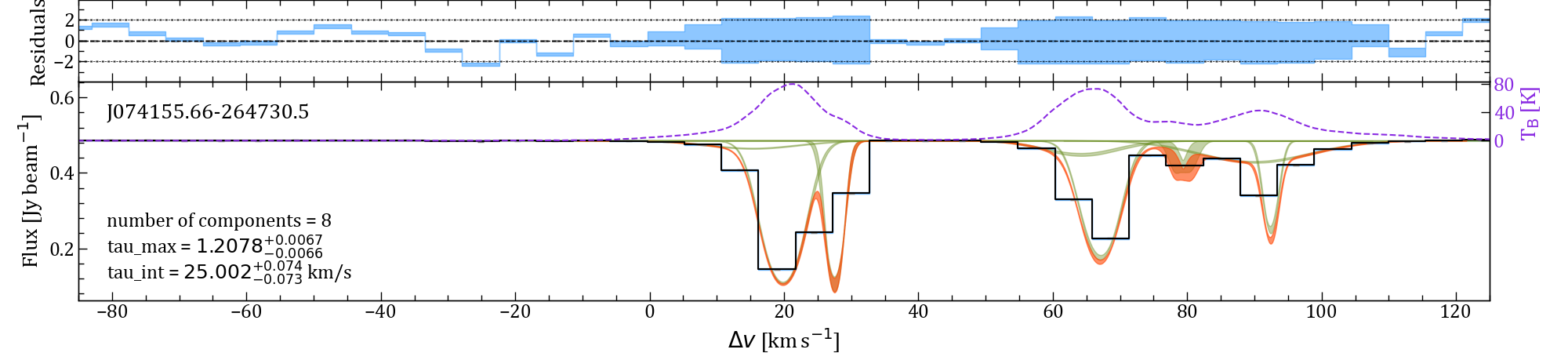}
         }
    \caption{
    Examples of \hi\ line profile fits to the spectra toward J012613.25$+$142013.6 (top), J022613.73$+$093726.6 (middle), and J074155.66$-$264730.5 (bottom). The black line shows the unsmoothed spectrum with a pixel size of 5.5\,\kms. The red and green filled regions indicate the total and individual components, respectively. The blue step-like regions indicate the total model profile integrated in bins. The model profiles are represented as 0.05..0.95 interquantile region of the unbinned model profile drawn from the posterior distribution function of the model parameters constrained during MCMC fit to the data (see Sect.~\ref{sec:auto}). The dashed violet line indicate the \hi\ emission profiles obtained using HI4PI survey near the position of the source. The upper panel show the 0.05..0.95 quantile regions of the residuals between the data and the model calculated using binned profiles. The text label in each panel indicates the derived global parameters of the profiles. 
    }
    \label{fig:HI_fit_example}
\end{figure*}
%
\begin{figure*}
    \centering
    \vbox{
        \includegraphics[trim=0 1.0cm 0 0, width=\hsize]{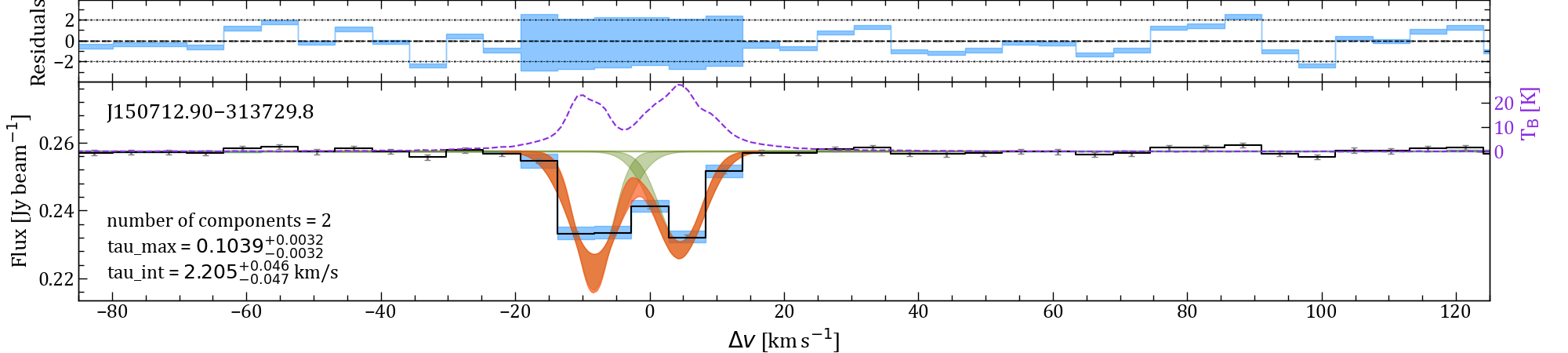} 
        \includegraphics[trim=0 0cm 0 0, width=\hsize]{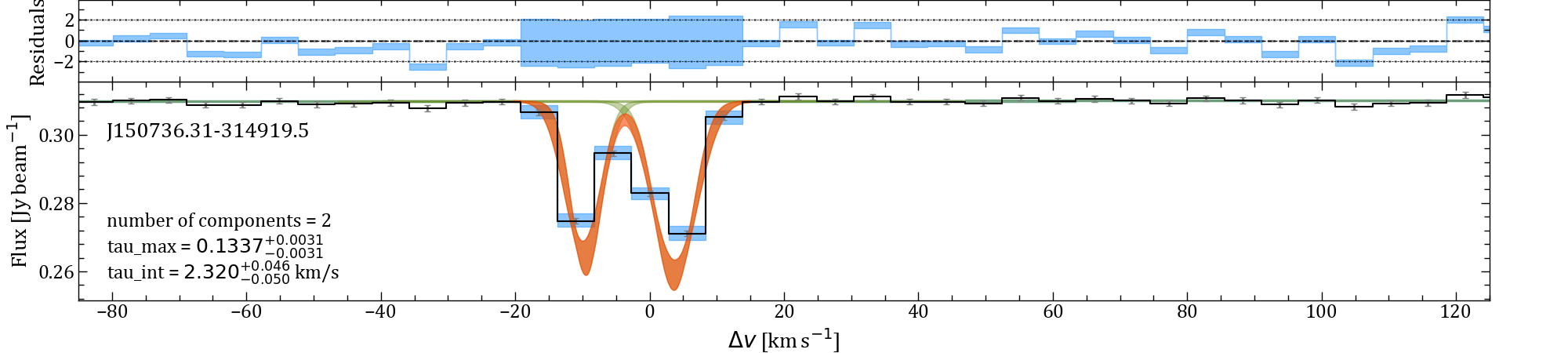} 
        }
    \caption{Comparison of the \hi\ line profile fit to the spectra toward the central and an off-axis (bottom) source belonging to the pointing -- J150712.88$-$313729.6. The graphical elements are same as in Fig.~\ref{fig:HI_fit_example}. 
    }
    \label{fig:HI_fit_example_comp}
\end{figure*}

We used an automatic procedure to detect \hi\ absorption lines and perform Gaussian decomposition of the observed profiles. This procedure involved the following steps. We selected a 500\,km\,s$^{-1}$-wide spectral window around the zero point velocity in the LSR system.  The selected velocity range may contain absorption lines either from the Galactic disk or the halo. We defined a model, $m$(v), consisting of the unabsorbed continuum, $C$(v), and an absorption line profile that may contain $n$ components contributing to the optical depth ($\tau$(v)) at a velocity (v), such that
\begin{equation}
    m({\text {v}}) = C({\text {v}}) e^{-\sum\limits_{i=0}^{n}\tau_i({\text v})}.
\end{equation}
%
For unabsorbed continuum we used a Chebyshev polynomial of the third degree, which was found to be adequate to represent the residual continuum subtraction uncertainties within the selected  region. We adopted a Gaussian profile to represent the optical depth of the individual absorption components, labeled as $i$ = 1, 2, .., $n$.  Naturally, each component is defined by three parameters: the amplitude ({\tt comp\_i\_amp}),  central position ({\tt comp\_i\_pos}) and dispersion ({\tt comp\_i\_sigma}). 
We note that Gaussian decomposition of complex profiles requiring multiple components may not be unique. But these are widely used as they provide, by experience \citep[][]{KalberlaHaud2018}, a useful representation of the \hi\ spectra to quantify the observed complexity of the absorption profile. They allow us to derive useful physical parameters such as kinetic temperature, and enable comparison with  various measurements across the literature.

Next, we calculated the model on a finer velocity grid to reduce the interpolation uncertainty and then summed the contributions of individual components within the spectral bins. The model spectrum was then compared with the observed spectrum using a standard likelihood function assuming that the pixel noise follows a Gaussian distribution. We also assumed the pixel uncertainty, $\sigma(\text{v})$, in the spectrum to be constant across the spectral region but allowed it to be an independent parameter in the fitting process. The assumption, $\sigma(\text{v})$ = a constant,  may not be strictly correct for all the sight lines. Especially at low latitudes, bright Galactic \hi\ emission may raise the system temperature, and hence increase the rms noise in certain velocity ranges.  We shall quantify and address this limitation appropriately in a future version of the catalog.

We fit the selected velocity range using a maximum likelihood optimization with Levenberg-Marquardt method. We iteratively increased the number of Gaussian components within the absorption profile, from 0 (i.e., no absorption line), until we found the best suited model using  Bayesian information criteria (BIC). The outcome may be a model with zero component, in which case we reported non-detection of \hi\ absorption. Finally, for the selected model with smallest BIC, we constrained the model parameters using sampling of the posterior probability function with affine-invariant Monte Carlo Markov Chain sampler, provided within {\sl emcee} package \citep{emcee2013}. We used a flat prior for all the model parameters, except the pixel uncertainty parameter, $\sigma$, for which we used appropriately chosen prior $\propto 1/\sigma$, assuming that it described the dispersion of the Gaussian noise in the spectral pixels. To report the point and interval estimates on the model parameters from the constrained posterior probability function, we used maximum a posteriori (MAP) probability estimate and the highest posterior density 0.683 credible interval, respectively. Examples of fit absorption profiles are shown in Figs.~\ref{fig:HI_fit_example} and \ref{fig:HI_fit_example_comp}.  These show a very good match between the absorption and emission line signals.  The latter provides a comparison between the central and an arbitrarily chosen off-axis sight line from the same pointing.

Using the constrained posterior functions of the model parameters and observed spectra, we estimated various spectral parameters, reported in the catalog (Table~\ref{tab:cat_cols}). The number of Gaussian components ({\tt n\_comp}) in column\,19 is directly derived from the fit. We defined the number of clumps ({\tt n\_clump}) in column\,20 detected along the sight lines as a number of grouped components that do not intersect each other's span. The span of a given component was simply defined as the MAP of the component's position $\pm3$ the MAP of its velocity dispersion. We constrained the maximal and integrated optical depth values -- {\tt tau\_max\_val} (column\,20) and {\tt tau\_int\_val} (column\,24) -- using the observed spectrum and sampling of the models obtained by the MCMC procedure as follows. We drew a random sample of the model parameters from the posterior distribution. For each model within the sample, we determined the spectral pixels that correspond to the spans of the fit components, as is described above. Then, we determined the width of the line profile as the maximal distance between these selected pixels, and using the spectrum we calculated the maximal and integrated optical depth, taking into account the pixel uncertainty. Finally, using the sampled values, we constrained the mean values of the maximal and integrated optical depth on the calculated samples, and derived the velocity span ({\tt delta\_v}; column\,27) of the absorber as 0.5 quantile of the sample of the widths. The errors on {\tt tau\_max\_val} and {\tt tau\_int\_val} are provided in columns\,21-22 and 25-26, respectively. By construction, since we sampled from the posterior distribution of the modes, the values obtained using the above procedure take into account continuum placement uncertainties.   We also determined the position of the maximal optical depth ({\tt pos\_max}; column\,23) as the position of the pixel that has the largest optical depth within the total line profile. 
The amplitude, position, and sigma of the individual Gaussian components are provided in columns\,28-36.  Naturally, these are repeated {\tt n\_comp} times.
%

\begin{figure*}
    \centering
    \vbox{
        \includegraphics[trim = {4cm 0cm 3cm 0cm}, width=1.0\textwidth,angle=0]{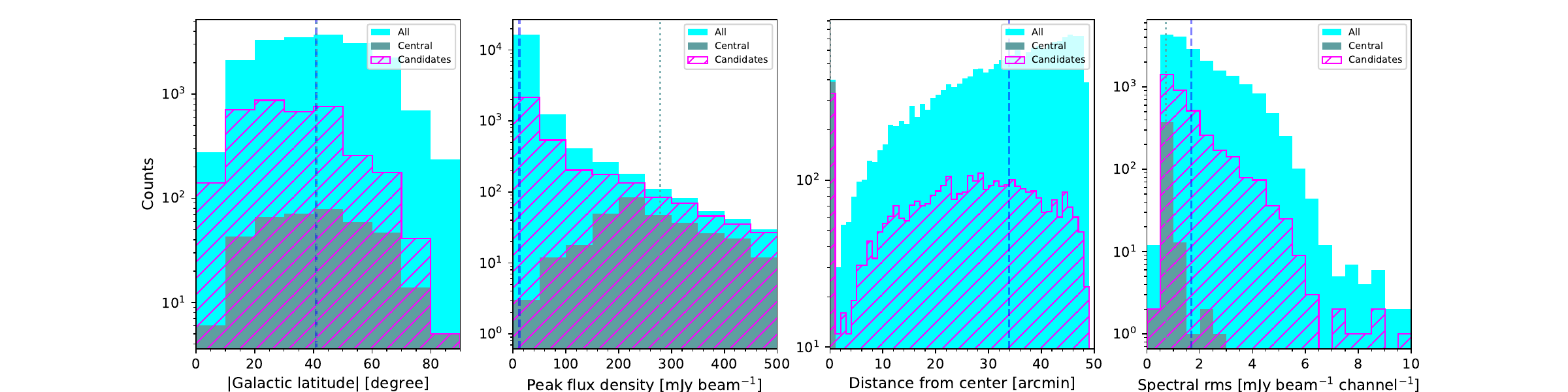}
    }
    \caption{
    Distributions of Galactic latitude ({\tt Peak\_pos\_b}), peak flux density ({\tt Peak\_flux}), distance from the pointing center ({\tt Distance\_pointing}) and  spectral rms ({\tt Spec\_rms}) for 19,130 sources (cyan) in logarithmic scales.  The gray histogram shows the bright central targets, within 60$^{\prime\prime}$ of the pointing center.    The vertical dashed and dotted lines mark median values for all sources and only central ones, respectively.
    The hatch-filled distributions represent candidates, i.e., detection counts uncorrected for false absorption features.
    For clarity in the second column, 176 targets with {\tt Peak\_flux} $>$ 500\,mJy\,beam$^{-1}$ have been omitted. 
    }
    \label{fig:samp}
\end{figure*}

Our automated procedure led to the detection of absorption features toward 3655 out of 19274 sight lines from 391 pointings.  The sight lines from one pointing (J213418.19-533514.7) are of poor quality.  Another 98 sight lines are duplicated due to overlapping pointings.  Excluding these, the number of sight lines and the detections considered hereafter are 19130 and 3640, respectively. 
The total number of Gaussian components corresponding to these is 4469.  In 2986 (82\%) cases, the feature is well fit with a single component,  with median FWHM = 7.8\,\kms\ ($\sigma$ = 3.3\,\kms).  Additionally, 529, 95, and 21 sight lines require two, three, and four or more components ({\tt n\_comp}), respectively.
Figure~\ref{fig:samp} shows basic statistical properties of these sight lines (cyan) and candidate detections (hashed) with respect to Galactic latitude, peak flux density of the background radio source, distance from the pointing center, and spectral rms noise.  
It may be noted that the detections presented in Fig.~\ref{fig:samp} and the full catalog are not examined for false or spurious detections arising from issues with the quality of the spectrum or simply statistical fluctuations.  Therefore, we refer to these as candidates to distinguish from the subset that is visually inspected and filtered for false positives as is described in Sect.~\ref{sec:purity}.  
Also shown in the figure are the distributions (gray histogram) of 390 sight lines toward the brightest source within $60^{\prime\prime}$ of the pointing center.  Owing to the survey design, these central sight lines generally represent the brightest source within the pointing. Consequently, these have the highest optical depth sensitivity and exhibit 322/390, a detection rate of $\sim$80\%, in comparison to $\sim$20\% for the full sample, which also includes off-axis lines of sight.
Besides this obvious dependence on sensitivity, a weaker dependence on Galactic latitude, in other words a higher detection rate at lower latitudes, is also apparent in Fig.~\ref{fig:samp}. Further, the sight lines with most complex absorbers, {\tt n\_comp}) $\geq$ 4, are at $-1.8^\circ$ latitude.  
We discuss the sky distribution of central and off-axis sight lines in Sect.~\ref{sec:sky}.

\subsection{Purity of the absorption line catalog}
\label{sec:purity}

A significant fraction of initial absorption-line candidates from the automated search may be false positives.  Here, we define a filtering approach to reduce the contamination by empirically identifying a S/N cutoff that maximizes the visually confirmed detections.  For this analysis, we consider 19130 sight lines from 390 pointings.

\subsubsection{Dependence on S/N}
\label{sec:purity-snr}

\begin{figure*}
    \centering
    \vbox{
    \includegraphics[trim = {0cm 0cm 0cm 0cm}, width=1.0\textwidth,angle=0]{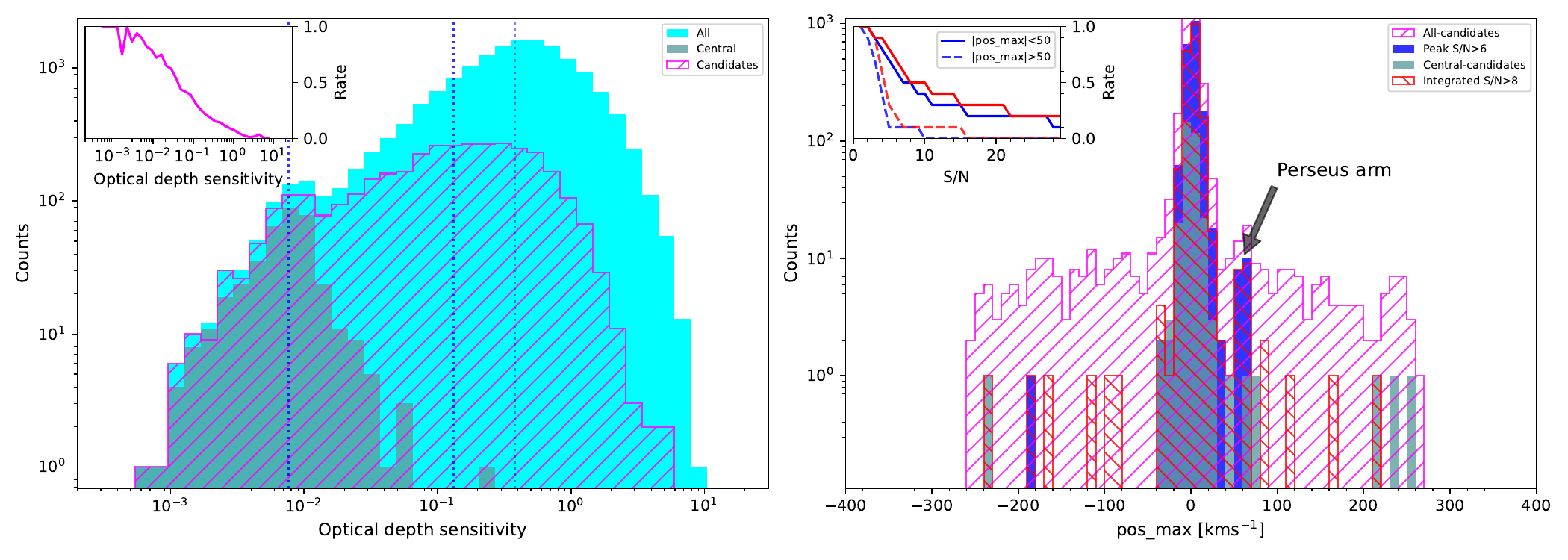}
    }
    \caption{
    Left: Optical depth sensitivity ($\tau_{3\sigma}$) for all (19130; cyan) and central (390; gray) sight lines. Hatch-filled (“/”) histogram is for sight lines (3640) with absorption candidates.  The vertical dotted lines represent median sensitivities for these three categories of sight lines.  The inset shows overall candidate detection rate. 
    Right: Positions of the absorption peak ({\tt pos\_max}) for all ('/'), peak S/N$>$6 (blue), integrated S/N $>$ 8 (red), and central (gray) candidates.  
    The inset shows detection rates as a function of S/N for candidates with $\vert${\tt pos\_max}$\vert$ smaller  and greater than 50\,\kms, respectively, for peak (blue) and integrated (red) optical depths. 
    The sharp cutoffs at |{\tt pos\_max}| = 250\,\kms\ correspond to the 500\,\kms\ width of line search window. An independent automated search over |{\tt pos\_max}| = 250-500\,\kms\ did not reveal any absorption features with peak S/N$>6$.  Only a marginal feature with $\int\tau$dv = 0.51\,\kms\ is detected toward J224543.63+024015.3 at $-$300\,\kms.
    }
    \label{fig:snr}
\end{figure*}

Fig.~\ref{fig:snr} (left panel) shows the distribution of 3$\sigma$ peak optical depth sensitivity ($\tau_{3\sigma}$), defined as $-log \rm (1 - 3 \times {\tt Spec\_rms} / {\tt Peak\_flux})$, \label{t3sig} for all the sight lines (median = 0.381). The inset shows that the candidate detection rate falls off with sensitivity. 
As was expected, higher sensitivity (median = 0.008) is achieved toward 390 central targets (gray histogram). The distribution of the position of peak optical depth, in other words {\tt pos\_max} of candidates detected toward these shows a prominent peak at $-0.9$\,\kms\  (median), mostly confined within +/-50\,\kms\ (Fig.~\ref{fig:snr}; right panel).  
The hatch-filled histogram (median = 0.131) in the left panel shows absorption candidates for all the sight lines.  The {\tt pos\_max} corresponding to these also exhibits a prominent peak at +0.8\,\kms\  (median) along with a uniform distribution of candidates over most of the velocity range searched for absorption features.
These candidates may hence include false detections due to noise fluctuations.

The inset in the right panel of Fig.~\ref{fig:snr} shows detection rate of candidates as a function of peak S/N defined as the ratio of {\tt tau\_max\_val} and 1$\sigma$ optical depth sensitivity.  Indeed, the occurrence of candidates beyond +/-50\,\kms\ falls off steeply with peak S/N. 
Based on this and the outcomes of visual inspection presented later, a peak S/N cutoff of 6 may be adopted as an optimal value.  This S/N cutoff reduces the number of candidates with $\vert${\tt pos\_max}$\vert$ $>$ 50\,\kms\ to 6\% at the expense of 40\% candidates with $\vert${\tt pos\_max}$\vert$ $<$ 50\,\kms.  The distribution of  2011 candidates with peak S/N $>$ 6 is also shown in the right panel of Fig.~\ref{fig:snr}. As was expected, the resultant distribution of {\tt pos\_max} is similar to that of the candidates toward the central targets.
This is statistically confirmed by a steady decrease in Wasserstein distance between the distributions of central and off-axis detections from 9.3 for no S/N cut to 2.4 for S/N $=$ 6.
We also examined the suitability of S/N based on integrated optical depth to detect absorption features.  The integrated S/N can be more robust against noise fluctuations and effective in detecting broad shallow absorption lines.  The inset in Fig.~\ref{fig:snr} (right panel) also shows variation in detection rate with respect to S/N based on integrated optical depth, estimated using {\tt tau\_int\_val} and rms in the spectrum smoothed by a kernel of the size of line FWHM.  It follows the same trend as the peak S/N except that the integrated S/N cutoff (8) is slightly higher.  The distribution of absorption systems selected using integrated S/N $>$8, also shown in Fig.~\ref{fig:snr}, is similar to those with peak S/N $>6$, with a Wasserstein distance between them of only 0.9.  The two-sample Kolomogorov-Smirnoff test (p-value = 0.98) implies that the null hypothesis that both the samples are drawn from the same distribution cannot be rejected.

Interestingly, the distribution of detections in Fig.~\ref{fig:snr} (right panel) exhibits a minor peak at $\sim$70\,\kms.  This peak can be attributed to two pointings, namely J074155.69-264729.8  ($l$ = 242.296012$^\circ$; $b$ = $-1.829431^\circ$) and J080622.15-272611.5  ($l$ = $245.653665^\circ$; $b$ = $2.497384^\circ$), at low galactic latitudes with 41 sight lines exhibiting significant absorption in multiple components over 0 - 100\,\kms.   In 18 cases, of which 14 belong to J080622.15-272611.5, the strongest component happens to be at $\sim$70\,\kms\ resulting in the above-mentioned minor peak.  The kinematic extent of \hi\ emission and absorption toward the center of the pointing, J074155.69-264729.8, is apparent in Fig.~\ref{fig:HI_fit_example}.  It is tempting to attribute the minor peak to the northern tip of the molecular ring between $0^{\circ} < l < 30^{\circ}$ \citep[][their Fig.~3]{Dame2001} that also has a  velocity of $\sim$70\,\kms.  However, MALS pointings in this longitude range have $+20^{\circ} < b < +90^{\circ}$.  This and the actual longitudes of J074155.69-264729.8 and J080622.15-272611.5 imply Perseus arm as the most likely origin of these features \citep[][their Fig.~3]{Vallee2017}. 
The isolated feature at $-180$\,\kms\ with peak S/N $>$ 6 is an HVC associated with the Magellanic bridge, and will be discussed in a future paper.  An automated search beyond 250-500\,\kms\ did not reveal any detections with peak S/N $>$ 6 (see Fig.~\ref{fig:snr} caption).
%

\begin{figure}
    \centering
    \vbox{
    \includegraphics[trim = {0cm 0cm 0cm 0cm}, width=0.45\textwidth,angle=0]{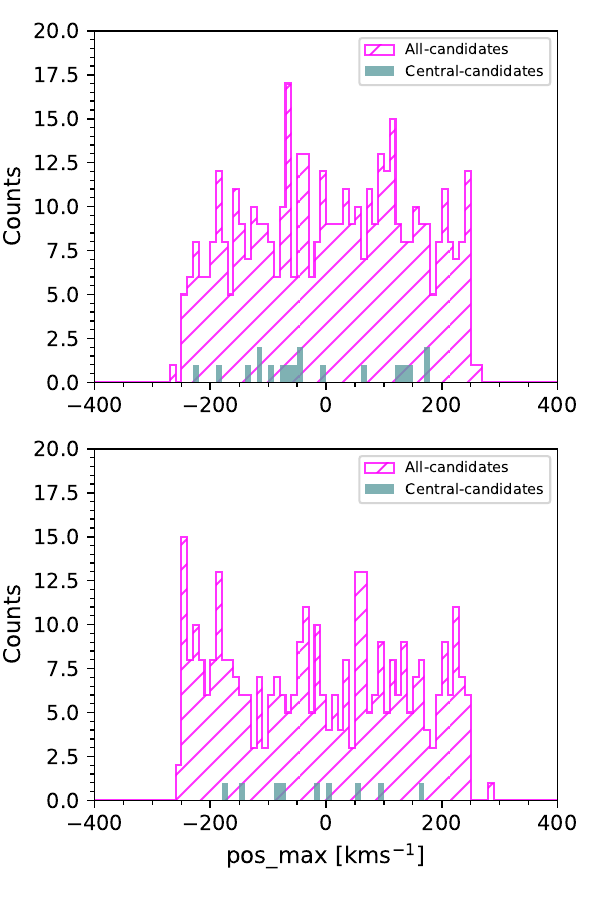}
    }
    \caption{Top: Positions of the absorption peak for an absorption line search centered on 1423.4\,MHz.  Only eight of these have peak S/N$>$6,  and none have S/N$>$7. Bottom: Same as the top panel but for the absorption lines searched in the negative (-1.0$\times$flux) spectra centered on 1423.4\,MHz. Less than 10\% of the absorption features have peak S/N$>$6 (see text for details).
    The blueshifted frequency range considered here is beyond all \hi\ 21-cm emission ever observed \citep{Putman2012}. 
    }
    \label{fig:false}
\end{figure}

We further demonstrate the efficiency of empirically determined  peak and integrated S/N cutoffs following two methods  using spectra covering $\pm$250\,\kms centered on 1423.4\,MHz; that is, away from the expected Galactic \hi\ signal. In this velocity range no Galactic and extragalactic \hi\ emission has been ever detected \citep[][; their Fig.~1]{Putman2012}. In the first method, we performed the automated line search on all the spectra. In the second method, we performed line search again centered on 1423.4\,MHz but on negative (i.e., -1.0$\times$flux) spectra (Fig.\,\ref{fig:false}).  In both cases, we do not expect any true absorption features. In the first case, 457 features overall and 18 toward the central sight lines are detected.  Only 15/457 have either peak S/N$>$6 or integrated S/N $>$ 8. In the second case, 361 and 9 absorption features are detected overall and toward central sight lines, respectively. Less than 10\% of these have S/N  above the cutoffs; in other words, the majority (90\%) are statistically insignificant. The slight excess of candidates in the first case is within statistical uncertainty.

Another interesting dependence on S/N is seen in the context of multiple detections within a pointing.  Without any S/N cut, in 306 pointings absorption is detected toward the central and at least one of the off-axis lines of sight.  In 14 pointings, no absorption is detected toward central and off-axis sight lines.  In 54 cases, while no absorption is detected toward the central source, at least one of the off-axis sources within the telescope's primary beam exhibits absorption.  In comparison, only 16 pointings exhibit the opposite; that is, the absorption is detected only toward the bright central source.   Indeed, considering only absorbers with S/N $>$6, as expected,  the number of pointings with no central absorption is 102.  These have, on average, fainter central targets (median peak flux density = 210\,mJy\,beam$^{-1}$) than the remaining sample. We note that 25 and 60 pointings show only an off-axis and central detection, respectively. In comparison,  228 pointings exhibit both central (median = 390\,mJy\,beam$^{-1}$) and off-axis detections. 
The central -- off-axis sight line pairs from these are of particular interest to investigate the small-scale structure of absorbing gas in Sect.~\ref{sec:galhioff}.

\subsubsection{Completeness fraction}
\label{sec:compl}

Applying a filter based on peak S/N$>$6 lowers the number of candidates from 3640 to 2011.  For integrated S/N $>$ 8, the number of candidates is 1842. To quantify the impact of S/N-based filtering on the completeness of the sample, we again consider the spectral range centered on 1423.4\,MHz.  We inject artificial absorption systems represented by single Gaussian components of widths, $\sigma$ = 1, 2, 4, 8, 12, 16 and 20\,\kms.  While the line width was chosen randomly, the peak optical depth was drawn from an exponential distribution such that at least 50\% of the injected absorbers have S/N in the range (1 $<$ peak S/N $<$ 10) relevant for the completeness analysis.   
We define the completeness fraction as the following,
\begin{equation}
    C(\sigma, {\rm S/N}) = \frac{1}{N_{\rm inj}}\sum_{i = 1}^{N_{\rm inj}} F(\sigma, {\rm S/N}),
    \label{eq:cmpl}
\end{equation}
where $N_{\rm inj}$ is the number of injected systems, and $F=1$ if the injected system is detected and 0 if not.
Figure~\ref{fig:absinj_stat} (left panel) shows distributions of line widths of injected systems and detections.  For the S/N cutoffs adopted for the MALS sample, the peak S/N performs better for systems with narrow line widths, $\sigma \leq$ 4\,\kms (FWHM $\leq$ 9.4\,\kms).  The distribution of line widths for detections in the MALS sample fit with single components is shown in right panel of Fig.~\ref{fig:absinj_stat}.

\begin{figure}
    \centering
    \vbox{
    \includegraphics[trim = {0cm 0cm 0cm 0cm}, width=0.45\textwidth,angle=0]{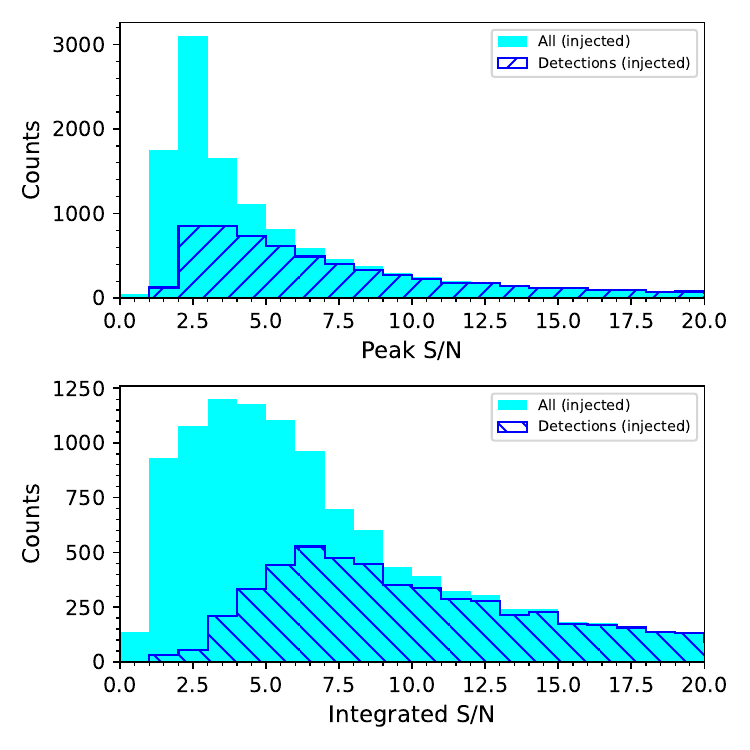}
    }
    \caption{Peak (top) and integrated (bottom) S/N of injected absorbers and detections.
    }
    \label{fig:inj_snr}
\end{figure}

\begin{figure}
    \centering
    \vbox{
    \includegraphics[trim = {0cm 0cm 0cm 0cm}, width=0.45\textwidth,angle=0]{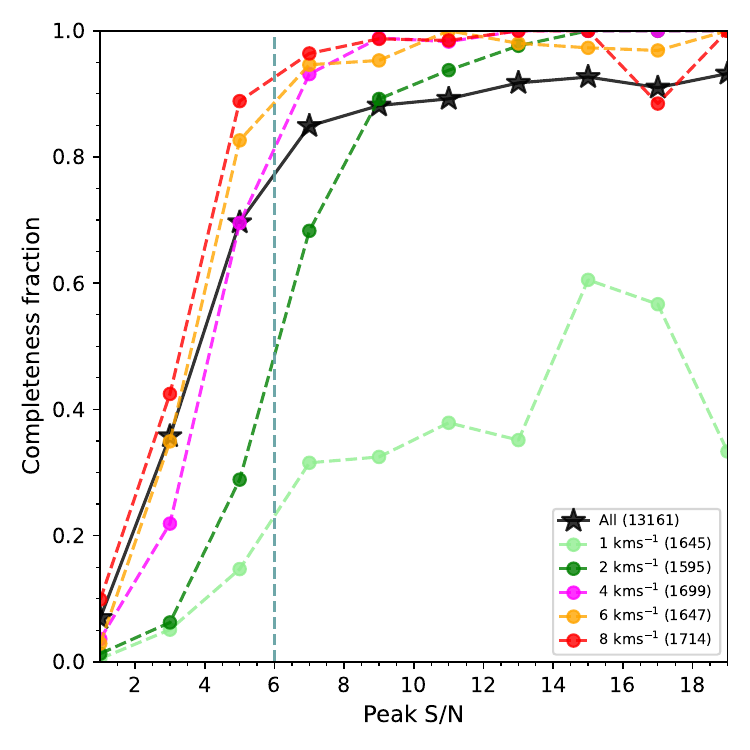}
    }
    \caption{Completeness fraction (Eq.~\ref{eq:cmpl}) as a function of peak S/N of injected absorbers. The number of injected absorbers are given in the legend specifying line width ($\sigma$). For clarity, the measurements for $\sigma >$8\,\kms, comprising $<10$\% of the MALS sample have been omitted.  The vertical dashed line corresponds to peak S/N cutoff of 6.
    }
    \label{fig:inj_cmp}
\end{figure}

About 80\% of injected absorbers with peak S/N $>$ 6  or integrated S/N $>$ 8 are detected  through our automated line search (see Fig.~\ref{fig:inj_snr}).  
In the MALS catalog, 82\% of the absorbers are fit with single Gaussian components with median $\sigma$ = 3.3\,\kms\ (FWHM = 7.8\,\kms), and only $\sim$10\% have $\sigma >$ 8\,\kms\ (FWHM $>$ 18.8).  Thus, the incompleteness in our catalog may primarily come from \hi\ absorption lines  close to or below the spectral resolution (5.5\,\kms) of the MALS setup.  These very narrow absorption lines ($\sigma\lesssim2$\,\kms\; ${\rm FWHM}\lesssim4.7$\,\kms) when combined with a low S/N are systematically confused with the thermal noise fluctuations in the spectra (see Fig.~\ref{fig:inj_cmp}).
Observationally, such narrow absorbers may be detected in higher-spectral-resolution observations but never alone. In 21-SPONGE survey, only 2 out of 57 sight lines have an absorber modeled as a single component with $\sigma\lesssim2$\,\kms \citep[][their Table~5]{Murray2018-sponge}.  MALS is optimized toward the high-galactic-latitude sky, and thus avoids massive molecular clouds hosting coldest gas phases. Therefore, such isolated narrow \hi\ absorption lines would be even less frequent within our sample.
Therefore,  on the basis of Fig.~\ref{fig:inj_cmp}, it is reasonable to assume an overall completeness of 90\% or better at peak S/N $>$6 for the MALS sample. 
An integrated S/N cutoff of 8 also leads to a similar completeness fraction but, as was noted above, 5\% less absorbers.  Otherwise, the subsets of detections in the MALS catalog with peak and integrated S/N cutoffs greater than 6 and 8, respectively, are nearly identical with a median $\sigma$ = 2.4\,\kms. 

\subsubsection{Visual inspection}
\label{sec:purity-vis}

The next step considered here is visual inspection of the candidates.  In principle, this is necessary for all the candidates,  even for sight lines marked as non-detections by the automated search. However, in order to keep the task manageable, we examine all 390 central sight lines but only off-axis sight lines with candidates. 
The key focus of visual inspection is to construct a “truth table” by simply displaying a spectrum and identifying absorption feature(s).  A candidate is marked 'True' if its visually identified position matches within $\pm$5\,\kms\ to the catalog value. 

First, we inspect 288 spectra corresponding to central candidates with peak S/N $>$ 6. All of these turn out to be credible detections and are marked as {\tt Vis\_flag} = {\tt True} in the catalog.  Notably, three sight lines are identified with as many as three clumps ({\tt n\_clump} = 3).  These are: J074155.66-264730.5 (see Fig.~\ref{fig:HI_fit_example}), exhibiting eight components spread over 0 - 100\,\kms; J021148.77+170722.8, in this case the third weak clump at -65\,\kms\ may be dubious, and J071046.99-381345.8, with the third weak clump at +40\,\kms.  The remaining sight lines exhibit simpler profiles, with 249 and 36 exhibiting {\tt n\_clump} = 1 and 2, respectively. The breakdown of an absorber into distinct clumps depends on the velocity resolution as well as the S/N. Therefore, in the catalog presented here we do not treat different clumps as physically distinct absorbers.  Consequently, the integrated optical depths ({\tt tau\_int\_val}) are summed over all the clumps. 
A side effect of this is that the velocity widths ({\tt delta\_v}) may be erroneous in the cases where the clumps are truly physically unassociated, or one of the clumps is a false detection. In principle, the contributions of different clumps can be easily segregated using the Gaussian component list.  However, caution is advised in using these estimates for cases (39/288) with {\tt n\_clump}$>$1.  The visual inspection of 68 central non-detections do not reveal any significant features.  These are marked as  {\tt Vis\_flag} = {\tt False} in the catalog.   Finally, we inspected 34 spectra with low-S/N (peak S/N$<$6) candidates.  Only four of these, with no specific dependence on S/N in the range of 2.6 - 4.0, appear to be misidentifications.  Three of these happen to have $\vert${\tt pos\_max}$\vert$ $>$ 70\,\kms.  The remaining 30 are marked {\tt True} in the catalog.  The overall exercise suggests that the S/N-based filtering may reduce the completeness by at least $\sim$10\%, reasonably consistent with the completeness fraction estimated in Sect.~\ref{sec:compl}.

\begin{figure}
    \centering
    \vbox{
    \includegraphics[trim = {0cm 0cm 0cm 0cm}, width=0.45\textwidth,angle=0]{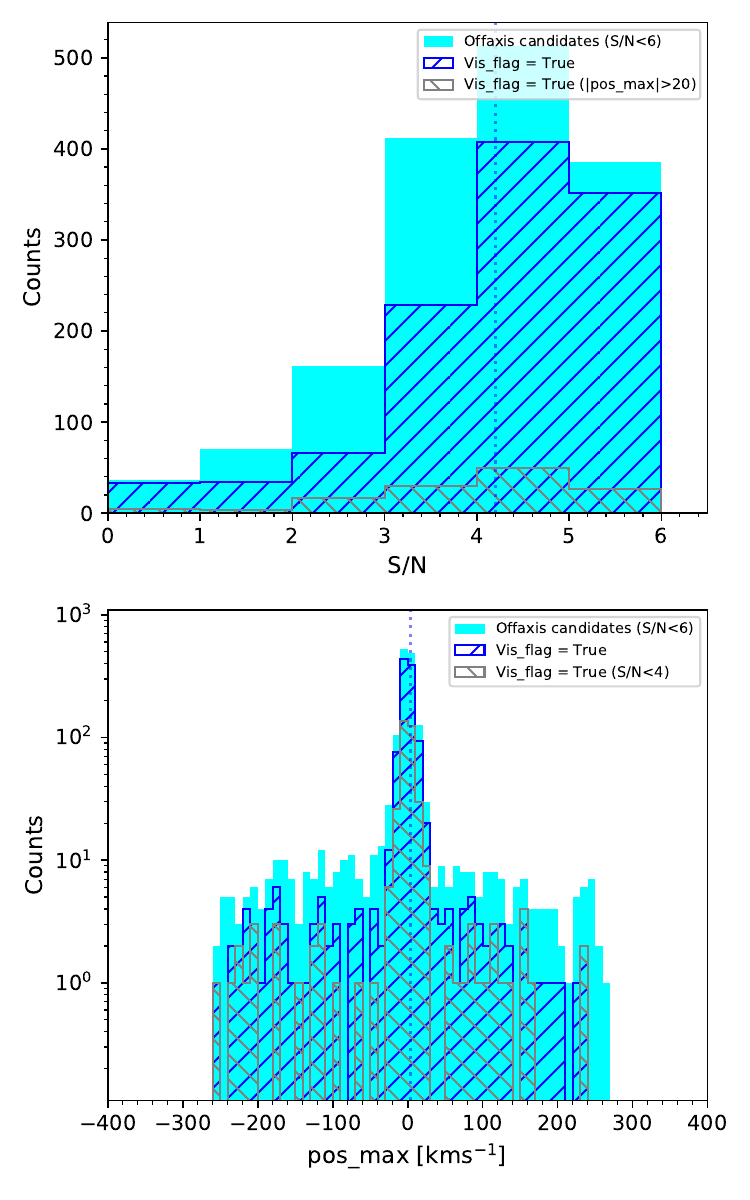}
    }
    \caption{Distribution of peak S/N (top) and position (bottom) for off-axis candidates with S/N$<$6, with vertical dotted lines representing the median values.  The subset of candidates with {\tt Vis\_flag = True} are shown as hatched histograms. The histograms in “$\backslash$” demonstrate that high-velocity ($\vert${\tt pos\_max}$\vert$ $>$ 20\,\kms) detections are not necessarily at low-S/N (see text for details).
    }
    \label{fig:offsnr}
\end{figure}

Next, we inspected off-axis sight lines.  We did not visually inspect 15421 off-axis non-detections, but left {\tt Vis\_flag} = {\tt Nan} for these.   Considering detections (3310), 1710 out of 1716 candidates with peak S/N $>$ 6 turn out to be credible upon visual inspection and are marked {\tt True} in the catalog. 
The remaining 1594 off-axis candidates have peak S/N $<$ 6.  Among these, 1130 (71\%) appear credible on visual inspection.  The top panel of Fig.~\ref{fig:offsnr} shows that the recovery of these candidates ({\tt Vis\_flag} = {\tt True}) is near perfect ($\sim$95\%) at S/N $>$ 4 but falls off for lower S/N except in the first bin (S/N: 0--1).
Figs.~\ref{fig:lowsnrdet-true} and \ref{fig:lowsnrdet-false} present examples of some of these low S/N detections.
With respect to the position of the absorption peak shown in the bottom panel of the figure, the recovery is in excess of 80\% near 0\,\kms\ but on average below 50\% at $\vert${\tt pos\_max}$\vert$ $>$ 20\,\kms. 
Based on the figure, it is interesting to note that the high-velocity “absorption features” are not necessarily at low-S/N.
In some cases, as discussed in Sect.~\ref{sec:purity-snr}, this could be due to a complex widespread absorption with strongest component at higher velocities.  However, the visual inspection revealed another aspect that at lower optical depth sensitivities and Galactic latitudes the peak S/N of absorption; that is, its detectability and parametrization may be affected by blending with the Galactic \hi\ emission.  This is consistent with the expectations from the simple radiative transport (Eq.~\ref{eq:radtrans}).  We shall model the underlying \hi\ emission within the MeerKAT beam and address this issue in a future revised version of the catalog. For now, caution is advised in using cataloged values for low-S/N detections. 

 Finally, we revisit the effectiveness of peak and integrated S/N cutoff in constructing reliable absorber samples.  Both cutoffs lead to absorber samples with  $\sim$99\% detections having {\tt Vis\_flag} = {\tt True}. But the former leads to a sample that is $\sim$10\% larger, possibly implying higher completeness.  Therefore, for further analysis, hereafter we consider the sample defined on the basis of peak S/N cutoff.

\section{Results}
\label{sec:res}   

\begin{table*}
\caption{\hi\ 21-cm absorption detections for different integrated optical depth limits.}
\vspace{-0.4cm}
\begin{center}
\begin{tabular}{cccccc}
\hline
\hline
 Optical depth     & Number of              & Number of           &   Percentage$^\ddag$ \\
 cutoff           &  sight lines$^\dag$    & detections$^\ddag$  &       \\
  (\kms)           &                 &             &    (\%)   \\   
 \hline          
0.02 &  71   &  63   (64)    &  89 $\pm$ 11      (90 $\pm$ 11) \\
0.05 & 320   & 268   (278)   &  84 $\pm$ 5       (87 $\pm$ 5)  \\
0.1  & 629   & 476   (509)   &  76 $\pm$ 4       (81 $\pm$ 4)  \\
0.2  & 1041  & 700   (777)   &  67 $\pm$ 3       (75 $\pm$ 3)  \\
0.5  & 2249  & 1113  (1327)  &  50 $\pm$ 2       (59 $\pm$ 2)   \\
1.0  & 4130  & 1497  (1900)  &  36 $\pm$ 1       (46 $\pm$ 1)   \\
2.0  & 7191  & 1794  (2475)  &  25.0 $\pm$ 0.6   (34.4 $\pm$ 0.7)  \\
5.0  & 12304 & 2001  (3010)  &  16.3 $\pm$ 0.4   (24.5 $\pm$ 0.5)  \\
10.0 & 14762 & 2011  (3137)  &  13.6 $\pm$ 0.3   (21.3 $\pm$ 0.4)  \\
20.0 & 15795 & 2011  (3149)  &  12.7 $\pm$ 0.3   (19.9 $\pm$ 0.4)  \\
40.0 & 15994 & 2011  (3151)  &  12.6 $\pm$ 0.3   (19.7 $\pm$ 0.4)   \\
\hline
\end{tabular}
\label{tab:detrate}
\end{center}
$\dag$: Number of sight lines with integrated 3$\sigma$ optical depth sensitivity better than the optical depth cutoff.  The integrated 3$\sigma$ optical depth sensitivity is estimated as $\tau_{3\sigma} \times 6.1$, where 6.1\,\kms\ is the median line FWHM of absorption detections. $\ddag$: Number and percentage of sight lines with peak S/N $>6$ absorbers.  The values in brackets are for absorbers with {\tt Vis\_flag} = {\tt True}, regardless of the peak S/N. 
\end{table*}

\begin{figure}
    \centering
    \vbox{
    \includegraphics[trim = {0cm 0cm 0cm 0cm}, width=0.45\textwidth,angle=0]{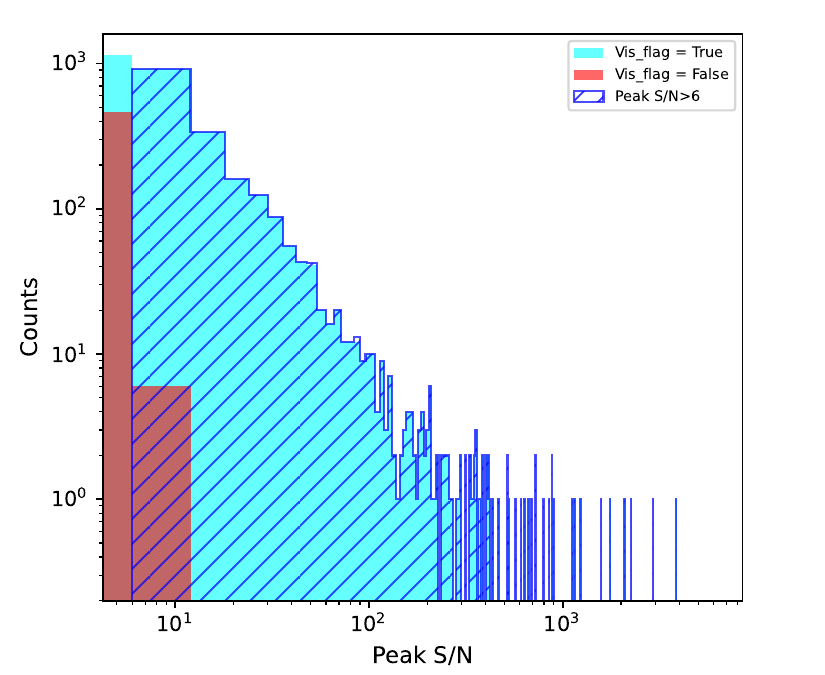}
    }
    \caption{Distribution of absorption detections with peak S/N$>$6 and, {\tt Vis\_flag} = {\tt True}  or {\tt False}.}
    \label{fig:detsnr}
\end{figure}

We detected 3640 unique Galactic \hi\ 21-cm absorption line features from 19130 lines of sight toward radio sources brighter than  1\,mJy at 1.4\,GHz.  
Of these, 3158 absorbers are visually confirmed and  2011 also have peak S/N $>$ 6 (Fig.~\ref{fig:detsnr}). The latter subset of absorbers has a purity of $\sim$95\%.
The median spatial resolution and 3$\sigma$ optical depth sensitivity, $\tau_{3\sigma}$, of the survey at the spectral resolution of 5.5\,\kms\ are $\sim9^{\prime\prime}$ and 0.38, respectively. 
The off-axis sight lines comprising the majority of the sample are toward fainter AGNs and also affected by primary beam attenuation.  Consequently, much higher optical depth sensitivity (median $\tau_{3\sigma}$ = 0.008) is achieved toward bright sources, typically brighter than 200\,mJy, at the center of the pointings.  The majority of the absorbers in the sample have simple profiles and are modeled using single Gaussian components (2986; 82\%).
The percentage of sight lines detected in absorption for a range of integrated optical depth cutoffs for absorbers with peak S/N$>$6 and {\tt Vis\_flag} = {\tt True} are provided in Table~\ref{tab:detrate}.  
 Overall only 18\% of the sight lines are detected in absorption, but 90\% are detected when only sight lines with highest sensitivity achieved in the survey are considered.  

In the following, we describe the sky distribution of the pointings and derive physical conditions in the gas.  Each MALS pointing offers several sight lines to explore Galactic \hi\ in absorption. 
We first focus on central sight lines to derive integrated properties of the neutral hydrogen gas over the telescope beam and toward the pointing center using the large scale detailed \hi\ 21-cm emission line and FIR radiation maps (Sect.~\ref{sec:galhicent}).  Later in Sect.~\ref{sec:galhioff}, using peak S/N$>$6 detections, we examine the differences in the properties of the absorbing gas toward the central and off-axis sight lines to investigate the parsec scale structure in the absorbing gas.

\subsection{Gas properties inferred from central sight lines}
\label{sec:galhicent}   

 \begin{figure*}
     \includegraphics[trim = {0cm 0cm 3cm 0cm}, width=0.75\textwidth,angle=0]{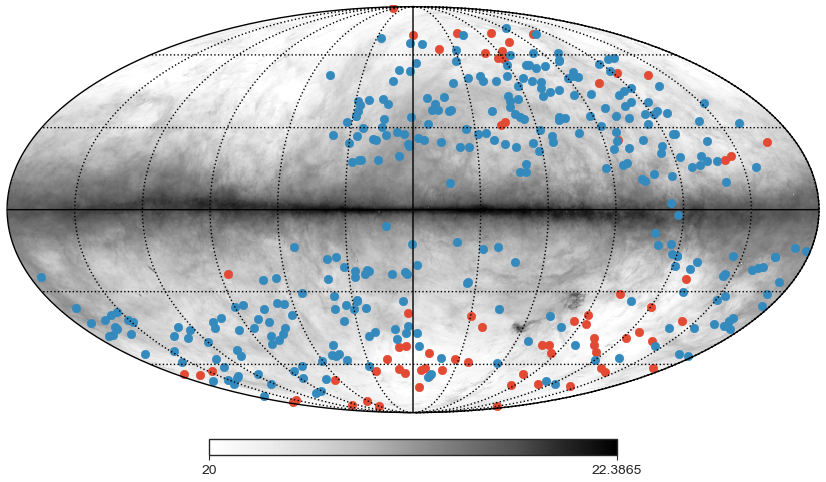}
     \caption{MALS pointings superposed on the HI4PI \hi\ column density distribution.  The map is centered on the Galactic center. For display purposes, the \hi\ column density is shown on a logarithmic scale. Each marker point in blue and red represents a MALS pointing with absorption line detection and non-detection toward central sight lines, respectively.
    }
     \label{Fig:MALS_HI4PI}
\end{figure*}

The key objective of this section is to utilize the absorption line properties derived in Sect.~\ref{sec:obscat} to quantitatively evaluate the physical conditions within the gas clouds; that is, volume density, gas temperature, and gas pressure. 
The optical depth is related to \hi\ column density ($N_{\mathrm{HI}}$) and $T_\mathrm {Spin}$ of the gas as $\tau$ $\propto$ $N_{\mathrm{HI}}$ / $T_\mathrm {Spin}$.
For a single homogeneous cloud, $N_{\mathrm{HI}}$ is given by  
\begin{eqnarray}
    N_\mathrm{HI} &=&  1.823\times 10^{18} \mathrm{cm^{-2}} \int T_\mathrm {Spin} \cdot \tau \cdot dv.
    \label{eq:nhi}
\end{eqnarray}
%

Phase transitions, either from neutral to the ionized or, more relevant here when exploring \hi\ absorption lines, to the molecular gas phase might deviate from the linearity, $\tau \propto \frac{\rho}{T_{\mathrm {Spin}}}$, implied by Eq.~\ref{eq:nhi} \citep{Fukui2014, Fukui2015, Murray2018}. 
Opacity effects can be investigated by cross-correlating the FIR radiation from interstellar dust grains and the \hi\ 21-cm line emission \citep{Clark2019, Lenz2019,Planck2014}. Dust grains exhibit the continuous spectrum of a modified black body \citep{Draine2003}. According to the Stefan--Boltzmann law, the radiation power, L,  depends strongly on dust temperature, T$_\mathrm{dust}$, as $L \propto T_\mathrm{dust}^4$. Therefore, we use the Schlegel, Finkbeiner and Davis survey data \citep[][hereafter SFD]{SFD1998} for calculating the optical extinction, $A_\mathrm{V} = R_\mathrm{V} E_\mathrm{B-V}$, where $E_\mathrm{B-V}$ is the difference in observed extinction between the Johnson B and V bands.  In the following we assume a standard Milky Way value of $R_\mathrm{V} =3.1$ \citep[][]{Gordon2003}.  The SFD reddening map is based on the FIR data from COBE/DIRBE and IRAS, covering the full sky at an angular resolution of $6\farcm1$. A dust temperature correction has already been applied to the SFD data.  We also applied the correction factor from \citet[][]{SchlaflyFinkbeiner2011} to the SFD data. In numerous studies, such as \cite{Lenz2019}, it has been shown that with the dust temperature correction, $\tau$ in the optically thin regime is linearly linked to the optical extinction $A_\mathrm{V}$, implying
\begin{equation}
     A_\mathrm{V} \propto N_\mathrm{HI}. 
     \label{Eq:DefExtinction}
\end{equation}
This may not be necessarily applicable to radio continuum sources luminous in FIR.  In such cases, one would overestimate the dust-to-gas ratio.  We note that the contributions of confirmed point sources have been removed from the SFD map \citep[see][]{SFD1998}.   Further, due to the low spatial resolution of the survey, the contribution from bright radio sources is smeared out.  We verified that even toward the MALS flux density calibrators the relationship in Eq.\,\ref{Eq:DefExtinction} is applicable.  Therefore, in general, the SFD map is suitable to calculate extinction toward MALS sight lines.

Single-dish high spectral resolution \hi\ emission line data may resolve the relative contributions of the different \hi\ phases along the line of sight. However, the coarse angular resolution does not allow for the spatial decomposition.
The earlier interferometric Milky Way absorption line studies used \hi\ emission data from the Leiden/Argentine/Bonn survey \citep[LAB; ][]{LAB2005}. 
LAB's advantage, even in comparison to surveys performed with much larger single-dish telescopes, was its correction against stray radiation \citep{Hartmann1996}, which significantly alters the inferred $N_\mathrm{HI}$ by observing date and  season.  Due to this, LAB has been identified as a valuable resource of information on the \hi\ column density toward the high Galactic latitude sky \citep[e.g.,][]{Kanekar2011, Roy2013, Roy2013b, Liszt2014a, Liszt2024}. The disadvantage of LAB is its coarse angular resolution. The sky has been sampled beam-by-beam according to the Shannon sampling theorem \citep[][]{Shannon1949} and the effective angular resolution is then only about a degree. The ISM structures of smaller angular scales are therefore interpolated. 
Here, we instead use \hi\ emission spectra from the HI4PI survey \citep[][]{HI4PI}, which combines the data from the Effelsberg–Bonn \hi\ Survey \citep[EBHIS;][]{Kerp2011, Winkel2016}  and the Galactic All-Sky Survey \citep[GASS;][]{Mcclure-Griffiths2009, Kalberla2015}.  HI4PI, like LAB, is also fully corrected for stray-radiation and the absolute flux density scale calibration has been performed by regularly observing the Milky Way's standard calibration areas S7, S8 and S9 \citep[][]{HI4PI}.
However, HI4PI's effective angular resolution is a factor of four higher than that of LAB. Even more importantly, the sky is fully sampled according to Shannon sampling theorem,  hence no spatial interpolation of structures larger than the beam is needed.  Consequently, all the ISM structures are measured down to the angular resolution limit of the Parkes 64-m aperture of $16\farcm4$. To determine the \hi\ column density we use data stored on a 1024 HEALPix grid \citep[][]{Gorski2005}, offering about $\Theta_\mathrm{pix}$ = $3\farcm44$ angular resolution per pixel. \citet[][]{KalberlaHaud2018} decomposed the \hi\ spectra into Gaussian components. Here we use their decomposition to evaluate the relative contribution of the individual \hi\ phases to a line of sight.  

We note that an obvious caveat of combining MALS absorption line measurements with HI4PI and SFD measurements is that the absorption is traced by AGN pencil-beam whereas $N_\mathrm{HI}$ and $A_\mathrm{v}$ measurements are smoothed over several arcminutes.  However, since the warmer phases in emission are more homogeneous than CNM, it is  still possible to reasonably compare the average column densities from emission and absorption. 

\subsubsection{Sky distribution of MALS pointings}
\label{sec:sky}

 \begin{figure*}
 \hbox{
   \includegraphics[width=8.5cm]{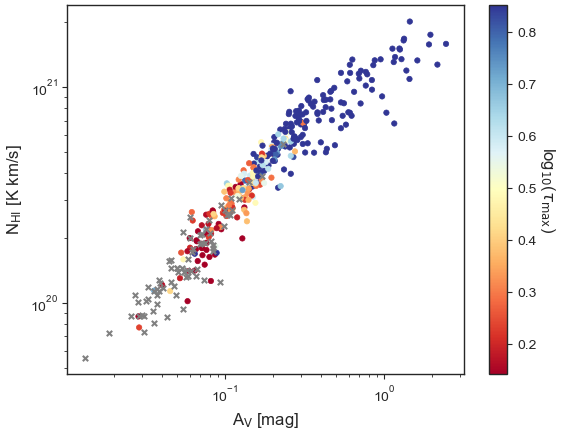 }
   \includegraphics[width=8.5cm]{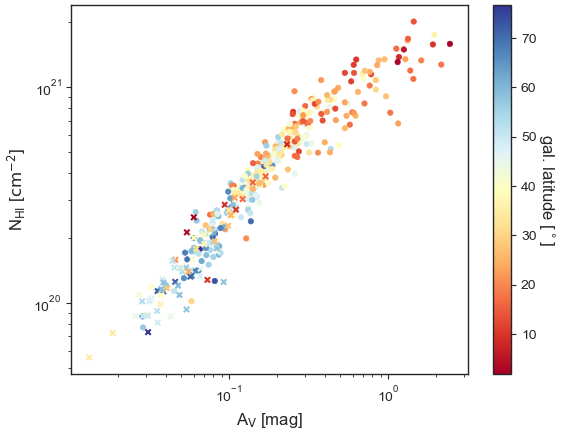 }
   }
    \caption{Double logarithmic plot displaying $N_\mathrm{HI}$ versus optical extinction $A_\mathrm{V}$ for central sight lines. The optical extinction is calculated with R(V) = 3.1 \citep{Gordon03} and applying correction factor from \cite{SchlaflyFinkbeiner2011} to the SFD data. $N_\mathrm{HI}$ is determined from HI4PI. The dots and crosses represent \hi\ absorption detections and non-detections, respectively. To guide the eye, the dashed line representing the best fit gas-to-dust ratio (Eq.~\ref{Eq:linreg}) is plotted in the left panel. The color coding represents the peak \hi\ optical depth ({\tt tau\_max\_val}; left panel) and the absolute value of Galactic latitude ({\tt Peak\_pos\_b}; right panel). 
    }
  \label{Fig:MALS_AV}
\end{figure*}

By design, extragalactic AGN surveys, in their exploration of temporal and spectral varying radiation, aim to avoid confusion by Galactic foregrounds. Consequently, AGN catalogs are generally biased toward the high Galactic latitude sky, thereby minimizing the possibility of luminous Milky Way synchrotron emission outshining fainter extragalactic radio sources \cite[e.g.,][]{Condon98}.  MALS, with its primary objective to detect extragalactic \hi\ and OH lines, has to compulsorily satisfy this requirement while finding a compromise for simultaneously fulfilling the necessities of other competing science objectives.  MALS footprint of $\sim$400 pointings is based on a larger pool of $\sim$650 potential targets described in \citet[][]{Gupta21salt}.  In general, the target pool tends to avoid low Galactic latitudes and is focused toward those gaseous structures where the “optically thin” approach for the Milky Way ISM as described above is applicable.  
We utilized the flexibility offered by the large pool of targets to favor the sight lines that explore the environment of the Magellanic Cloud system \cite[][]{Bruens2005} and the large scale Galactic structures, such as the \textit{eRosita\/} bubbles \cite[][]{Predehl2020} and the north polar spur \citep{Panopoulou2021}, without compromising the extragalactic science objectives. As part of this process, we also inspected the distributions of $N_\mathrm{HI}$  and $A_\mathrm{V}$  derived using the HI4PI \citep{HI4PI} and the SFD \citep{SFD1998} surveys.  In the end, we have a footprint that samples both the diffuse and translucent ISM phases of the Galaxy (see Sect.~\ref{sec:dtg}), but avoids the Galactic plane and dense molecular clouds. 

Figure\,\ref{Fig:MALS_HI4PI} displays the sky distribution of 390 MALS pointing centers. Blue dots mark pointings with central sight lines detected in \hi\ absorption while red dots mark non-detections. In total 322 ($83\pm5$\%) out of 390 central sight lines have been detected in absorption.  This is consistent with the 88\% detection rate of 21-SPONGE survey consisting of 57 sight lines from 48 targets observed with optical depth sensitivity, $\tau_{1\sigma}$ $<$ 0.001 per 0.42\,\kms\ \citep[][]{Murray2018-sponge}. In fact, applying the corresponding 3$\sigma$ optical depth  sensitivity cut ($\tau_{3\sigma}$ = 0.010) scaled to the MALS' spectral resolution of 5.5\,\kms, we get 253 out of 286 detections (89\%) in excellent match with 21-SPONGE.
68 central sight lines show no absorption lines.
Figure~\ref{Fig:MALS_HI4PI} shows that these are preferentially located toward low hydrogen column density regions, with an apparent preference toward the southern Galactic polar region. 
As will be illustrated below, these non-detections help us to obtain insights into the physical composition of the gaseous clouds detected in \hi\ absorption. 

Interestingly, \citealt[][]{Wakker2006} using a dedicated FUSE survey obtained a similar detection rate of  $\sim80$\% for H$_2$ absorption ($N$(H$_2$) = $10^{14-20}$\,\cmsq) in the ultraviolet range toward high galactic latitude ($|\rm b|>20^{\circ}$) extragalactic targets. As was discussed earlier, the presence of H$_2$ indicates  cold, $T\sim100\,$K, gas \cite[constrained using ortho-para ratio, see e.g.][]{Balashev2019}. While a one-to-one comparison of the targets from \citet[][]{Wakker2006} and our sample, their sky distribution and velocity profiles, is out of the scope of this paper, the similar detection rates for \hi\ and H$_2$ absorption from the two surveys suggest that in the solar vicinity (see Sect.~\ref{sec:distgas}) the sub-percent level of H$_2$ present in the gas traces well the CNM probed by \hi, at least for the quantities integrated along the line of sight. 
%

\subsubsection{The gas-to-dust ratio}
\label{sec:dtg}
%
Fig.\,\ref{Fig:MALS_AV} displays the scaling relation between the optical extinction $A_\mathrm{V}$ (SFD) and $N_\mathrm{HI}$ (HI4PI) for MALS pointing centers in a double--logarithmic plot according to Eq.\,\ref{Eq:DefExtinction}.  As was expected, the non-detections are preferentially located toward the low extinction and low column density portion of the diagram while at the other end of this linear relation only absorption line detections are observed. There is a significant overlap of detections and non-detections toward low column densities.  This holds true even for the subset of 286 sight lines with sensitivity, $\tau_{3\sigma}$ $<$ 0.010, same as 21-SPONGE (see Sect.~\ref{sec:sky}). 
%

In comparison to previous searches \citep[e.g.,][]{Kanekar2011}, MALS has enabled the detection of \hi\ 21-cm absorption lines to much lower $N_\mathrm{HI}$ and $A_\mathrm{V}$. Two lines of sight with $N_\mathrm{HI} \simeq 8\times 10^{19}\,\mathrm{cm^{-2}}$ or equivalent $A_\mathrm{V} \simeq 0.03$\,mag are detected in \hi\ 21-cm absorption (see also off-axis sight lines in Sect.~\ref{sec:galhioff}). In contrast, the highest column density of gas without an \hi\ absorption detection is $N_\mathrm{HI} \simeq 6.2\times 10^{20}\,\mathrm{cm^{-2}}$ or $A_\mathrm{V} \simeq 0.23$\,mag. While in the MALS sample of central sight lines the probability for detection of an \hi\ absorption line is on average about five times higher than for a non-detection, at the low $N_\mathrm{HI}$ and $A_\mathrm{V}$ end the non-detection probability is significantly enhanced (Fig.\,\ref{Fig:MALS_AV}).

We evaluate the correlation between $A_\mathrm{V}$ and $N_\mathrm{HI}$ according to Eq.\,\ref{Eq:DefExtinction}. As is displayed in Fig.\,\ref{Fig:MALS_AV}, in the range of $0.03 \leq A_\mathrm{V}\,\mathrm{[mag]} \leq 0.7$, we find 
\begin{equation}
     A_\mathrm{V} = \frac{N_\mathrm{HI}}{(2.21\,\pm 0.20)\times 10^{21} \,\mathrm{cm^{-2}}} - (0.04\,\pm 0.10)\,\,\mathrm{mag}
\label{Eq:linreg}
\end{equation}
with a correlation coefficient of 0.84 with a probability value of 0.72.  The slope corresponds to the gas-to-dust ratio of $N_\mathrm{HI} = (2.21 \pm 0.20)\times 10^{21}\,A_\mathrm{V} \, {\rm mag^{-1}\,cm^{-2}}$. 
This value is, within the uncertainties, equal to the one obtained by \cite{GueverOezel2009}. Within the 3$\sigma$ uncertainties, this value is also consistent with the measurements of \cite{Liszt2014a} and \cite{Lenz2017}.
\cite{Liszt2014a} set the limit for the maximal $A_\mathrm{v}$ beyond which the linear gas-to-dust relationship deviates to 0.22\,mag while \citet[][their Fig.~1]{Lenz2017} implied a much lower value of $A_\mathrm{v} = 0.14$\,mag. Both these values are much lower than the optical extinction range probed in MALS, with up to $E_\mathrm{B-V} \sim 0.8$\,mag ($A_\mathrm{V} \sim 2.5$\,mag). Below, we investigate deviations from the linear relation presented in Eq.\,\ref{Eq:linreg}. 

It is important to note that Fig.\,\ref{Fig:MALS_AV} displays, marked by crosses, the values of optical extinction and \hi\ column density for the sight lines with no \hi\ absorption detection.
Figure\,~\ref{Fig:Av_vs_NHI} displays these separately for detections and non-detections. The lines of sights with detection and non-detection follow the same general scaling law between optical extinction and \hi\ column density. However, some lines of sight with \hi\ absorption line detections have lower \hi\ column density than those with the non-detections.  This implies that $N_\mathrm{HI}$ and $A_\mathrm{V}$ may not be unique markers for the detectability of \hi\ absorption.  The different spatial resolutions of MALS, HI4PI and SFD, and the clumpiness of the CNM may also contribute to some mismatches and the overall scatter.  Nevertheless, the common scaling law suggests that the \hi\ gas in emission is exposed to comparable environmental conditions but the gas detected in absorption has a different density structure (see also Sect.~\ref{sec:turb}). 
%

\subsubsection{The distance to the absorbing gas}
\label{sec:distgas}

The gas and dust are known to be well mixed in the Galactic ISM \citep[e.g.,][]{Lenz2017}.  High Galactic latitude sight lines, such as those covered in MALS, preferentially pass through the LISM. These possibly probe the gas density structures exposed to a similar radiation field \citep[][their Fig.~5]{Wolfire03}, resulting in the remarkable linearity of $N_\mathrm{HI} \propto A_\mathrm{V}$ noted in Fig.\,\ref{Fig:MALS_AV}.
To further test the local origin hypothesis of the absorbing gas, we investigate two quantities associated with our dataset: first, the radial velocity distribution of the absorption features, and second, an increase in the column density and optical extinction as a function of galactic latitude. 

The radial velocity distribution of the absorption lines from central sight lines, is displayed in the right panel of Fig.~\ref{fig:snr}. With the exception of a handful of sight lines, all absorption features are within $-25 < v_\mathrm{LSR}[$\kms$] < 25$ range. 
In only 9 cases a significant (S/N$>$6) absorption component is detected at velocities, $\vert v_\mathrm{LSR} \vert$ $>$ 25\,\kms. Only 4 of these are at $\vert v_\mathrm{LSR} \vert$ $>$ 50\,\kms. 
This implies that the probed gaseous medium in general orbits the Milky Way's center with similar LSR velocity. 
The narrow width of the velocity distribution is remarkable and argues strongly for a local origin of the ISM probed by the MALS \hi\ absorption lines. A re-inspection of the high-latitude ($\vert b \vert > 10^\circ$) Nan\c{c}ay \hi\ line survey of \cite{Crovisier1978} by \cite{Wenger2024} revealed a nearly identical radial velocity distribution \citep[][their Fig.\,2]{Wenger2024}. From this narrow velocity distribution of the absorption features, for a Gaussian vertical distribution of \hi\ absorbing clouds they infer a scale height of $\sigma_\mathrm{z} = 61 \pm 9$\,pc. For an exponential distribution, also consistent with the data, the scale height, $\lambda_\mathrm{z} = 32 \pm 5$\,pc. The similar narrowly confined alignment of the radial velocity distribution of MALS absorbers indeed argues for a local origin of the \hi\ absorbing medium outside the local cavity of low-density hot gas. 

The right panel of Fig.~\ref{Fig:MALS_AV} represents the modulus of galactic latitude color-coded on top of the correlation between the \hi\ column density and the optical extinction. The high column density lines are observed toward low Galactic latitudes, while the low column densities are observed toward the Galactic poles. Clearly, the MALS data exhibit a smooth transition from high- to low-Galactic latitudes without an apparent break or discontinuity. This further confirms that the different lines of sight indeed form a homogeneous sample of neutral hydrogen clouds in the solar neighborhood. 
In future, the dataset presented here will be used to improve constraints on the vertical distribution of cold \hi\ gas in the local ISM of the Galaxy. 

\subsubsection{\hi\ opacity as a function of $N_\mathrm{HI}$ and $A_\mathrm{V}$}
\label{Sect:N_HI and A_V}

The high degree of linearity in the scaling relation between $A_\mathrm{V}$ (SFD) and $N_\mathrm{HI}$ (HI4PI), discussed in Sect.~\ref{sec:dtg}, further underlines the hypothesis that the MALS absorption lines are predominantly caused by gas exposed to very comparable physical conditions. At the lower and the upper extremes of $A_\mathrm{V}$ some inherent scatter is obvious. \citet[][their Fig.~2]{Basu2022} combined their Galactic plane measurements with those of the high Galactic latitude sky obtained by \cite{Roy2013} to investigate this.  While the integrated 21-cm optical depth, $\int \tau dv$  and the $N_\mathrm{HI}$ inferred from \hi\ emission exhibit a linear scaling relation, the situation is significantly different when considering $A_\mathrm{V}$ instead of $N_\mathrm{HI}$ \citep[see Fig.~3 of][]{Basu2022}. \cite{Basu2022} attribute this difference to the onset of the formation of molecular hydrogen out of the neutral atomic gaseous phase, which is especially relevant for the \hi\ absorption line data based on the THOR Galactic plane survey \citep[][]{Beuther2016}. 

\begin{figure*}
    \centerline{
    \hbox{
    \includegraphics[width=8cm]{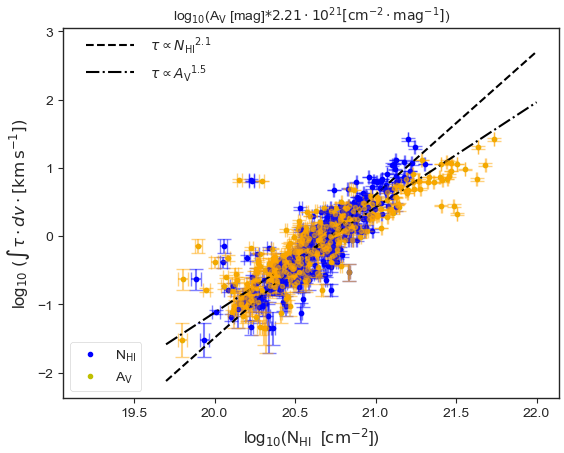}}
    \includegraphics[width=8.2cm]{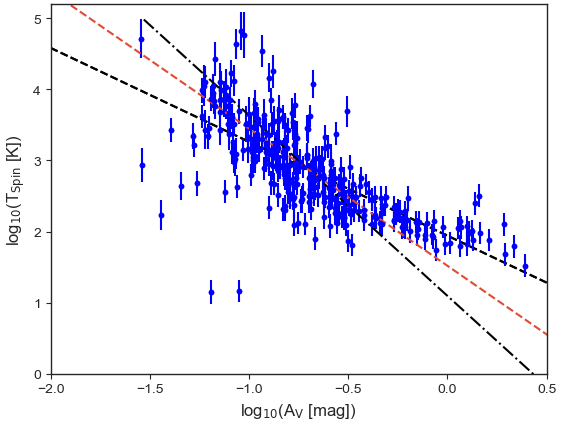}
    }
     \caption{Left: $\int \tau dv$\,(MALS) versus $N_\mathrm{HI}$ (HI4PI; blue dots) or $N^{\rm ext}_{\rm HI}$ = $A_\mathrm{V} \mathrm{[mag]}\cdot2.21\times10^{21}\,\mathrm{cm^{-2}}$ (yellow dots; based on Eq.\,\ref{Eq:linreg}) is plotted on a double logarithmic scale. The two regression lines, both with correlation coefficients of 0.89 and 0.85 respectively, are plotted as dashed and dash-dotted lines (see text for details). The error bars on \hi\ column density correspond to HI4PI $3\sigma$ detection levels, whereas the error bars in the y direction are based on uncertainties in {\tt tau\_int\_val}.
     Right: $T_\mathrm{Spin}$ (see Eq.~\ref{eq:nhimals}) versus $A_\mathrm{V}$ on a double logarithmic scale.  The dashed and dash-dotted lines mark the range of uncertainty by fitting a linear function to the data values. The dashed red line represents the average of both least-square approximations. 
     }
     \label{Fig:Tau_int_vs_AV}
\end{figure*}

Figure~\ref{Fig:Tau_int_vs_AV} (left panel; blue dots) shows $\int \tau dv$  and $N_\mathrm{HI}$ from \hi\ absorption and emission, respectively.   We note that $N_\mathrm{HI}$ from HI4PI used here are estimated in the optically thin limit and may represent true \hi\ column density only at low optical depths ($\int \tau dv$ <1).  At higher optical depths, specifically for $1< \int \tau dv < 10$ relevant here, the column densities are likely underestimated by as much as 30\% \citep[][their Fig.~6]{Kim2014}.
In the same panel $\int \tau dv$ versus $N^{\rm ext}_{\rm HI} \equiv A_\mathrm{V}\mathrm{[mag]} \cdot2.21\times10^{21}\,\mathrm{cm^{-2}}$\,(Eq.\,\ref{Eq:linreg}) is also shown (yellow dots). Consistent with \citet[][]{Basu2022} we find that the $\int \tau dv$ scales differently for $N_\mathrm{HI}$ (blue dots) and $A_\mathrm{V}$ (yellow dots).  Overall, we find  $\int~\tau~dv \propto N_\mathrm{HI}^{2.10\pm0.22}$ and $\int~\tau~dv \propto A_\mathrm{V}^{1.50\pm0.13}$ with correlation coefficients of 0.89 and 0.85, respectively (see Fig.~\ref{Fig:Tau_vs_NHI_AV}). 
Both the regression lines cross each other at $\mathrm{log_{10}} (N_\mathrm{HI}\mathrm{[cm^{-2}]}) = 20.7$, corresponding to $A_\mathrm{V} = 0.23$\,mag.  At the low $N_\mathrm{HI}$ or equivalent $A_\mathrm{V}$ end the distributions between the $N_\mathrm{HI}$ and $A_\mathrm{V}$ do not differ, and we find  $\int~\tau~dv \propto N_\mathrm{HI}^2 \propto A_\mathrm{V}^2$. The situation is different at the upper column density or optical extinction end, specifically, above $N_\mathrm{HI} \simeq 1\times 10^{21}\,\mathrm{cm^{-2}}$ or $A_\mathrm{V}\simeq1$\,mag, inferred by-eye. It implies that this is the threshold in column density and optical extinction when dust starts to trace not only $N_\mathrm{HI}$ but total hydrogen column $N_\mathrm{H}$, in the local ISM.  The inferences drawn here remain valid when we consider only the 286 most sensitive ($\tau_{3\sigma}<$0.010) central sight lines. 
%

The combination of \hi\ emission ($T_\mathrm{B}(v)$) and absorption ($\tau(\nu)$) spectra may be used to derive line-of-sight average spin temperatures.  The most common method is to calculate the optical-depth-weighted mean spin temperature <$T_\mathrm{Spin}$> as
\begin{equation}    
    <T_\mathrm{Spin}> = \frac{\int \tau(v)  T_\mathrm{s}(v) dv}{\int \tau(v) dv} \nonumber = \frac{\int \tau(v)  \frac{T_\mathrm{B}}{(1-e^{-\tau(v)})} dv} {\int \tau(v) dv}. \nonumber \\
    \label{eq:tspin}
\end{equation}
This has has been shown to be a reasonable estimator of the gas spin temperature \citep[][]{Kim2014}.
Here, we rather estimate line-of-sight average spin temperature $T_\mathrm{Spin}$ for each sight line using  
\begin{equation}
    T_\mathrm{Spin} = \frac{N_\mathrm{HI}(\mathrm{HI4PI})}{1.823\cdot 10^{18}\cdot \int\tau. dv(\mathrm{MALS})},
    \label{eq:nhimals}
\end{equation}
where the $N_\mathrm{HI}$ measurements from HI4PI have not been corrected for the optical depth effects. 
%
We note that neither $<T_\mathrm{Spin}>$ nor $T_\mathrm{Spin}$ represents physical temperature of the gas.  Nevertheless, for a two-phase medium $T_\mathrm{Spin}$ estimated this way can be related to the fraction of cooler phase responsible for the absorption.  The apparent relationship between $T_\mathrm{Spin}$ and $A_\mathrm{V}$  in Fig.\,\ref{Fig:Tau_int_vs_AV} (right panel) implies that the CNM fraction as expected increases for higher $A_\mathrm{V}$.  
In a future paper, \hi\ emission line spectra in the vicinity of absorption sight lines from the MALS spectral line cubes will be used to determine $<T_\mathrm{Spin}>$ and investigate CNM fraction as a function of dust extinction.  This will also address systematics in measurements due to the vastly different spatial resolutions of emission and absorption line measurements used here, and small-scale structure in the absorbing gas presented in Sect.~\ref{sec:galhioff}).  In the following section, we use gas phase decompositions of HI4PI spectra by \cite{KalberlaHaud2018} to investigate the detectability of \hi\ absorption.

\subsubsection{Phase transitions}
\label{sec:phase}

Fig.~\ref{Fig:Tau_int_vs_AV} exhibits a tight linear relation between the small scale \hi\ absorbing gas column and the more largely extended gas structure measurable with the coarse beam of a single dish radio telescope. A spatially tiny, isolated clump with a high volume density has a very small beam--filling factor for such a single dish telescope ($\Omega_\mathrm{Clump}/\Omega_\mathrm{beam}$). For emission-absorption comparison to be reasonable, the gas volume causing the \hi\ absorption line needs to share similar physical conditions with its more largely extended enveloping medium. Therefore, observed stronger linear correlation can be interpreted as evidence that the CNM and LNM form out of the WNM \citep{KalberlaKerp2009} and thus show up with comparable excitation conditions for \hi\ 21-cm line \citep[][]{Field1959}.  The narrowly confined radial velocity spread displayed in Fig.~\ref{fig:snr} (right panel) is also a consequence of this. This is also evident from the very strong correlation of \hi\ absorption and emission profiles for the majority of lines of sight, as is shown in  Figs.~\ref{fig:HI_fit_example} and \ref{fig:HI_fit_example_comp}. However, one would expect that the medium causing the \hi\ absorption lines ``freezes out'' of the warm enveloping gas because of momentum conservation with the same radial velocity of the gas bulk motion. To further investigate this hypothesis, we studied the CNM, LNM, and WNM composition of gas using the single-dish \hi\ spectra in greater detail.

\cite{KalberlaHaud2018} decomposed \hi\ 21-cm lines detected in the HI4PI spectra using the Gaussian components and deduced averaged values of doppler or maximum kinetic temperature ($T_\mathrm{D}$) in the absence of non-thermal broadening  for WNM ($T_\mathrm{D} \sim 2000\,-10000$\,K, FWHM $\simeq 24\,\mathrm{km\,s^{-1}}$) and CNM ($T_\mathrm{D} \sim 20\,-400$\,K, FWHM $\simeq 5\,\mathrm{km\,s^{-1}}$ ). In addition to these two stable phases of the ISM they deduced that 41\% of the sky shows up with LNM \citep[][]{Heiles03}, an unstable gaseous phase which is never observed as an isolated single phase \citep{KalberlaHaud2018}.  The LNM's FWHM, determined to be $\simeq 11\,\mathrm{km\,s^{-1}}$, is in between the FWHMs for CNM and WNM. According to \citet{KalberlaHaud2018}, LNM forms a gaseous interface located spatially in between the WNM and the CNM. Consequently, these share the environmental conditions with respect to heating and cooling  but posses different volume densities \citep{Wolfire03}.  Indeed, a negative spatial correlation of CNM and LNM with WNM is observed \citep[][their Fig.~13]{KalberlaHaud2018}.
%

\begin{figure*}
    \centering
    \includegraphics[width=8cm]{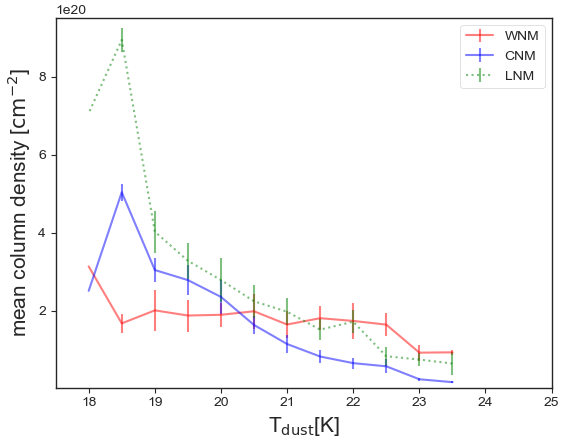}
    \includegraphics[width=8cm]{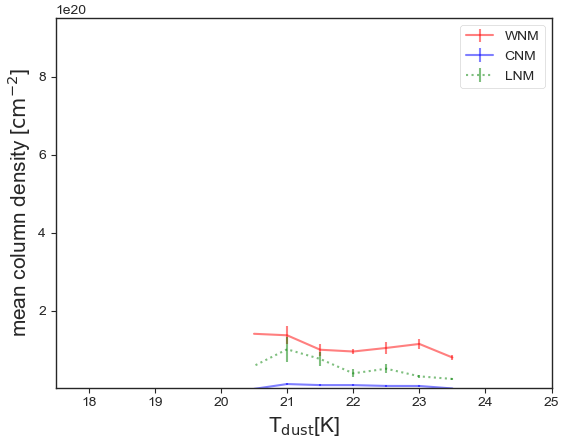}
     \caption{$N_\mathrm{HI}$ of WNM, CNM, and LNM over the column density interval $0.03\cdot10^{20} \leq N_\mathrm{HI}\,\mathrm{cm^{-2}}\leq 9.5\cdot10^{20}$ according to \cite{KalberlaHaud2018} versus $T_\mathrm{dust}$ in bins of 0.5\,K for central sight lines with \hi\ 21-cm absorption detections (left) and non-detections (right).  The error bars represent the scatter as the median absolute deviation (MAD), corresponding to 0.68$\sigma$ for normally distributed data. 
     }
     \label{Fig:Phases}
\end{figure*}

Since opacity, $\tau$, is related to $N_\mathrm{HI}$/ ${T_{\mathrm {Spin}}}$, one would expect that a high CNM fraction is essential to detect an absorption line for the same total columns of the gas. According to \cite{Wolfire03}, a high CNM fraction is associated with a low kinetic gas temperature, a high optical extinction and a low dust temperature. 
The temperature of dust grains is linked to the strength of the ultraviolet field (U) as $T_\mathrm{dust} \propto {U^{1/6}}$ \citep[see e.g.][]{Draine2011}.   At lower $T_{\rm dust}$, the integrated ultraviolet flux is also lower which thereby lowers the pressure limit and hence the number density limit at which the CNM can exist \citep[see e.g.][]{Bialy2019}.
Thus, we relate in the following the fractions of different atomic gas phases; that is, $N_\mathrm{CNM/LNM/WNM}/N_\mathrm{HI}$ as a function of dust temperature ($T_\mathrm{dust}$).  The latter is based on the full sky dust temperature map determined by the Planck consortium \citep{Planck2014dust}. 
Thus, for each MALS central sight line, we calculated the dust temperature, $T_\mathrm{dust}$, from the Planck data and the fractions of the WNM, LNM, and CNM from the decomposition of HI4PI spectra by \cite{KalberlaHaud2018}.

Figure\,\ref{Fig:Phases} displays the relative contributions of the three \hi\ gas phases to $N_\mathrm{HI}$ as a function of $T_\mathrm{dust}$ for central sight lines. We calculated these values for the dust temperature range relevant to our \hi\ data, $ 18 \leq T_\mathrm{dust}\,\mathrm{[K]} \leq 25$, in steps of $\Delta T_\mathrm{Dust} =$\,0.5\,K.  Both, detections (Fig.\,\ref{Fig:Phases}, left panel) and non-detections (right panel) share the same upper limit for the dust temperature of $T_\mathrm{dust} = 23.5$\,K. At this upper dust temperature limit, for both detections and non-detections, the WNM dominates over the LNM and CNM in column density.
\cite{Roy2013} reported that some of the gas detected in \hi\ 21-cm absorption must arise from gas warmer than CNM; that is, LNM or WNM. This is consistent with our findings here that the CNM is not the dominant phase in low column density (Fig.\,\ref{Fig:Phases}, left panel) and rather warm environments (Fig.\,\ref{Fig:dusttemp}). Notably, all the MALS sight lines with \hi\ absorption have $N_\mathrm{HI}({\rm CNM})\,+\,N_\mathrm{HI}({\rm LNM}) \geq N_\mathrm{HI}({\rm WNM})$. This is markedly different for the sight lines with no \hi\ absorption detections.
In view of the phase transition from atomic to molecular hydrogen at the lowest dust temperatures, Fig.\,\ref{Fig:Phases} might be interpreted in the way that the CNM is the donor phase for molecular hydrogen but not the LNM. The dominating LNM column density may be only an apparent feature, caused by the fact that the \hi\ line measurements are limited only to the atomic phase. In future, MALS \hi\ and OH absorption line measurements will be used to test these assertions, and derive mass distributions of CNM, LNM, and WNM  currently based on the decomposition of emission lines.

For completeness, we note that toward the flux density and bandpass calibrators -- that is, PKS\,1939–638 and PKS\,0408–658 -- discussed in Sect.~\ref{sec:obs}, the WNM/LNM/CNM column densities are 1.6/1.6/1.8 $[\cdot 10^{20}\,cm^{-2}]$ with $A_\mathrm{V}$ = 0.23\,mag at $T_\mathrm{dust}$ = 20.4\,K and 1.6/0.8/0.4 $[\cdot 10^{20}\,cm^{-2}]$ with $A_\mathrm{V}$ = 0.14\,mag at $T_\mathrm{dust}$ 20.3\,K, respectively.  This indicates that there is not a one-to-one correspondence between $A_\mathrm{V}$ or  $T_\mathrm{dust}$ measurements and gas phase. 

\subsubsection{The lines of sight with no \hi\ absorption}
\label{sec:noabs}

\begin{figure}
    \includegraphics[width=8cm]{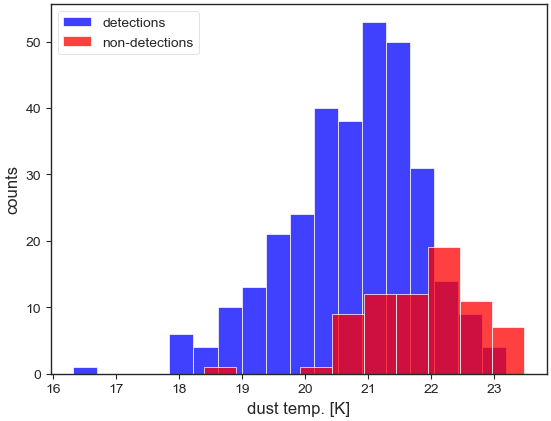}
     \caption{Distribution of \hi\ absorption detections (blue) and non-detections (red) as a function of the dust temperature. }
     \label{Fig:dusttemp}
\end{figure}
%
\begin{figure}
    \centering
    \hbox{
    \includegraphics[trim = {0cm 0cm 0cm 0cm}, width=0.45\textwidth,angle=0]{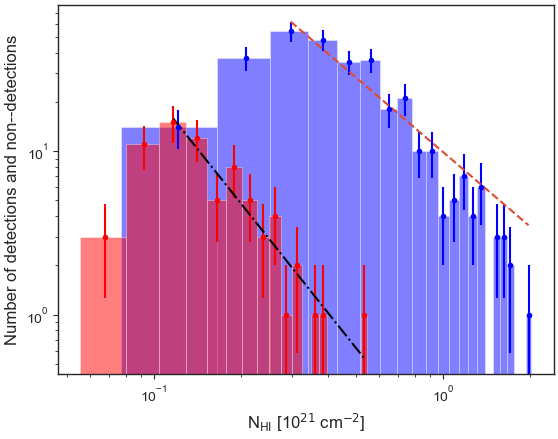}
    }
    \caption{
    Number (N) of \hi\ absorption line detections (blue) and non-detections (red) as a function of $N_\mathrm{HI}$ column density. The error bars correspond to $\sqrt{N}$. The dashed red and dash-dotted black lines represent the best fits. 
    } 
    \label{fig:t3sig}
\end{figure}

There are 72 central lines of sight that do not show an \hi\ absorption line. As Fig.\,\ref{Fig:MALS_AV} displays, these tend to occupy low $N_\mathrm{HI}$ and $A_\mathrm{V}$ end of the distributions.  These also fit the tight scaling relation between $N_\mathrm{HI}$(HI4PI) -- $A_\mathrm{V}$ determined from sight lines with absorption detections (Fig.~\ref{Fig:Av_vs_NHI}).
As is shown in Fig.\,\ref{Fig:Phases}, the dominance of the WNM across the whole dust temperature range is the major difference. The dominance of warmer environment toward sight lines with absorption non-detections is also supported by the distribution of $T_\mathrm{dust}$ presented in Fig.\,\ref{Fig:dusttemp}. As was discussed in the previous subsection, the higher dust temperatures for non-detections in comparison with detections can be explained using the factors associated with the ultraviolet flux variations.

\subsubsection{\hi\ absorption line detections and turbulence}
\label{sec:turb}

Fig.~\ref{Fig:MALS_AV} shows that the majority of central absorption line detections are toward rather low column densities with a median of $N_\mathrm{HI}$(HI4PI) = $0.45\times 10^{21}\,\mathrm{cm^{-2}}$. 
They also show up with low \hi\ opacities, with a median $\int\tau dv = 0.69$\,\kms  (Fig.~\ref{Fig:Tau_int_vs_AV}).  Notably for $\tau_{3\sigma}$ as quantified in Sect.\,\ref{t3sig}, all except five central sight lines with absorption non-detections have $\int\tau_{3\sigma}$dv $<$ 0.69\,\kms, implying sufficient sensitivity to detect the typical \hi\ 21-cm absorption line with $N_\mathrm{HI}$ = 1.3$\times10^{20}$\,(100\,$\mathrm{K}$/$T_{\rm Spin}$)\,\cmsq.
Figure~\ref{fig:t3sig} shows the  $N_\mathrm{HI}$ distributions for central sight lines with absorption  detections (blue) and  non-detections (red). The uncertainties are estimated according to $\sqrt{N}$. We approximated the two distributions as power laws with slopes of $-1.53 \pm 0.13$ and $-2.22 \pm 0.22$ for the detections and non-detections, respectively. 
The power law slopes of the detections and non-detections are different at only $>2\sigma$ level. Interestingly, a similar difference is also seen in the slopes of $N_\mathrm{HI}$ for high-$z$ ($z\gapp$1.8) damped Lyman-alpha systems (DLAs) with and without H$_2$ detections \citep[see e.g.,][]{Balashev2018}, with a steeper slope of the latter. Since H$_2$ as well as \hi\ 21-cm absorption are tracers of the CNM, this indicates an important similarity in factor(s) driving the physical conditions in the gas governing the detections of these absorption lines.
We note that the DLA samples by definition trace only $N_\mathrm{HI}$(Ly$\alpha$) $\geq 2\times10^{20}$\,\cmsq.

In case of high-$z$ DLAs, the differences in slopes and  normalizations of  $N_\mathrm{HI}$ are usually attributed to the properties of the \hi-to-H$_2$ transition \citep{Noterdaeme2015}. The above-mentioned similarity between $N_\mathrm{HI}$ distributions of \hi\ 21-cm and H$_2$ absorption suggests that both may have a common driver: the WNM-to-CNM transition.
Indeed, as was mentioned in Sect.~\ref{sec:sky}, while the detection rate of H$_2$ in the Galaxy is similar to the MALS \hi\ 21-cm absorptions, the typical H$_2$ column densities concerned in the Galactic and DLA samples are certainly lower than that needed for \hi-to-H$_2$ transition. This leads to the conclusion that H$_2$ is likely only a tracer of the CNM in high-$z$ DLAs rather than governing the physical state of the gas over the column density ranges considered here.  
In the scenario that the extended warm gas is exposed to similar physical conditions, the most reasonable factor driving the WNM-to-CNM transition is interstellar turbulence \citep{FalcetaGon2014}. The turbulence may compress gas at various physical scales, expediting the cooling and hence freezing-out of CNM clouds from the WNM. 
The $N_\mathrm{HI}$ distributions may represent the turbulent density structure of CNM clouds provided the turbulence correlation lengths of the CNM and the WNM are same.  Another requirement is that the detectability of \hi\ or H$_2$ absorption not be a limiting factor.  Based on Fig.\,\ref{fig:t3sig}, this is certainly the case for \hi\ 21-cm absorption toward central sight lines.   However, a one-to-one comparison of slopes and normalizations of  $N_\mathrm{HI}$ distributions of the Galactic and high-$z$ absorber samples involving investigation of departures from these requirements is beyond the scope of this paper.

\subsection{Small-scale structure inferred from off-axis absorbers}
\label{sec:galhioff}   
%
\begin{figure*}
    \centering
    \hbox{
    \includegraphics[trim = {0cm 0cm 0cm 0cm}, width=0.45\textwidth,angle=0]{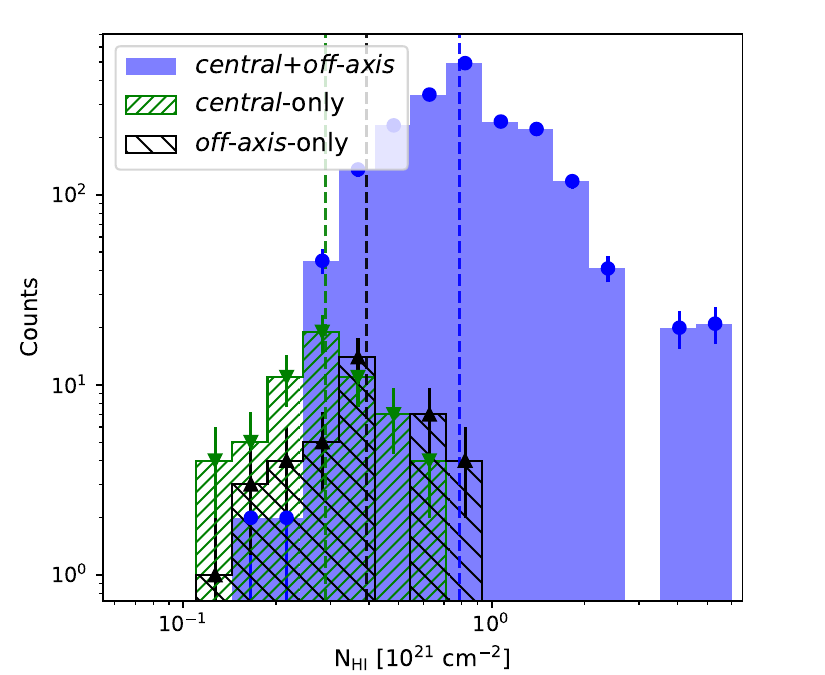}
    \includegraphics[trim = {0cm 0cm 0cm 0cm}, width=0.45\textwidth,angle=0]{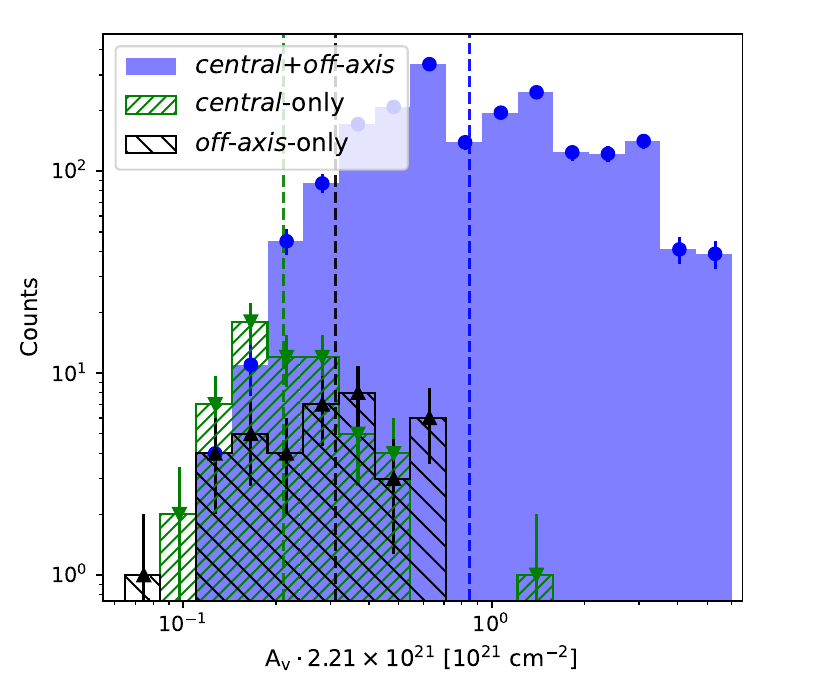}
    }
    \caption{Distributions of $N_\mathrm{HI}$ from HI4PI (left) and 2.21$\times10^{21}$A$_{\rm V}$ (right;  based on Eq.~\ref{Eq:linreg}) for sight lines with peak S/N$>$6 absorption line detections from the  
    central+off-axis (228), off-axis-only (25), and central-only (60) pointing subsets (see text for details).  The vertical dotted lines from left to right represent median values, respectively. The error bars correspond to $\sqrt{counts}$. 
    }
    \label{fig:pnt_gnhiav}
\end{figure*}

\begin{figure}
    \centering
    \vbox{
    \includegraphics[trim = {0cm 0cm 0cm 0cm}, width=0.45\textwidth,angle=0]{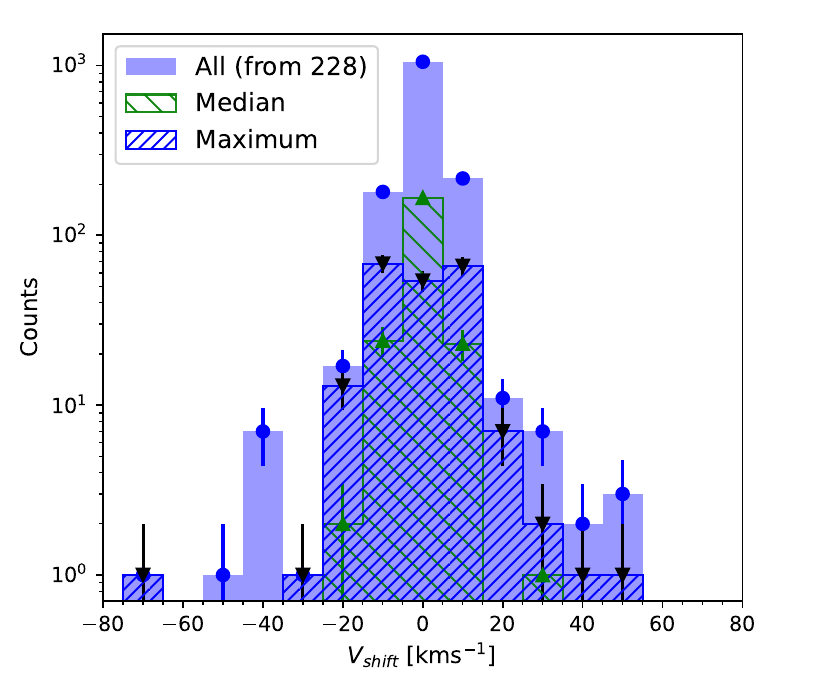}
    }
    \caption{Distributions of change in absorption peak ($V_{shift}$; Eq.~\ref{eq:vshift}) for the subset of 228 pointings.  Also shown are median and maximum $V_{shift}$ per pointing.  Two measurements at 150 and -180\,\kms\ have been excluded for clarity (see text for details).  The error bars correspond to $\sqrt{count}$. Of interest is that the velocity distribution mainly extends out only to the intermediate-velocity (IVC) regime. LVCs and IVC are both objects of the infrared cirrus. This implies that they belong to the region below the disk halo interface \citep[][]{Crovisier1985}.
    }
    \label{fig:pnt228_vshift}
\end{figure}

MeerKAT's excellent sensitivity across its large field of view has led to the detection of 3310 absorption lines from 18740 off-axis lines of sight within $48\farcm5$ from the pointing center. Out of these,  1716 have peak S/N $>$6, with 1710 also visually confirmed.  From the remaining with peak S/N $<$6,  1130 are visually confirmed, resulting in a total of 2840 visually confirmed detections.  
From Fig.~\ref{fig:samp}, first panel, it is obvious that the fraction of off-axis sight lines detected in \hi\ 21-cm absorption decreases toward the Galactic pole.  The figure also shows a lower number of detections below $10\degree$ because of the target selection strategy. It is also noticeable that off-axis sight lines, essentially due to the lower optical depth sensitivity, exhibit a much lower (15.2$\pm$0.3\%) detection rate compared to central sight lines (83$\pm$5\%). When optical depth variation is taken into account, in other words considering only sight lines with $\tau_{3\sigma} <$ 0.010 cutoff based on 21-SPONGE (see Sect.~\ref{sec:sky}), the detection rates from our off-axis ($88\pm9$\%; 106/120), central and 21-SPONGE sight lines are on par (see also Table~\ref{tab:detrate}).

\begin{table*}
\caption{Details of  central+off-axis (228), central-only  (60), and off-axis-only  (25) pointing subsets.}
\vspace{-0.4cm}
\begin{center}
\begin{tabular}{lccccc}
\hline
\hline
                    & central+off-axis & central-only  & off-axis-only \\
 \hline          
Total sight lines (central + off-axis)             & 11008   & 2956   & 1277        \\
central sight lines (with peak S/N$>6$ detection)        & 228    & 60     & 0           \\
off-axis sight lines (with peak S/N$>6$ detection)       & 1678   & 0      & 38          \\
off-axis sight lines (with $\int\tau^{off}_{6\sigma}$dv $<$ $\tau^{cen}_{int}$)$^\dag$   & 1822   & 77 & --  \\
$V_{shift}$[\kms] (median, $\sigma$)      &    (0, 8.9)       &  --    &    -- \\
Line width, $\sigma_{line}$ [\kms] (median, $\sigma$)      &    (2.7, 2.3)             &  --    &    -- \\
|$\Delta\tau$| (median, $\sigma$)         &    (0.10, 0.28)   &   --     &   --    \\   
|$\Delta\tau_{rel}$| (median, $\sigma$)  &    (0.18, 0.16)  &    --    &   --    \\     
|$\Delta\tau_{int}$| (median, $\sigma$)$^\ddag$         &    (0.77, 2.33)   &   --     &  --     \\   
|$\Delta\tau_{rel, int}$| (median, $\sigma$)$^\ddag$  &    (0.15, 0.16)  &    --    &   --    \\     
%
\hline
\end{tabular}
\label{tab:pntsubset}
\end{center}
$\dag$: off-axis sight lines sensitive to detect central absorption. $\ddag$: Same as |$\Delta\tau$| and |$\Delta\tau_{rel}$| but estimated using integrated optical depths.
\end{table*}

\begin{figure}
    \centering
    \hbox{
    \includegraphics[trim = {0cm 0cm 0cm 0cm}, width=0.5\textwidth,angle=0]{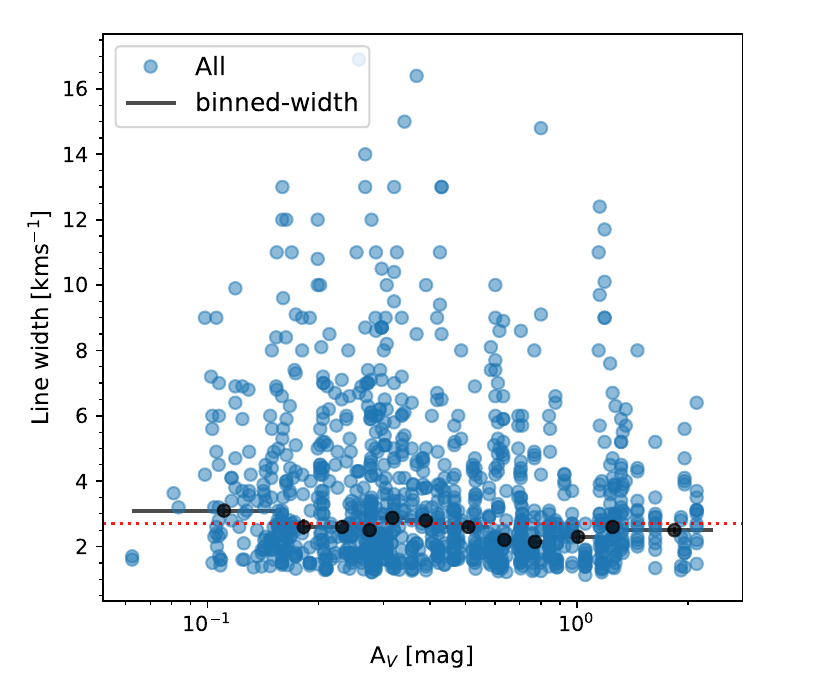}
    }
    \caption{Line width ($\sigma_{line}$) versus A$_V$ for the central+off-axis subset, with a horizontal dashed line showing the median value.  Note that A$_V$ is measured toward pointing centers.  Also shown is the median line width in equal frequency bins.  The 1$\sigma$ uncertainty on averaged width per bin is too small and not shown.  
    }
    \label{fig:pnt228_linewidth_av}
\end{figure}

\begin{figure*}
    \centering
    \vbox{
    \hbox{
    \includegraphics[trim = {0cm 0cm 0cm 0cm}, width=0.45\textwidth,angle=0]{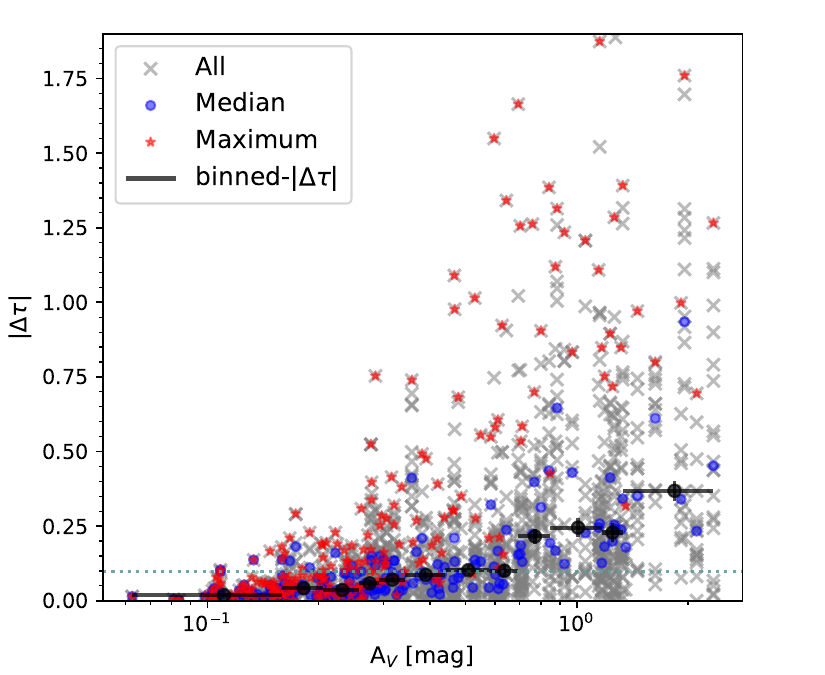}
    \includegraphics[trim = {0cm 0cm 0cm 0cm}, width=0.45\textwidth,angle=0]{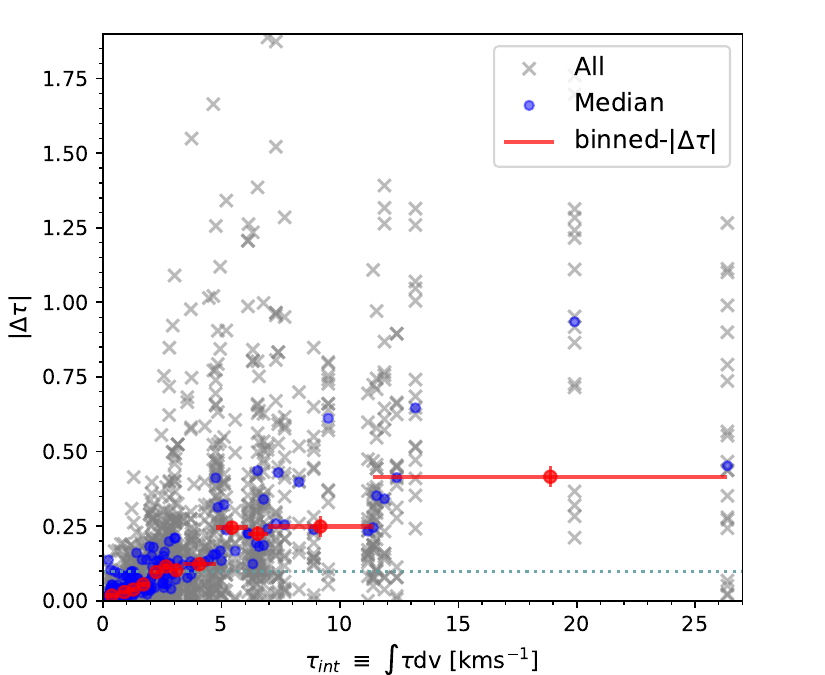}
    }
    \hbox{
    \includegraphics[trim = {0cm 0cm 0cm 0cm}, width=0.45\textwidth,angle=0]{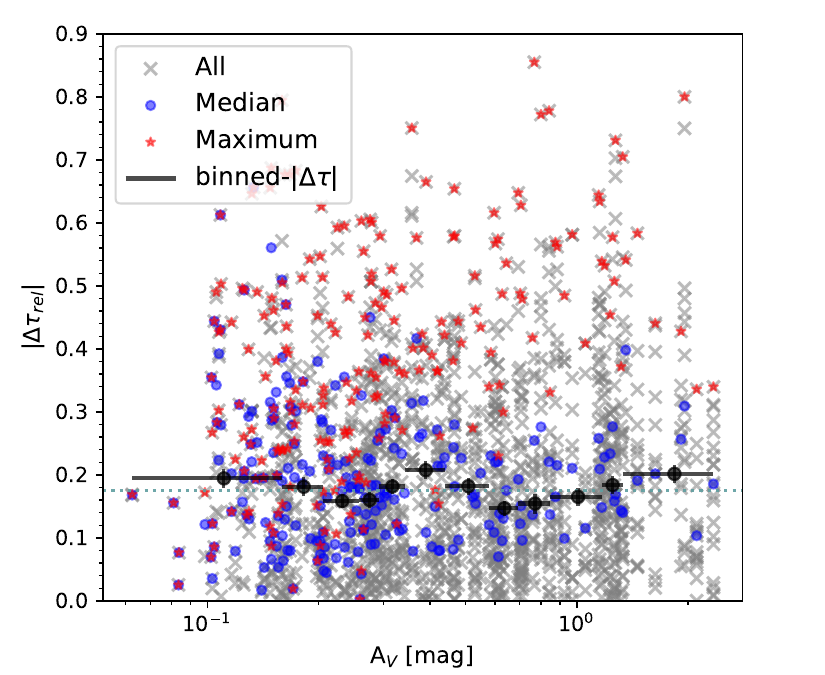}
    \includegraphics[trim = {0cm 0cm 0cm 0cm}, width=0.45\textwidth,angle=0]{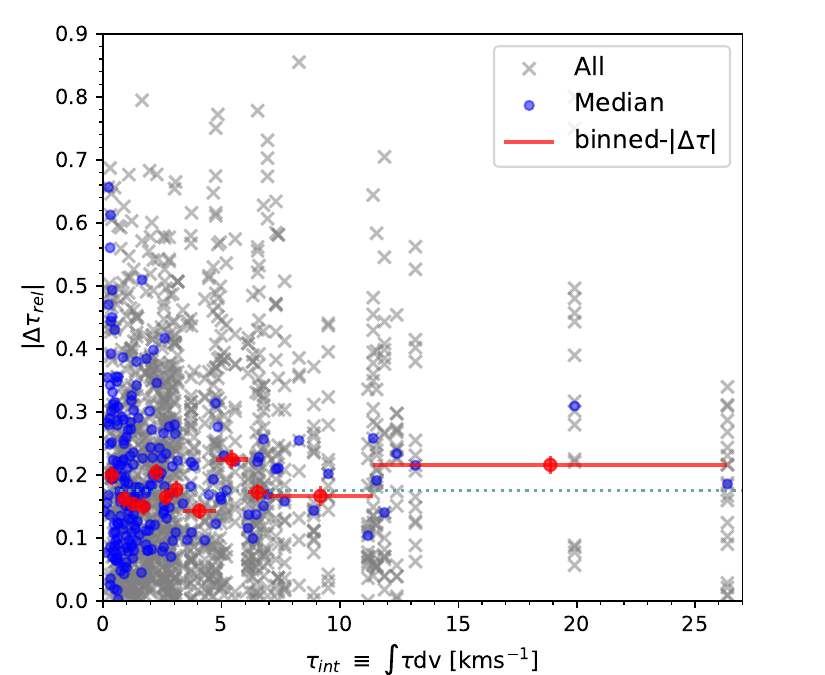}
    }
    }
    \caption{Top: |$\Delta\tau$| (Eq.~\ref{eq:deltatau}) versus A$_V$ (left) and $\tau_{int}$ $\equiv$ $\int\tau$dv (right) for the central+off-axis subset, with the horizontal dashed line showing the median value.  Note that A$_V$ and $\tau_{int}$ are measured toward pointing centers. Also shown are median and maximum of |$\Delta\tau$| per pointing, and median |$\Delta\tau$| in equal frequency bins.  The 1$\sigma$ uncertainty on averaged |$\Delta\tau$| per bin is too small and not shown.  For clarity, one measurement with |$\Delta\tau$| = 2.53 is omitted from the plot.  
    Bottom: |$\Delta\tau_{rel}$| (Eq.~\ref{eq:reltau}) versus A$_V$ (left) and $\tau_{int}$ $\equiv$ $\int\tau$dv (right). For clarity, one measurement with |$\Delta\tau_{rel}$| = 1.66 is omitted from the plot.  The remaining details are same as top panels.
    }
    \label{fig:pnt228_deltatau_fractau}
\end{figure*}

Naturally, the investigation of the small-scale structure through the comparison of central and off-axis sight lines need to take variations in sensitivity into account.  For reliability, we consider only absorbers with peak S/N$>$6 (see Sect.~\ref{sec:purity-snr}).  This limits our analysis to  313 pointings that have at least one such absorber.  The properties of these pointings, split into subsets of 228, 60, and 25 pointings with absorption toward both central and off-axis, only central and only off-axis sight lines, respectively, are summarized in Table~\ref{tab:pntsubset}. 
From Fig.~\ref{fig:pnt_gnhiav} it is apparent that the central-only (60) and the off-axis-only (25) subsets mostly occupy the low $N_\mathrm{HI}$(HI4PI) and $A_\mathrm{_V}$(SFD) region, dominated by WNM (see also Sect.~\ref{sec:phase}).
In fact, the off-axis-only subset has higher $N_\mathrm{HI}$ and $A_\mathrm{_V}$ compared to the central-only subset.  The two-sample Kolomogorov-Smirnoff test probability of these two samples being drawn from the same parent sample is very small, with p-values of $10^{-4}$ and $10^{-3}$ for $N$(\hi) and $A_\mathrm{V}$ variables, respectively. However, the off-axis-only subset, as noted in Sect.~\ref{sec:purity-snr}, has fainter central targets\footnote{In this subset 50\% of the central sight lines have a weak (peak S/N$<$6) but visually confirmed absorption.} resulting in absorption non-detections.  Indeed, the detection rates across these three  subsets, when matched in \hi\ column density and optical depth sensitivity, are comparable. 
Therefore, to investigate the optical depth variations with respect to dust extinction, gas column density and spatial separation it is pragmatic to focus only on the central+off-axis subset.  The  central-only and the off-axis-only subsets are not discussed hereafter.

\subsubsection{Variations with dust extinction and gas column density}
\label{sec:var_column}

The obvious limitation in the analysis presented here is that $N_\mathrm{HI}$ and $A_\mathrm{_V}$ have been evaluated toward the central sight line.
For now, we proceed with the reasonable hypothesis based on the finding of Sect.~\ref{Sect:N_HI and A_V} that the absorbing gas toward central sight lines shares same excitation conditions with more spatially extended \hi\ emission from the HI4PI survey.  
A detailed analysis involving independent higher angular resolution \hi\ emission measurements for central and off-axis sight lines will be presented in a forthcoming paper.
Furthermore, we use the properties of absorbers determined from the Gaussian fits obtained in Sect.~\ref{sec:auto}.  In general, the Gaussian decomposition is not unique but the majority of absorbers ($\sim$82\%) in our sample are simple and fit with a single component.  Therefore, it is reasonable to expect that the Gaussian decomposition adequately represent the properties of the bulk of absorbers in our sample, and can be used to match and compare properties of gas across different sight lines. The decomposition may not be unique for a small subset of complex absorbers.  We discuss the complications due to these on the physical interpretation of the results.
For each pointing, we calculate the differences in the position of absorption peak ($V_{LSR} \equiv$ {\tt pos\_max}), peak optical depth ($\tau_{peak} \equiv$  {\tt tau\_max\_val}), and relative optical depth for all the central--off-axis sight line pairs as   
\begin{equation}
    V_{shift} = V_{LSR}^{off} - V_{LSR}^{cen},
    \label{eq:vshift}
\end{equation}
\begin{equation}
    \Delta\tau = \tau_{peak}^{off} - \tau_{peak}^{cen}, \,\, {\textrm{and}}
    \label{eq:deltatau}
\end{equation}
\begin{equation}
    \Delta\tau_{rel} = \Delta\tau / (\tau_{peak}^{off} + \tau_{peak}^{cen}),
    \label{eq:reltau}
\end{equation}
respectively.  We also measure $\Delta\tau_{int}$ and $\Delta\tau_{rel, int}$ using integrated optical depth ($\tau_{int} \equiv$  {\tt tau\_int\_val}) instead of peak optical depth.

In case, as hypothesized above, the absorption features toward central and off-axis sight lines within a pointing trace a coherent physical gas structure, then both the opacity, or volume density, and even more importantly, the radial velocities of the absorption features need to match within the velocity spread of the WNM. 
For the central+off-axis subset, 1822 off-axis sight lines  have sufficient integrated optical depth sensitivity ($\int\tau^{off}_{6\sigma}$dv $<$ $\tau^{cen}_{int}$) to detect the central absorption.
Figure~\ref{fig:pnt228_vshift} shows the distribution of $V_{\rm shift}$ (Eq.~\ref{eq:vshift}) for 1499 visually confirmed detections from these.  In general, the shifts of individual pairs as well as the median and maximum of  $V_{\rm shift}$ per pointing are small (median = 0\,\kms; Table~\ref{tab:pntsubset}).   The WNM temperature range of 4000 - 8000\,K corresponds to a thermally broadened line with a FWHM of 14 - 19\,\kms \citep[$\sigma$ = 6 - 8\,\kms;][]{Heiles03, Haud2007}. The observed small velocity shifts of $\sigma$($V_{shift}$) = 8.9\,\kms (Table~\ref{tab:pntsubset}) support the scenario of CNM/LNM clouds freezing out from the extended WNM phase.  
There are some outliers  -- specifically, 18 pairs have |$V_{\rm shift}$| $>$ 30\,\kms.  Sixteen out of eighteen are associated with four pointings, two of these being J074155.69$-$264729.8 and J080622.15$-$272611.5 discussed in Sect.~\ref{sec:purity-snr},  exhibiting complex absorption profiles with multiple components.  In the remaining two cases with $V_{\rm shift}$ $=$ 150 and $-180$\,\kms, the absorber toward off-axis sight line has peak S/N $<6$ and is likely due to noise fluctuations. In conclusion, the median and maximum of $V_{\rm shift}$ per pointing, also shown in Fig.~\ref{fig:pnt228_vshift}, remain small. These measurements are unaffected if we consider only 1276/1499 pairs with peak S/N $>$ 6.

The typical (median) line FWHM  of the absorbers in the MALS sample is 6.1\,\kms\ ($\sigma$ = 2.7\,\kms) and may have contributions from both thermal and turbulence broadening. 
The absorbers with more than three components are primarily at $A_\mathrm{V}$ $>$ 1.0\,mag and low Galactic latitude (median |$b$| = $2.4^\circ$), indicating complexity usually associated with gas that has high dust content and CNM fraction.  However, as was previously noted, the majority of the profiles are simple and mostly fit with a single Gaussian component.  Although the \hi\ absorption lines appear to be slightly broader at lower extinctions (Fig.~\ref{fig:pnt228_linewidth_av}), we do not find a significant relationship between the line width and optical extinction ($\sigma_{line}$ $\propto A_\mathrm{V}^{-0.08\pm0.03}$).  This is not surprising considering the low spectral resolution (5.5\,\kms) of our data.  

\begin{figure}
    \centering
    \vbox{
    \includegraphics[trim = {0cm 0cm 0cm 0cm}, width=0.45\textwidth,angle=0]{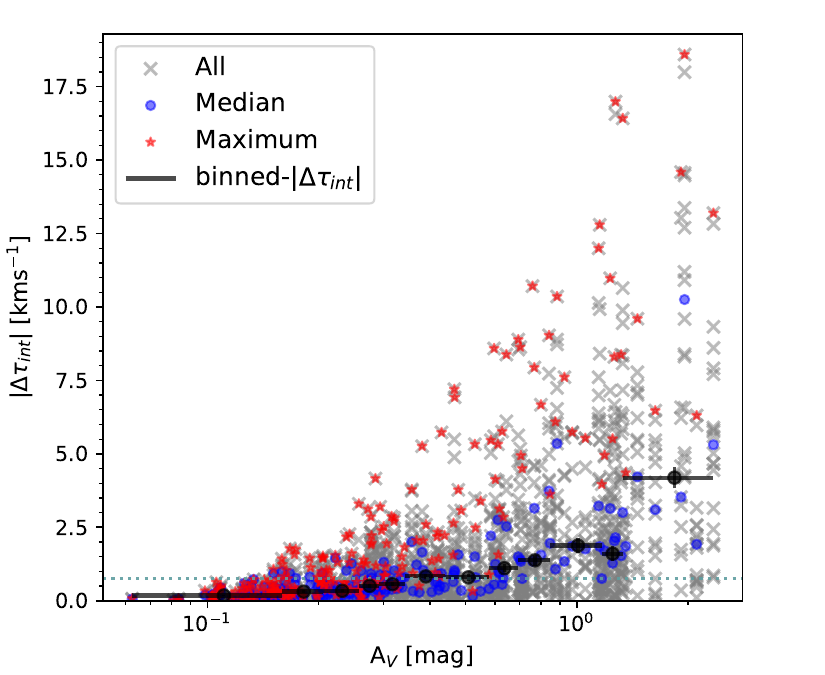}
    }
    \caption{|$\Delta\tau_{int}$|  versus A$_V$ for the central+off-axis subset, with a horizontal dashed line showing the median value.  Note that A$_V$ is measured toward pointing centers. Also shown are the median and maximum of |$\Delta\tau_{int}$| per pointing, and the median |$\Delta\tau$| in equal frequency bins.  The 1$\sigma$ uncertainty on averaged |$\Delta\tau_{int}$| per bin is too small and not shown.  
    }
    \label{fig:pnt228_deltainttau_av}
\end{figure}

\begin{figure*}
    \centering
    \hbox{
    \includegraphics[trim = {0cm 0cm 0cm 0cm}, width=0.45\textwidth,angle=0]{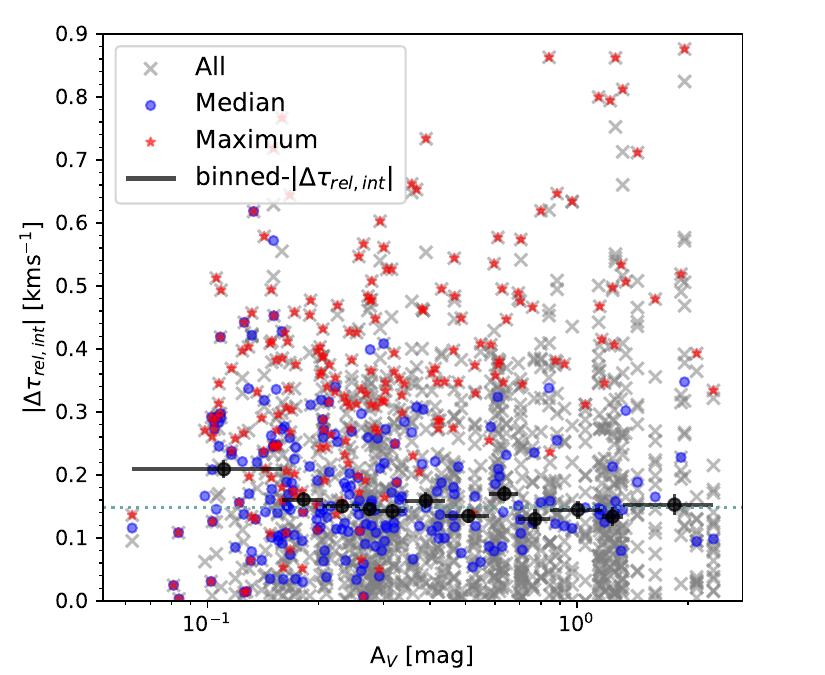}
    \includegraphics[trim = {0cm 0cm 0cm 0cm}, width=0.45\textwidth,angle=0]{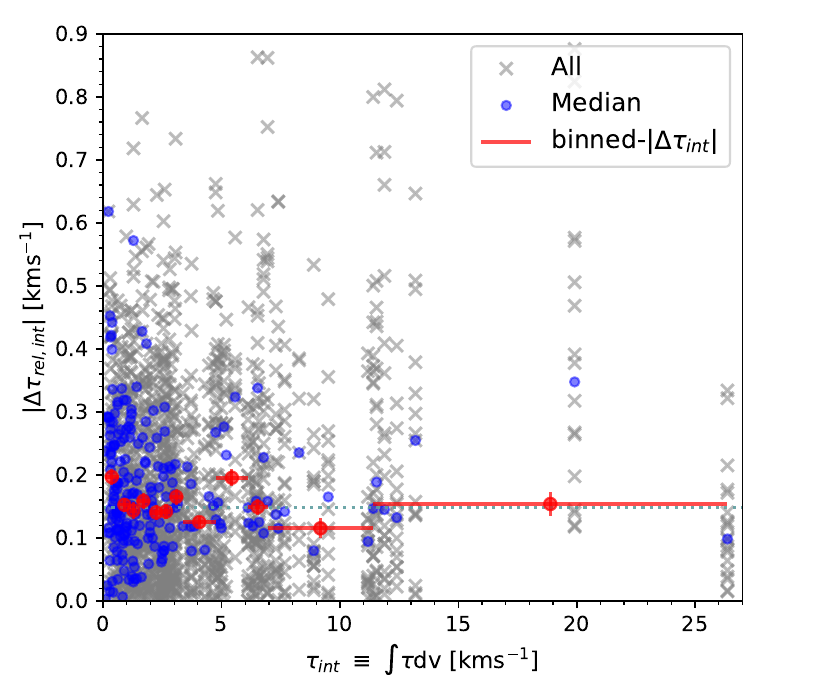}
    }
    \caption{|$\Delta\tau_{rel,int}$| versus A$_V$ and $\tau_{int}$ $\equiv$ $\int\tau$dv for the central+off-axis subset, with a horizontal dashed line showing the median value.  Note that A$_V$ and $\tau_{int}$ are measured toward pointing centers.  Also shown are the median and maximum of |$\Delta\tau_{int}$| per pointing, and median |$\Delta\tau$| in equal frequency bins.  The 1$\sigma$ uncertainty on averaged |$\Delta\tau_{int}$| per bin is too small and not shown.  
    }
    \label{fig:pnt228_deltarelinttau_av}
\end{figure*}

Figure~\ref{fig:pnt228_deltatau_fractau} (top panels) show optical depth differences, |$\Delta\tau$|, for 1499 sight line pairs with absorptions that have peak S/N $\geq$ 6.  For each pointing, we also estimate the median and maximum of |$\Delta\tau$|.  The overall median |$\Delta\tau$| = 0.10.  There is a significant scatter ($\sigma = $0.27) but clearly the optical depth variation inferred from the median values in equal frequency bins is higher for pointings with high extinction (|$\Delta\tau$| $\propto$ $A_\mathrm{V}^{0.95\pm0.08}$).  Since, optical depth and extinction are correlated (Fig.~\ref{Fig:Tau_int_vs_AV}) a similar but weaker dependence (|$\Delta\tau$| $\propto$ $\int\tau dv^{0.71\pm0.06}$) is also apparent in the top-right panel of Fig.~\ref{fig:pnt228_deltatau_fractau}.  Both these correlations imply that optical depth variations are higher for pointings with higher \hi\ column density and CNM fraction.  
The MALS sample on average shows similar changes in peak optical depth as the work of \citet[][]{Crovisier1985} presenting measurements based on the observations of 7 extended radio sources with a velocity resolution of $\sim$4\,\kms.  We note that \citet[][]{Crovisier1985} probe similar physical scales (0.1 - 10\,pc) as our sample (discussed below in Sect.~\ref{sec:var_separ}).  In comparison, the variations in the sample of \citet[][]{Rybarczyk2020} based on 14 sources are much larger (average $\Delta\tau$ = 0.35).  This is likely due to the much higher ($\sim$0.4\,\kms) spectral resolution used in their observations, even though they probe much smaller ($<$0.1\,pc) scales. 
Unlike peak optical depth, integrated optical depth does not depend on spectral resolution.  Indeed, the differences in these for the MALS sample (median = 0.77; Table~\ref{tab:pntsubset}) shown in  Fig.~\ref{fig:pnt228_deltainttau_av} are comparable to  \citet[][]{Rybarczyk2020}, but with much larger scatter, again possibly due to larger scales probed in our sample.
We find the |$\Delta\tau_{int}$| $\propto$ $A_\mathrm{V}^{1.23\pm0.13}$ dependence to be steeper than the relationship for |$\Delta\tau$|.  This is primarily due to absorbers at $A_\mathrm{V} > 1$ that are complex and better represented by the integrated than the peak optical depth measurements.  The exponents in the two cases match within 1$\sigma$ when only sight lines with $A_\mathrm{V} < 1$ are considered.

The bottom panels of Fig.~\ref{fig:pnt228_deltatau_fractau} show relative variation in optical depth (Eq.~\ref{eq:reltau}) with respect to extinction and integrated optical depth, respectively.  We find  |$\Delta\tau_{rel}$| $\propto$ $A_\mathrm{V}^{0.00\pm0.04}$ and $\propto$ $\int\tau dv^{0.04\pm0.04}$, respectively.  
Using integrated optical depths, we find |$\Delta\tau_{rel,int}$| $\propto$ $A_\mathrm{V}^{-0.10\pm0.04}$ and $\propto$ $\int\tau dv^{-0.06\pm0.04}$.  In both cases, we do not see any significant systematic trends with respect to $A_\mathrm{V}$ or $\int\tau$dv (Fig.~\ref{fig:pnt228_deltarelinttau_av}).   
This may appear inconsistent with the findings reported in the literature, suggesting that fractional optical depth variations at sub-parsec scales are larger at smaller integrated optical depths \citep[e.g.,][]{Rybarczyk2020, Hacker2013}.  These are interpreted as overall small covering fraction of small-scale structure, making it easier to detect at small integrated optical depths.  
We note that the MALS sight lines do not predominantly trace variations at $\int\tau$dv < 0.5\,\kms where such differences have been observed \citep[][ their Fig. 6]{Rybarczyk2020}.   
Indeed, in the MALS sample for |$\Delta\tau_{rel,int}$| there is a tentative indication that variation may be larger at the lowest extinction; that is, $A_\mathrm{V} <$ 0.2\,mag, or, equivalently, $\int\tau$dv < 0.5\,\kms\ (Fig.~\ref{fig:pnt228_deltarelinttau_av}). 
%

It is clear from the comparison with \citet[][]{Rybarczyk2020} that spectral resolution plays an important role in the measured optical depth variations. 
The simplistic analysis presented here has used $N_\mathrm{HI}$ and $A_\mathrm{_V}$ toward central sight lines smoothed over several arcminutes.  Despite this, the observed reasonable correlations are consistent with the findings of Sect.~\ref{sec:galhicent} that the absorbing gas is exposed to the same physical conditions as the extended warm gas.

\begin{figure*}
    \centering
    \hbox{
    \includegraphics[trim = {0cm 1.0cm 0cm 0.0cm}, width=0.95\textwidth,angle=0]{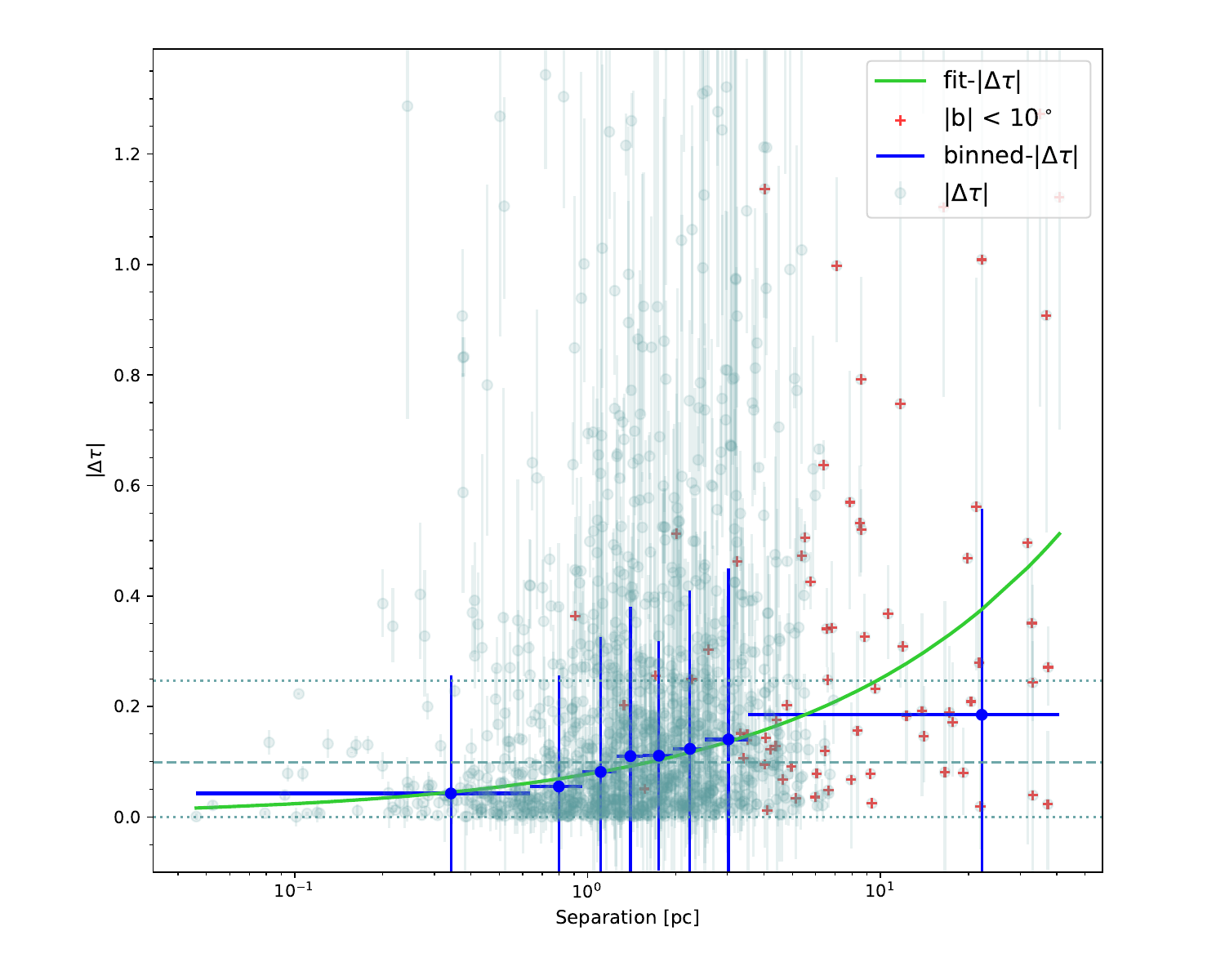}    }
    \caption{
    |$\Delta\tau$| versus separation in parsec between central and off-axis sight lines for absorption lines matched within $\pm$6\,\kms. The horizontal dashed line marks median |$\Delta\tau$|, and dotted lines represent 1$\sigma$ scatter based on the median absolute deviation.  The median |$\Delta\tau$| in equal frequency bins with 1$\sigma$ scatter which are essentially $\Delta\tau_{rms}$, are also shown.  The 1$\sigma$ uncertainty on averaged |$\Delta\tau$| per bin is 0.02 and not shown.
    The measurements at |b|$<10^\circ$ are marked as '+'.   The power law fit to |$\Delta\tau$| at |b|$>10^\circ$ is also shown.
    For clarity, 9 measurements in the range 1.39 $<$ |$\Delta\tau$| $<$ 2.53 have been omitted from the plot.
    }
    \label{fig:taudiffpc}
\end{figure*}

\subsubsection{Variations with linear separation}
\label{sec:var_separ}

%
The spatial distances to the \hi\ absorbing clouds are more difficult to ascertain.  Based on the CNM scale height inferred from the vertical distribution of \hi\ 21-cm absorbers \citep[][]{Crovisier1978, Wenger2024}, we adopt 100\,pc/sin|$b$| corresponding to the edge of the local cavity as the approximate distance to the clouds at a given latitude ($b$).   
For investigating optical depth variations with linear separation, we consider central - off-axis pairs matched in velocity.
Figure~\ref{fig:taudiffpc} shows |$\Delta\tau$| versus linear separation for 1381/1499 pairs with absorption peaks matching within 6\,\kms\ ; that is, the median FWHM of the central absorption lines.  About $\sim$80\% of these are fit with single Gaussian components.  
Indeed toward low Galactic latitudes (|b|$<10^\circ$) the distances become larger but also more uncertain (shown by “+” symbols in Fig.~\ref{fig:taudiffpc}).   The median $\vert b\vert$ of 37 measurements with linear separation larger than 7\,pc is $2.3^\circ$.  This corresponds to a linear distance of 2.5\,kpc, which already probes the gaseous environment of the next spiral arm (Scutum-Centaurus). It would be surprising for gas at this distance to have the same radial velocity as the LISM, represented in Fig.~\ref{fig:pnt228_vshift}.
The best fit modeled as a simple power law, |$\Delta\tau$($x$)| =  $\Delta\tau^o$ $\times$ $x^{a}$ where $x$ is the linear separation in parsecs, with a = 0.510 $\pm$ 0.099 and $\Delta\tau^o$ = 0.077 $\pm$ 0.008, considering only measurements at |b|$>10^\circ$ is also shown in Fig.~\ref{fig:taudiffpc}. 
The optical depth variations\footnote{The exponent matches within 1$\sigma$ for |$\Delta\tau_{int}$|.} are greater at larger angular scales and the span is consistent with the range covered by \citet[][]{Crovisier1985}.
In comparison, the relative variations in the peak and integrated optical depths are constant (a = 0.06 $\pm$ 0.03) over the scales covered in our sample.

At the median linear separation ($\sim$1.5\,pc) of the sample, the median |$\Delta\tau$| $\sim$ 0.1 and |$\Delta\tau_{int}$| $\sim$ 0.8.  For a CNM temperature of 100\,K, these imply a change in column density, $\Delta N_{\mathrm{HI}}$, of 1.1-1.5$\times10^{20}$\,\cmsq.  For a spherical geometry, the calculated $\Delta N_{\mathrm{HI}}$ at 1.5\,pc implies a density of  $n_\mathrm{HI}=$ 20-30\,cm$^{-3}$.   These are consistent with typical densities of diffuse CNM ($<$100\,cm$^{-3}$).    Thus, in general, the optical depth variations in the MALS sample imply that \hi\ absorbing gas in the LISM primarily represents sustained structures in pressure equilibrium with the warmer gas.
However, 10\% of the sight lines in the sample have 0.5 $<$  |$\Delta\tau$| $<$ 1.9, which corresponds to 100$<n$(cm$^{-3}$)$<$500\,.  We note that the criteria of  |b|$>10^\circ$  has already eliminated 27/29 sight lines with {\tt n\_comp} $>3$ from the analysis.  Therefore, the sight lines with |$\Delta\tau$| $>$ 0.5 are not unusually complex with respect to the rest of the sample.  These may actually represent over-pressured structures in the ISM. Or alternatively, as suggested by \citet[][]{Heiles97} these belong to sheets or filaments and we have generally overestimated densities in the MALS sample.  

Next, we investigate the rms value of the opacity difference $\Delta\tau_{rms}$($x$) at a linear separation $x$ in parsecs.  This is simply given by the square root of the structure function of the optical depth at $x$.  For optical depth variation plotted in Fig.~\ref{fig:taudiffpc}, the structure function is simply the variance of |$\Delta\tau$| per bin, almost monotonically increasing from 0.21 to 0.33. 
Following \citet[][]{Deshpande2000a}, we model rms fluctuations in |$\Delta\tau$| for |b|$>10^{\circ}$ measurements as $\Delta\tau_{rms}$($x$) =  $\Delta\tau_{rms}^o$ $\times$ $x^{(\alpha -2)/2}$. The resultant fit has $\alpha$ = 2.327 $\pm$ 0.153 and $\Delta\tau_{rms}^o$ = 0.233 $\pm$ 0.016. 
We find that the measurements of \citet[][their Table 3]{Crovisier1985} yield $\alpha$ = 2.31 $\pm$ 0.20 and $\Delta\tau_{rms}^o$ = 0.25 $\pm$ 0.05.  Within the measurement uncertainties these earlier data from \citet[][]{Crovisier1985} are consistent with the rms fluctuations observed in the MALS sample.  
Interestingly, the exponent in the quiescent LISM as sampled by MALS is shallower than the values of 2.75$\pm$0.08 and 2.5 observed for the Perseus arm toward Cas\,A and the outer arm toward Cygnus A over 0.01 - 4\,pc  at a spectral resolution of $\sim$0.5\,\kms \citep[][]{Deshpande2000a}.  
It will be interesting to extend these measurements to large samples at smaller scales and test the slope of the power law in different environments.

\section{Summary and concluding remarks}    
\label{sec:summ}  

MALS observed 391 pointings at declinations $\lesssim +20^\circ$ and centered on radio sources brighter than 200\,mJy at 1\,GHz.  The radiation from the bright central source as well as numerous other off-axis radio sources within the telescope's pointing inevitably pass through the Galaxy.
This paper presents spectra covering 1419.4 - 1421.4\,MHz toward 19130 central and off-axis radio sources brighter than 1\,mJy at 1.4\,GHz within the $48\farcm5$ of the centers of 390 pointings\footnote{One pointing with poor quality spectra is excluded.}.  The spectral resolution, the median spatial resolution, and the median 3$\sigma$ optical depth sensitivity ($\tau_{3\sigma}$) are 5.5\,\kms, $\sim9^{\prime\prime}$, and 0.381, respectively. 
Due to the extragalactic nature of the survey, the MALS sight lines mostly pass through the LISM at high Galactic latitudes (median absolute latitude = $41^{\circ}$) and trace diffuse and translucent ISM phases of the Galaxy.

Through an automated search, we have detected 3640 \hi\ 21-cm absorption features. Of these, 3158 absorbers have been visually confirmed and 1870 also have peak S/N $>$ 6. In 2986, in other words about 82\% of the cases, the absorption feature is  well fit with a single Gaussian component, whereas 529, 95, and 21 sight lines require two, three, and four or more components.  The overall \hi\ 21-cm absorption detection rate is $\sim$18\%. The off-axis sight lines, comprising the majority of the sample, are not only toward fainter AGN, but are also affected by primary beam attenuation.  Therefore, much higher optical depth sensitivities (median $\tau_{3\sigma}$ = 0.008) and detection rates (82$\pm$5\%) are achieved toward central sight lines. 
Overall, the absorber sample at peak S/N $>$6 has a completeness of 90\% or better, and a purity of $\sim$95\%.
The main results from the survey using peak S/N$>$6 absorbers, complemented by publicly available \hi\ 21-cm emission \citep[HI4PI;][]{HI4PI}, optical emission \citep[SFD; ][]{SFD1998}, and dust temperature \citep[][]{Planck2014dust} maps, are summarized below.  We first (1 -- 4) focus on central sight lines to derive integrated properties of gas over the telescope's primary beam and then (5 -- 10) on the differences in the properties of the absorbing gas toward central and off-axis sight lines.

\begin{enumerate}
    
    \item The remarkable linear correlation exhibited by $N_{\mathrm{HI}}$(HI4PI) and $A_{\rm V}$(SFD) over 0.03 - 0.7\,mag, along with the confinement of \hi\ 21-cm absorption features toward central sight lines to a narrow range of velocities (-25$<v_{\rm LSR}$[\kms]$<$+25), imply that the detected absorption lines form a homogeneous sample of neutral hydrogen clouds in the solar neighborhood. 

    \item For absorbers toward central sight lines, at $N_{\mathrm{HI}}$(HI4PI) $\lapp$ $2\cdot10^{21}$\,\cmsq\ and $A_{\rm V}$(SFD)$\lapp$1\,mag the integrated optical depth is $\int\tau$dv $\propto$ $N_{\mathrm{HI}}^2$ $\propto$ $A_{\rm V}^2$.  However, at higher $N_{\mathrm{HI}}$ and $A_{\rm V}$, likely due to the onset of formation of molecules, a different scaling relation, $\int\tau$dv $\propto$ $N_{\mathrm{HI}}^{2.1}$ and $A_{\rm V}^{1.5}$, is observed.

    \item Using neutral gas phase decomposition from \citet[][]{KalberlaHaud2018}, we find that  all the MALS sight lines with \hi\ absorption have $N_\mathrm{HI}({\rm CNM})\,+\,N_\mathrm{HI}({\rm LNM}) \geq N_\mathrm{HI}({\rm WNM})$. This is markedly different for the sight lines with no \hi\ absorption detections.

    \item The slopes of $N_{\mathrm{HI}}$ distributions for central sight lines with \hi\ 21-cm absorption detections and non-detections are $-1.53\pm0.13$ and $-2.22\pm0.2$, respectively.  A similar difference is observed for H$_2$ detections and non-detections in damped Lyman-alpha systems at $z\gapp$1.8. This suggests that a WNM-to-CNM transition through turbulence is the common governing factor for the presence of absorption in both cases.

    \item For 228 pointings with both central and off-axis detections, we find that the shift in the off-axis absorption peak with respect to the central absorption is well within the thermally broadened line for WNM.  This confirms the hypothesis of CNM and LNM clouds freezing out of extended WNM.  

    \item The optical depth variations between central and off-axis detections are higher for pointings centered on regions with high extinction (|$\Delta\tau$| $\propto$ $A_\mathrm{V}^{0.95\pm0.08}$) and integrated 21-cm absorption optical depth (|$\Delta\tau$| $\propto$ $\int\tau dv^{0.71\pm0.06}$).  These imply that optical depth variations are higher for regions with a higher $N_{\mathrm{HI}}$ and CNM fraction. 
    
    \item Most likely due to the low spectral resolution of our data, we do not find a significant relationship between the absorption line widths and central optical extinction ($\sigma_{line}$ $\propto A_\mathrm{V}^{-0.08\pm0.03}$).

    \item For relative variations in peak and integrated optical depths, representing the covering fraction of small-scale structures, we do not find any dependence with respect to central visual extinction and the integrated optical depth. Specifically, |$\Delta\tau_{rel}$| $\propto$ $A_\mathrm{V}^{0.00\pm0.04}$ and $\propto$ $\int\tau dv^{0.04\pm0.04}$, and |$\Delta\tau_{rel,int}$| $\propto$ $A_\mathrm{V}^{-0.10\pm0.04}$ and $\propto$ $\int\tau dv^{-0.06\pm0.04}$, respectively.  

    \item Over 0.1 - 10\,pc, the optical depth variations between central and off-axis sight lines are larger at larger scales |$\Delta\tau$($x$)| =  ($0.077 \pm 0.008$) $\times$ $x^{0.510 \pm 0.099}$, where $x$ is the linear separation in parsecs.  The densities (20-30\,cm$^{-3}$) inferred from the optical depth variations at the median separation (1.5\,pc) of the sample are typical of the CNM values.  Larger variations observed toward a small number (10\%) of sight lines imply higher densities (100 $<n$ (cm$^{-3}$) $<$500), but overall the MALS sample implies that \hi\ absorbing gas in the LISM primarily represents sustained structures in pressure equilibrium with the warmer gas.

    \item The rms fluctuations in optical depth variations in the relatively quiescent LISM sampled by MALS sight lines have a slope of $2.327\pm 0.153$.  As was expected, this is much shallower than the values of 2.75 and 2.5 observed for the Perseus arm toward Cas\,A and the outer arm toward Cygnus A over 0.01 - 4\,pc, respectively.  

\end{enumerate}
%

The Galactic \hi\ 21-cm absorption spectral line catalog presented here represents an increase of more than two orders of magnitude in the sight lines probing the local ISM of the Galaxy.  This represents the largest Galactic \hi\ absorption line catalog to date.  
The analysis presented in this paper demonstrates the potential of MALS, primarily an extragalactic survey, to enable improved constraints on the distribution of \hi\ gas in the local ISM of the Galaxy and, in future, its deviations from the disk geometry.  
We used the decomposition of neutral gas phases from the analysis of HI4PI spectra by \citet[][]{KalberlaHaud2018}.
The versatility of the MeerKAT array configuration allows for exploration of \hi\ 21-cm emission at scales much smaller than the single dish surveys used here.   These will be presented in future papers to address the limitations arising from different spatial resolutions of MALS, HI4PI (16\farcm4), and SFD (6\farcm1). 
This will quantitatively test the conclusions that used $N_{\mathrm{HI}}$(HI4PI) and $A_{\rm V}$(SFD) smoothed over several arcminutes under the assumption that the warmer gas is homogenized over scales larger than several arcminutes.
While the conclusions 1--4 and observed trends in conclusions 5--10 support this assumption and imply that the CNM is indeed exposed to similar excitation conditions as the extended warmer gas, a detailed analysis based on independent $N_{\mathrm{HI}}$ corrected for optical depth effects and $A_{\rm V}$ estimates toward all the sight lines will reveal the turbulent density structure and the limits of physical scales over which \hi\ gas phase structures are homogenized.
In this context, the \hi\ absorption line catalog will also be updated to take into account the underlying \hi\ emission and its influence on spectral rms, which may affect the reliability and completeness of the samples at low S/N.

The MALS L-band observations also covered the OH 18-cm main and the 1612\,MHz satellite lines.  These will be presented in a forthcoming paper and may provide insights into the formation of H$_2$ in the diffuse ISM. These will also be key inputs to interpret results from extragalactic molecular absorption line surveys \citep[][]{Balashev2021} with the Square Kilometre Array (SKA) and its precursors \citep[e.g., the Galactic ASKAP survey;][]{Dickey2013, Nguyen2024}, and upcoming large optical surveys such as the 4MOST–Gaia Purely Astrometric Quasar Survey \citep[4G-PAQS;][]{Krogager2023}, which will study atomic and molecular hydrogen in absorption-selected galaxies at cosmic noon.


\section*{Data availability}

The Galactic \hi\ 21-cm absorption spectral line catalog, the Stokes-$I$ spectra and the associated continuum images are publicly available at \url{https://mals.iucaa.in}.  
%

\begin{acknowledgements}
We thank the anonymous referee for useful and detailed comments.
The MeerKAT telescope is operated by the South African Radio Astronomy Observatory, which is a facility of the National Research Foundation, an agency of the Department of Science and Innovation. 
The MeerKAT data were processed using the MALS computing facility at IUCAA (\url{https://mals.iucaa.in/releases}).
The Common Astronomy Software Applications (CASA) package is developed by an international consortium of scientists based at the National Radio Astronomical Observatory (NRAO), the European Southern Observatory (ESO), the National Astronomical Observatory of Japan (NAOJ), the Academia Sinica Institute of Astronomy and Astrophysics (ASIAA), the CSIRO division for Astronomy and Space Science (CASS), and the Netherlands Institute for Radio Astronomy (ASTRON) under the guidance of NRAO.
NG acknowledges NRAO for generous financial support for the sabbatical visit at Socorro during which a part of this work was done. 
The National Radio Astronomy Observatory is a facility of the National Science Foundation operated under cooperative agreement by Associated Universities, Inc.  This research has made use of the NASA/IPAC Extragalactic Database (NED), which is funded by the National Aeronautics and Space Administration and operated by the California Institute of Technology. 
\end{acknowledgements}

\bibliographystyle{aa}
\bibliography{mybib}

\begin{appendix}

\onecolumn

\section{MALS Galactic \hi\ absorption line catalog }
\label{sec:listpoint}

\begin{figure*}
    \centering
    \vbox{
    \includegraphics[trim = {0cm 0cm 0cm 0cm}, width=0.95\textwidth,angle=0]
    {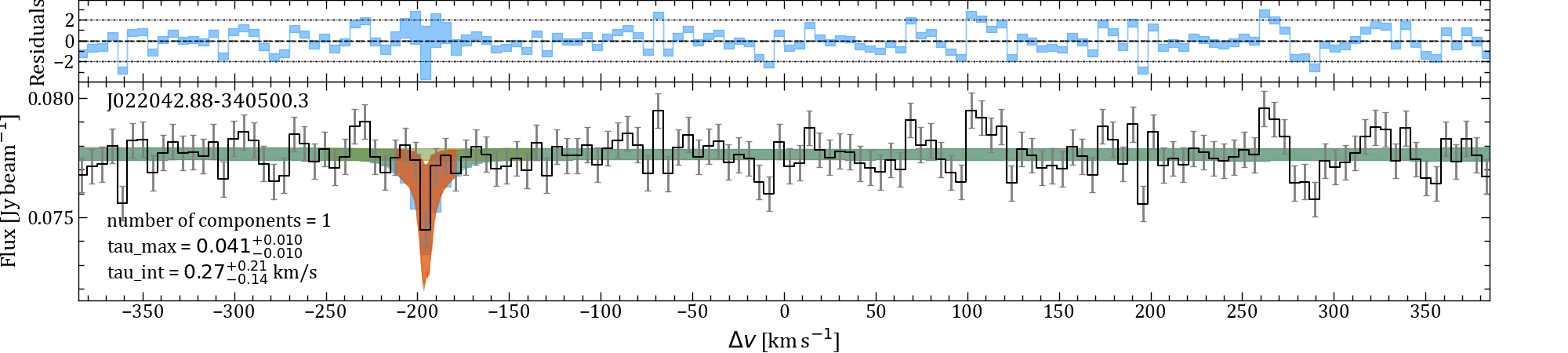}
    \includegraphics[trim = {0cm 0cm 0cm 0cm}, width=0.95\textwidth,angle=0]
    {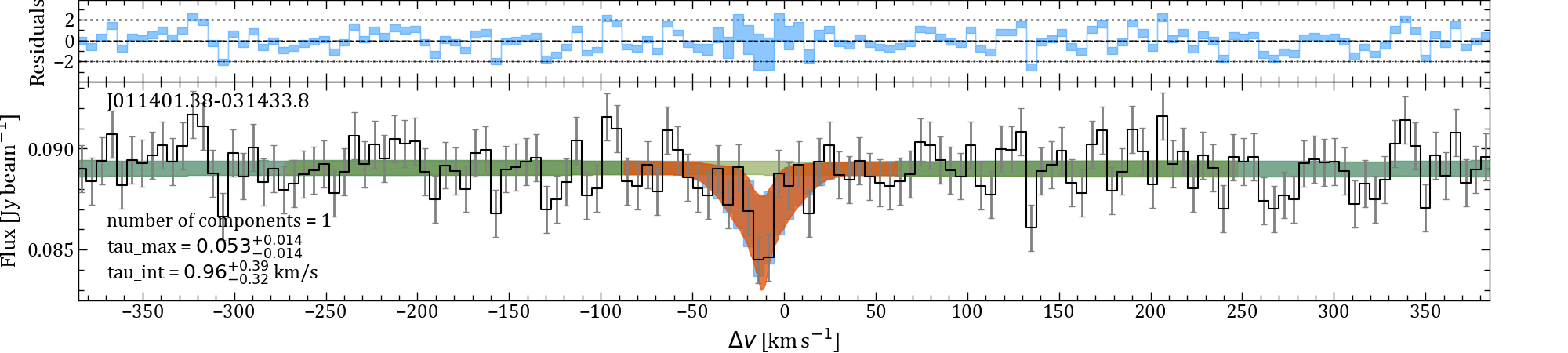}
    \includegraphics[trim = {0cm 0cm 0cm 0cm}, width=0.95\textwidth,angle=0]
    {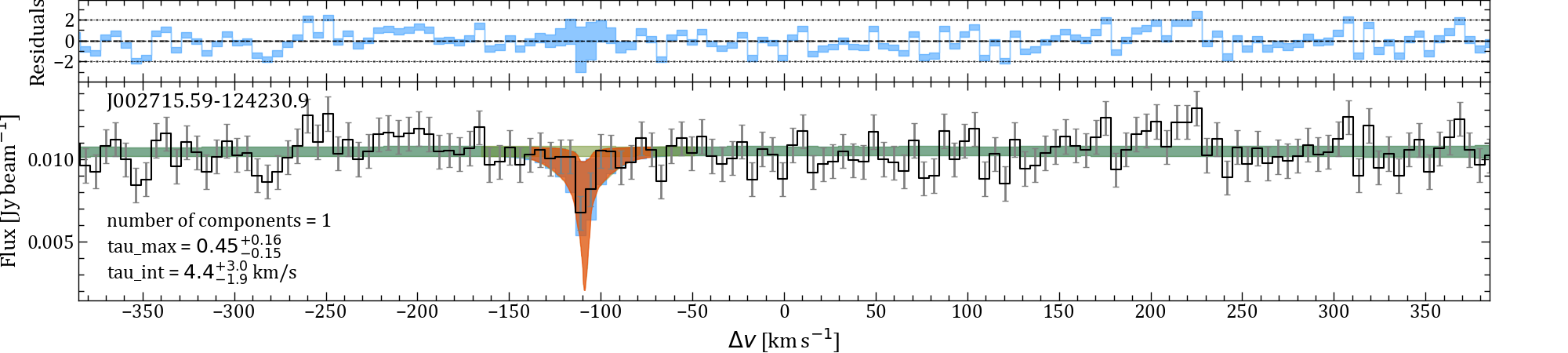}
    \includegraphics[trim = {0cm 0cm 0cm 0cm}, width=0.95\textwidth,angle=0]
    {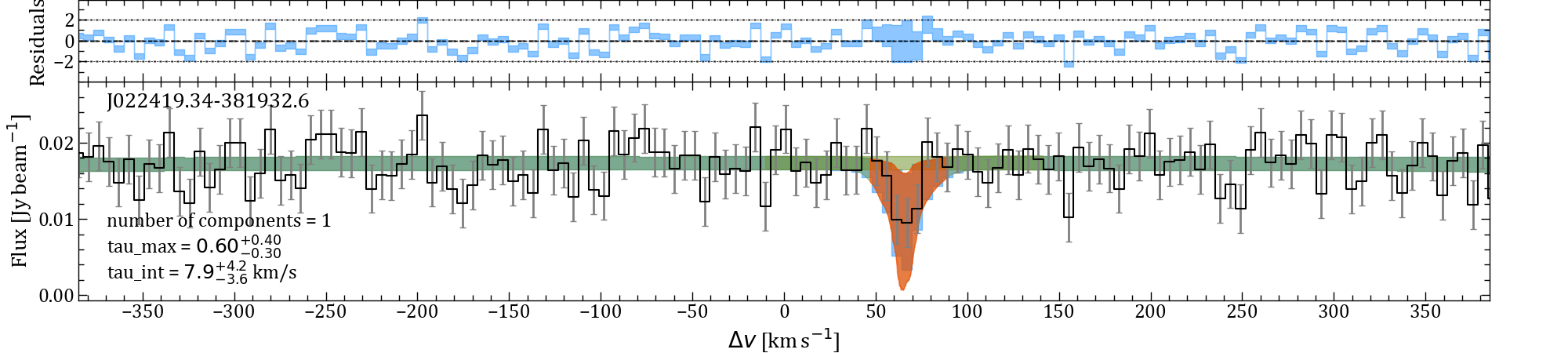}
    \includegraphics[trim = {0cm 0cm 0cm 0cm}, width=0.95\textwidth,angle=0]
    {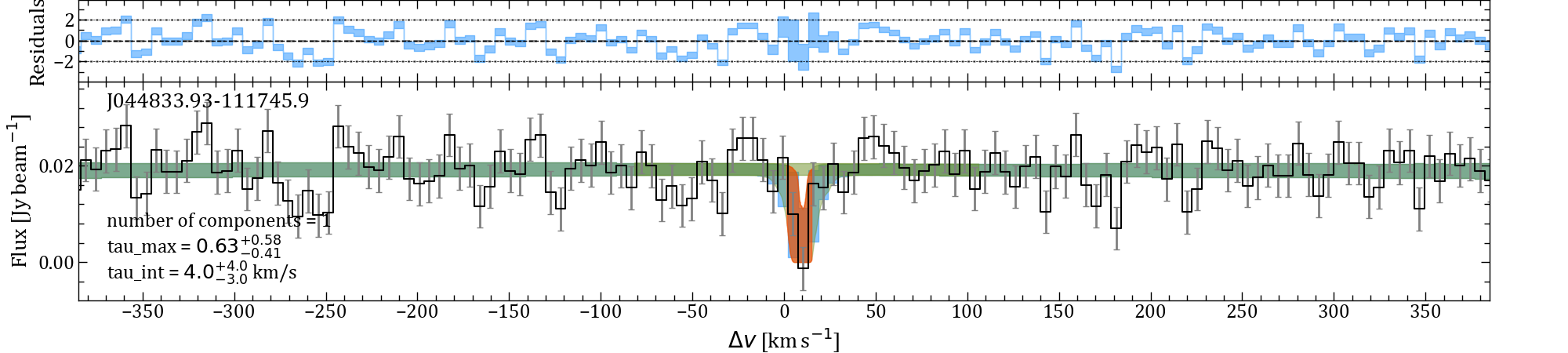}
    }
    \caption{Examples of \hi\ line profile fits with peak S/N of detections = 4.6, 4.0, 3.2, 2.9, and 1.5, and  {\tt vis\_flag} = {\tt True}. The graphical elements are same as in Fig.~\ref{fig:HI_fit_example}   
    }
    \label{fig:lowsnrdet-true}
\end{figure*}

\begin{figure*}
    \centering
    \vbox{
    \includegraphics[trim = {0cm 0cm 0cm 0cm}, width=0.95\textwidth,angle=0]
    {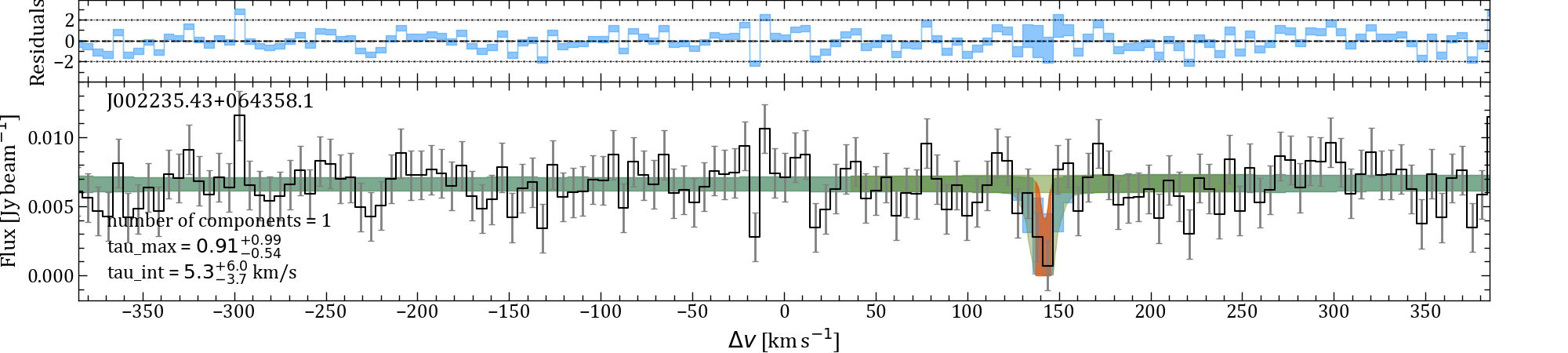}
    \includegraphics[trim = {0cm 0cm 0cm 0cm}, width=0.95\textwidth,angle=0]
    {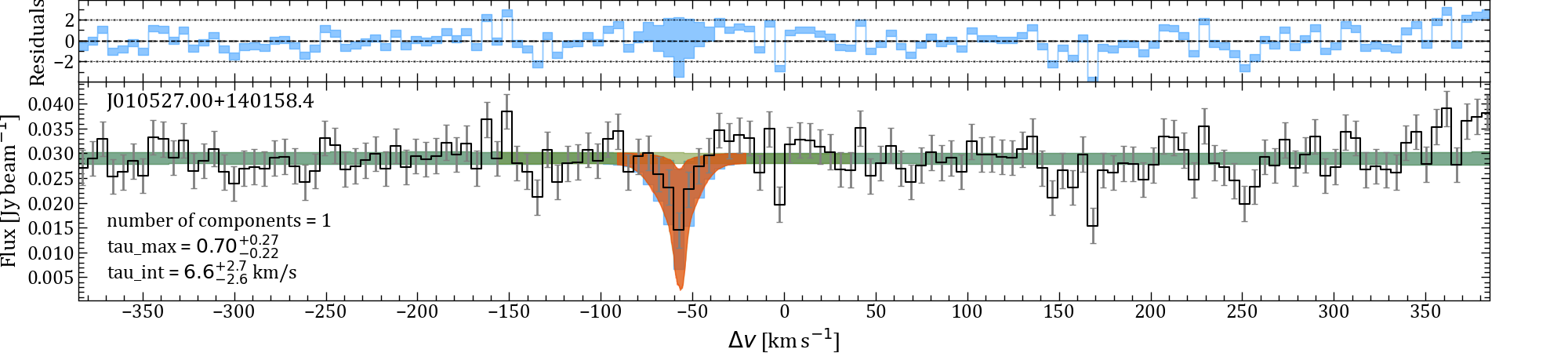}
    \includegraphics[trim = {0cm 0cm 0cm 0cm}, width=0.95\textwidth,angle=0]
    {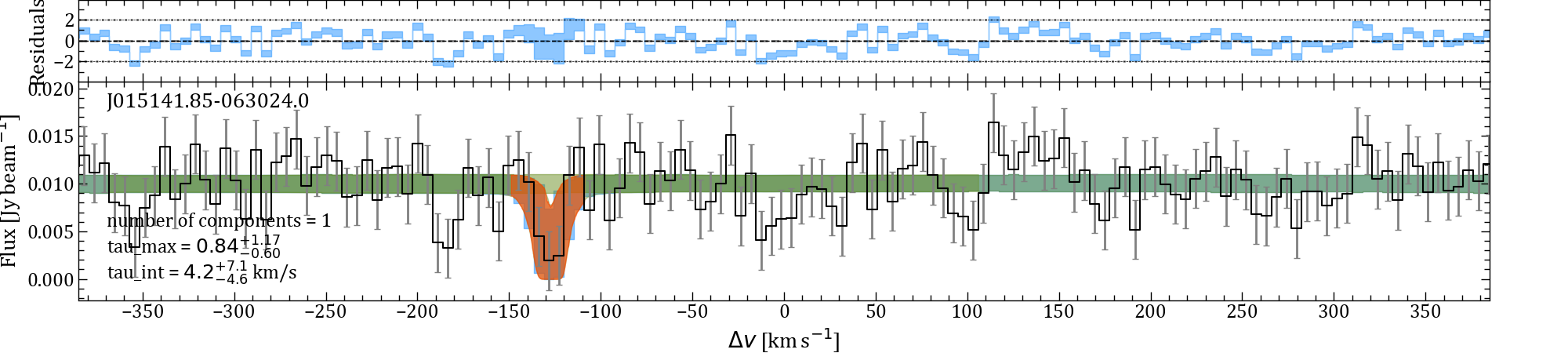}
    \includegraphics[trim = {0cm 0cm 0cm 0cm}, width=0.95\textwidth,angle=0]
    {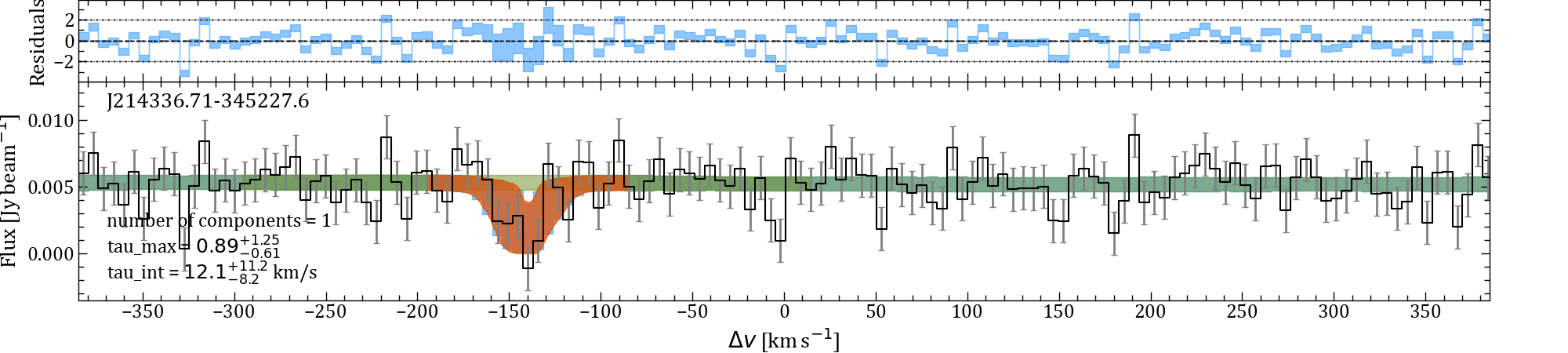}
    \includegraphics[trim = {0cm 0cm 0cm 0cm}, width=0.95\textwidth,angle=0]
    {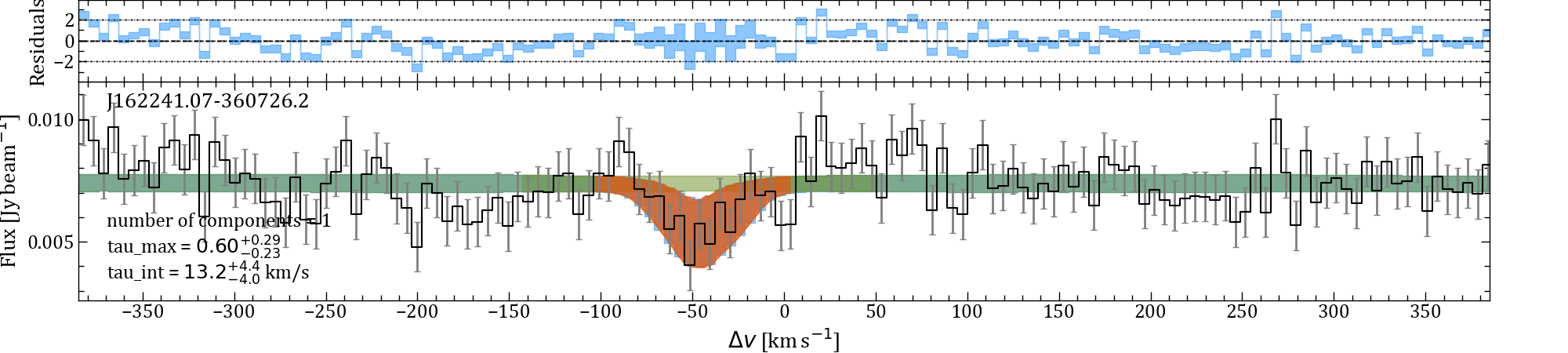}
    }
    \caption{Examples of \hi\ line profile fits with peak S/N of detections =   4.8, 4.7, 3.3, 2.0, and 1.6, and   {\tt vis\_flag} = {\tt False} due to the presence of multiple positive or negative features of similar strength in the spectrum.  The graphical elements are same as in Fig.~\ref{fig:HI_fit_example}.  
    }
    \label{fig:lowsnrdet-false}
\end{figure*}

Here we present the column definitions and initial rows of the absorption line catalog.    The complete catalog with the continuum images and the spectra are available at \url{https://mals.iucaa.in/}.  
Figs.~\ref{fig:lowsnrdet-true} and \ref{fig:lowsnrdet-false} shows examples of peak S/N $<$6 detections that were marked as {\tt vis\_flag} = {\tt True} and {\tt False} based on visual inspection, respectively. 
%


\begin{table*}[h]
\caption{Catalog column descriptions.}
\label{tab:cat_cols}
\begin{tabular}{c  l  p{10.5cm} }
\hline
Number & Name & Description \\
\hline
    1 & {\tt Source\_name}    & MALS name of the source ({\tt JHHMMSS.ss+DDMMSS.s}) based on its right ascension and declination (J2000) in the continuum image from {\tt SPW\_id} specified in column\,6. \\
    2 & {\tt Pointing\_id}   & The MALS pointing ID ({\tt JHHMMSS.ss$\pm$DDMMSS.s}) based on the position (J2000) of the central source in NVSS or SUMSS. \\
    3 & {\tt Obs\_date\_U} &   The date and time (UTC) of the start of UHF-band observing block(s) in the format YYYY-MM-DDThh:mm. \\
    4 & {\tt Obs\_date\_L} &   The date and time (UTC) of the start of L-band observing block(s) in the format YYYY-MM-DDThh:mm. \\
    5 & {\tt Obs\_band} &  The observing band: L = L-band and U = UHF-band.\\
    6 & {\tt SPW\_id}  &   This identifies the SPW corresponding to the spectrum.  The  possible values for extragalactic spectra in the barycentric frame are LSPW\_\textit{i} and USPW\_\textit{i}; here \textit{i} goes from 0 to 14. For Galactic spectra in the LSRK frame, the possible values are LSPW\_10G, LSPW\_13G and LSPW\_14G corresponding to  \hi\ 21-cm, OHSAT1 and OHMAIN line, respectively. \\ 
   7 & {\tt Maj\_cbeam}  & The major axis (arcsec) of the restoring beam of the cube. \\
    8 & {\tt Min\_cbeam}  & The minor axis (arcsec) of the restoring beam of the cube. \\
    9 & {\tt PA\_cbeam}   & The position angle (degrees) of the restoring beam of the cube. \\
    10 & {\tt Peak\_pos} & The right ascension and declination (J2000) of the pixel corresponding to flux density peak at which the spectrum is extracted. \\
    11 & {\tt Peak\_pos\_l} & The Galactic longitude (degrees) of the {\tt Peak\_pos}. \\
    12 & {\tt Peak\_pos\_b} & The Galactic latitude  (degrees) of the {\tt Peak\_pos}. \\
    13 & {\tt Distance\_pointing} & The distance of the source (arcmin) from the pointing center. \\
    14 & {\tt Scale\_factor} & The scale factor applied for the primary beam correction. The primary beam attenuation is corrected using the {\tt katbeam} (version\,0.1) model.\\
    15 & {\tt Peak\_flux} & The peak flux density (mJy\,beam$^{-1}$) of the source.  \\
    16 & {\tt Spec\_rms} & The spectral rms (mJy\,beam$^{-1}$\,channel$^{-1}$) in the unsmoothed spectrum. \\
    17 & {\tt Vis\_flag} & This is a boolean (True or False) indicating whether an absorption feature based on visual inspection is detected in the spectrum; {\tt Nan} implies no visual inspection has been performed. \\
    18 & {\tt n\_clump} & The number of well detached absorption features detected in the spectrum. \\
    19 & {\tt n\_comp} & The total number of Gaussian components fit to the absorption features detected in the spectrum. \\
    20 & {\tt tau\_max\_val} &  The peak optical depth of the absorption feature. \\
    21 & {\tt tau\_max\_+err} & The positive 1$\sigma$ error on {\tt tau\_max\_val}. \\
    22 & {\tt tau\_max\_-err} & The negative 1$\sigma$ error on {\tt tau\_max\_val}. \\
    23 & {\tt pos\_max} & The position of the {\tt tau\_max\_val}; redshift (barycentric frame) for extragalactic absorption and in \kms\ (LSRK frame) relative to the rest frequency of \hi\ 21-cm, OHSAT1 or OHMAIN line for Galactic absorption. \\
    24 & {\tt tau\_int\_val} & The integrated optical depth (\kms). \\
    25 & {\tt tau\_int\_+err} & The positive 1$\sigma$ error on {\tt tau\_int\_val}.\\
    26 & {\tt tau\_int\_-err} & The negative 1$\sigma$ error on {\tt tau\_int\_val}.\\
    27 & {\tt delta\_v} & The velocity width containing 50\% of the {\tt tau\_int\_val}. \\
    28 & {\tt comp\_i\_amp$^\dag$} & The amplitude of the $i^{th}$ Gaussian component fit to the absorption line. \\
    29 & {\tt comp\_i\_amp\_+err$^\dag$} & The positive 1$\sigma$ error on {\tt Comp\_i\_amp}. \\
    30 & {\tt comp\_i\_amp\_-err$^\dag$} & The negative 1$\sigma$ error on {\tt Comp\_i\_amp}. \\
    31 & {\tt comp\_i\_pos$^\dag$} & The position of the $i^{th}$ Gaussian component relative to {\tt pos\_max} (\kms). \\
    32 & {\tt comp\_i\_pos\_+err$^\dag$} & The positive 1$\sigma$ error on {\tt Comp\_i\_pos}. \\
    33 & {\tt comp\_i\_pos\_-err$^\dag$} & The negative 1$\sigma$ error on {\tt Comp\_i\_pos}.  \\
    34 & {\tt comp\_i\_sigma$\dag$} &  The width ($\sigma$ in \kms) of the $i^{th}$ Gaussian component. \\
    35 & {\tt comp\_i\_sigma\_+err$^\dag$} & The positive 1$\sigma$ error on {\tt Comp\_i\_sigma}.  \\
    36 & {\tt comp\_i\_sigma\_-err$^\dag$} & The negative 1$\sigma$ error on {\tt Comp\_i\_sigma}. \\
\hline
\end{tabular}
\parbox[t]{170mm}{
\textbf{Notes:} $\dag$: For a spectrum, columns 28 - 36 are repeated {\tt n\_comp} times. \\
}  
\end{table*}

\setlength{\tabcolsep}{4pt}
\scriptsize{
\begin{landscape}
\setlength\LTcapwidth{\linewidth}
\label{tab:cat_rows}
\begin{longtable}{lccccccccccccccccccccccccc}
\caption{First six rows from MALS spectral line catalog.  The column definitions are given in Table~\ref{tab:cat_cols}. The complete catalog and the spectra are available at \url{https://mals.iucaa.in/}.} \\
\hline
Source\_name & Pointing\_id & Obs\_date\_U & Obs\_date\_L & Obs\_band & SPW\_id  & Maj\_cbeam & Min\_cbeam & PA\_cbeam \\ 
        &      &   &   &   &   &  (arcsec)  & (arcsec)  & (degree)  \\
   ~~~~~~~~(1)  &   (2)  & (3) & (4) & (5) & (6) & (7) & (8) & (9)  \\
    &  Peak\_pos  & Peak\_pos\_l & Peak\_pos\_b & Distance\_ponting & Scale\_factor & Peak\_flux & Spec\_rms &  Vis\_flag & n\_clump & n\_comp \\ 
            &   (J2000) &   (degree) &   (degree) & (arcmin) &   & (mJy\,beam$^{-1}$) & (mJy\,beam$^{-1}$) &  &  \\ 
        &   (10) &   (11) &   (12) & (13) & (14) & (15) & (16) & (17) & (18) & (19) \\  
    & tau\_max\_val & tau\_max\_+err  & tau\_max\_-err & pos\_max & tau\_int\_val & tau\_int\_+err & tau\_int\_-err & delta\_v \\
            &     &    &    & (\kms)  & (\kms) & (\kms) \\ 
        &   (20) &   (21) &   (22) & (23) & (24) & (25) & (26) & (27) \\ 
    & Comp\_1\_amp & Comp\_1\_amp\_+err & Comp\_1\_amp\_-err & Comp\_1\_pos & Comp\_1\_pos\_+err & Comp\_1\_pos\_-err & Comp\_1\_sigma & Comp\_1\_sigma\_+err & Comp\_1\_sigma\_-err \\
            &    &   &    & (\kms) & (\kms) & (\kms) & (\kms) & (\kms) & (\kms)   \\   
        &   (28) &   (29) &   (30) & (31) & (32) & (33) & (34) & (35) & (36)   \\ 
\hline
J000450.96-153224.1 & J000141.57-154040.6 & [] & ['2020-07-16T01:40'] & L & LSPW\_10G & 9.3 & 7.5 & -6.3\\ & 00:04:50.89 -15:32:23.5 & 77.53523 & -74.145749 & 46.3 & 0.2 & 16.05 & 3.72 &False & $\cdots$ & 0.0\\ & $\cdots$ & $\cdots$ & $\cdots$ & $\cdots$ & $\cdots$ & $\cdots$ & $\cdots$ & $\cdots$\\ & $\cdots$ & $\cdots$ & $\cdots$ & $\cdots$ & $\cdots$ & $\cdots$ & $\cdots$ & $\cdots$ & $\cdots$\\ & $\cdots$ & $\cdots$ & $\cdots$ & $\cdots$ & $\cdots$ & $\cdots$ & $\cdots$ & $\cdots$ & $\cdots$\\
J000426.61-152106.7 & J000141.57-154040.6 & [] & ['2020-07-16T01:40'] & L & LSPW\_10G & 9.3 & 7.5 & -6.3\\ & 00:04:26.25 -15:21:06.7 & 77.712445 & -73.93895 & 44.3 & 0.2 & 5.75 & 3.01 & False & $\cdots$ & 0.0\\ & $\cdots$ & $\cdots$ & $\cdots$ & $\cdots$ & $\cdots$ & $\cdots$ & $\cdots$ & $\cdots$\\ & $\cdots$ & $\cdots$ & $\cdots$ & $\cdots$ & $\cdots$ & $\cdots$ & $\cdots$ & $\cdots$ & $\cdots$\\ & $\cdots$ & $\cdots$ & $\cdots$ & $\cdots$ & $\cdots$ & $\cdots$ & $\cdots$ & $\cdots$ & $\cdots$\\
J000421.87-153741.2 & J000141.57-154040.6 & [] & ['2020-07-16T01:40'] & L & LSPW\_10G & 9.3 & 7.5 & -6.3\\ & 00:04:21.90 -15:37:40.9 & 77.002393 & -74.135021 & 38.7 & 0.3 & 19.02 & 1.91 & True & 1.0 & 1.0\\ & 0.4666 & 0.1617 & 0.1437 & -117.5 & 7.6064 & 2.5949 & 2.4545 & 60.6\\ & 6.900 & 2.400 & 2.000 & -112.0 & 4.0 & 5.0 & 10.0 & 3.0 & 4.0\\ & $\cdots$ & $\cdots$ & $\cdots$ & $\cdots$ & $\cdots$ & $\cdots$ & $\cdots$ & $\cdots$ & $\cdots$\\
J000402.65-151550.1 & J000141.57-154040.6 & [] & ['2020-07-16T01:40'] & L & LSPW\_10G & 9.3 & 7.5 & -6.3\\ & 00:04:02.68 -15:15:49.7 & 77.66258 & -73.810379 & 42.1 & 0.2 & 15.67 & 3.35 & False & $\cdots$ & 0.0\\ & $\cdots$ & $\cdots$ & $\cdots$ & $\cdots$ & $\cdots$ & $\cdots$ & $\cdots$ & $\cdots$\\ & $\cdots$ & $\cdots$ & $\cdots$ & $\cdots$ & $\cdots$ & $\cdots$ & $\cdots$ & $\cdots$ & $\cdots$\\ & $\cdots$ & $\cdots$ & $\cdots$ & $\cdots$ & $\cdots$ & $\cdots$ & $\cdots$ & $\cdots$ & $\cdots$\\
J000359.91-155715.8 & J000141.57-154040.6 & [] & ['2020-07-16T01:40'] & L & LSPW\_10G & 9.3 & 7.5 & -6.3\\ & 00:03:59.97 -15:57:15.8 & 75.95749 & -74.318261 & 37.2 & 0.3 & 5.14 & 1.57 & False & $\cdots$ & 0.0\\ & $\cdots$ & $\cdots$ & $\cdots$ & $\cdots$ & $\cdots$ & $\cdots$ & $\cdots$ & $\cdots$\\ & $\cdots$ & $\cdots$ & $\cdots$ & $\cdots$ & $\cdots$ & $\cdots$ & $\cdots$ & $\cdots$ & $\cdots$\\ & $\cdots$ & $\cdots$ & $\cdots$ & $\cdots$ & $\cdots$ & $\cdots$ & $\cdots$ & $\cdots$ & $\cdots$\\
J000349.07-151734.2 & J000141.57-154040.6 & [] & ['2020-07-16T01:40'] & L & LSPW\_10G & 9.3 & 7.5 & -6.3\\ & 00:03:49.01 -15:17:34.2 & 77.445711 & -73.796206 & 38.4 & 0.3 & 5.23 & 1.77 &False & $\cdots$ & 0.0\\ & $\cdots$ & $\cdots$ & $\cdots$ & $\cdots$ & $\cdots$ & $\cdots$ & $\cdots$ & $\cdots$\\ & $\cdots$ & $\cdots$ & $\cdots$ & $\cdots$ & $\cdots$ & $\cdots$ & $\cdots$ & $\cdots$ & $\cdots$\\ & $\cdots$ & $\cdots$ & $\cdots$ & $\cdots$ & $\cdots$ & $\cdots$ & $\cdots$ & $\cdots$ & $\cdots$\\
\hline
\end{longtable}
\end{landscape}
}

%



\FloatBarrier

\section{Completeness analysis: Additional details}
%
\begin{figure*}[h]
    \centering
    \hbox{
    \includegraphics[trim = {0cm 0cm 0cm 0cm}, width=0.46\textwidth,angle=0]{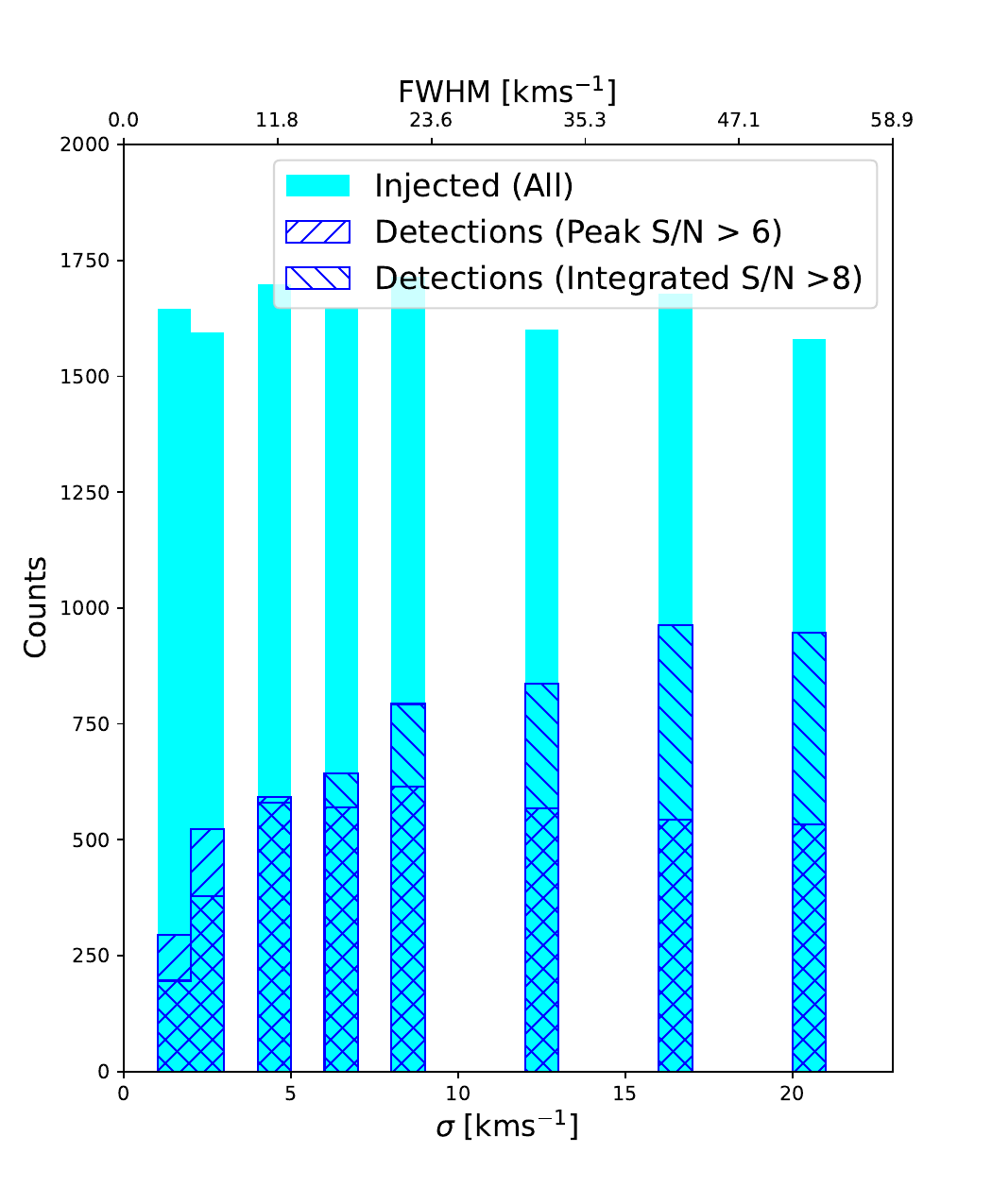}
    \includegraphics[trim = {0cm 0cm 0cm 0cm}, width=0.43\textwidth,angle=0]{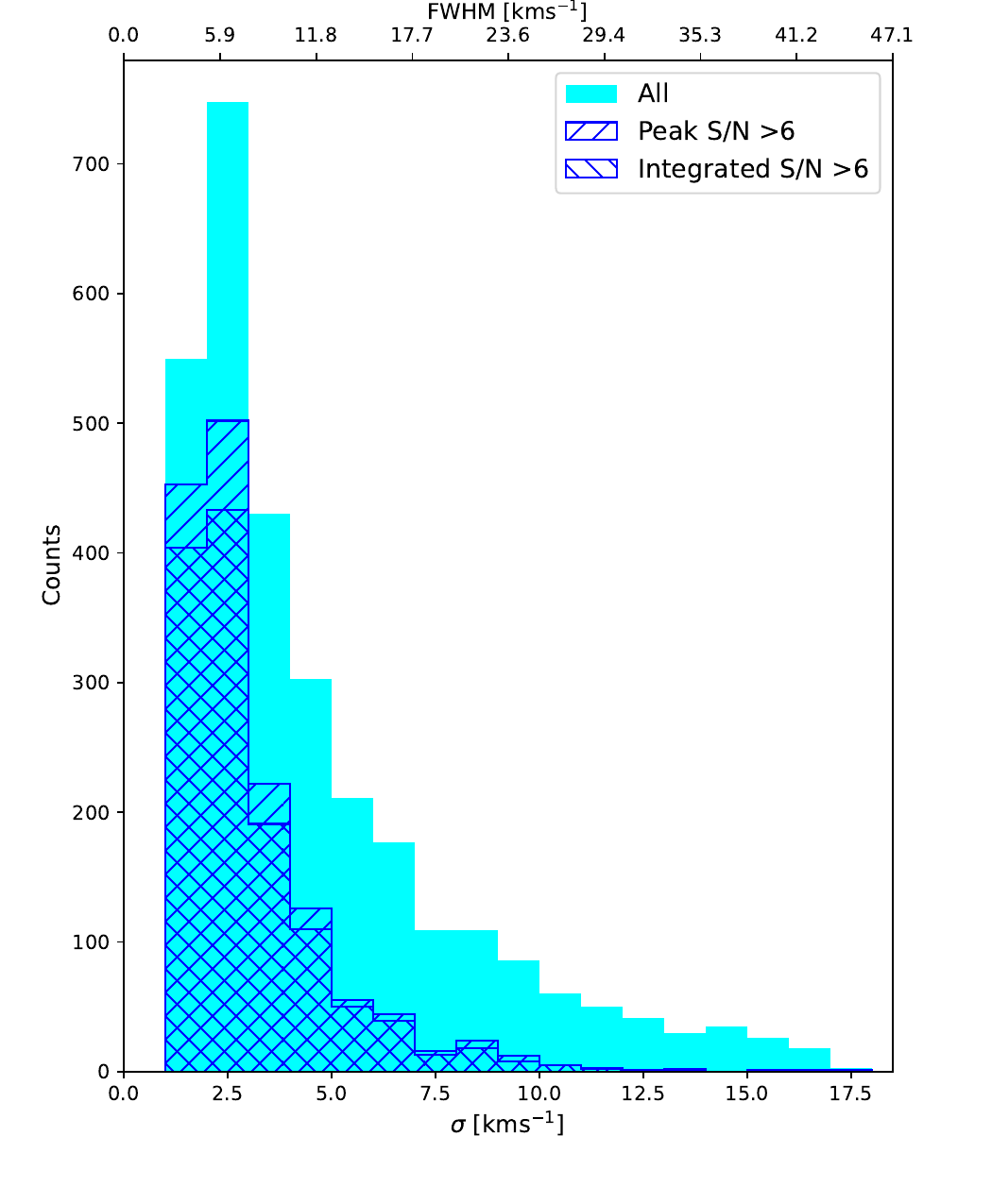}
    }
    \caption{Left: Distributions of line width ($\sigma$) of injected absorption lines and detections for peak and integrated S/N cutoffs adopted for the MALS sample.  In general, for absorption features with single Gaussian component fits, integrated S/N = peak S/N $\times$ $\sqrt{\frac{\text{Line FWHM}}{ \text{Spectral resolution}}}$. Right: Distributions of line width of 2986 MALS absorbers fit with single Gaussian components.
    }
    \label{fig:absinj_stat}
\end{figure*}

\FloatBarrier

\section{Correlation studies: Additional details}
%
\begin{figure*}[h]
    \centering
    \vbox{
    \includegraphics[width=16.0cm]{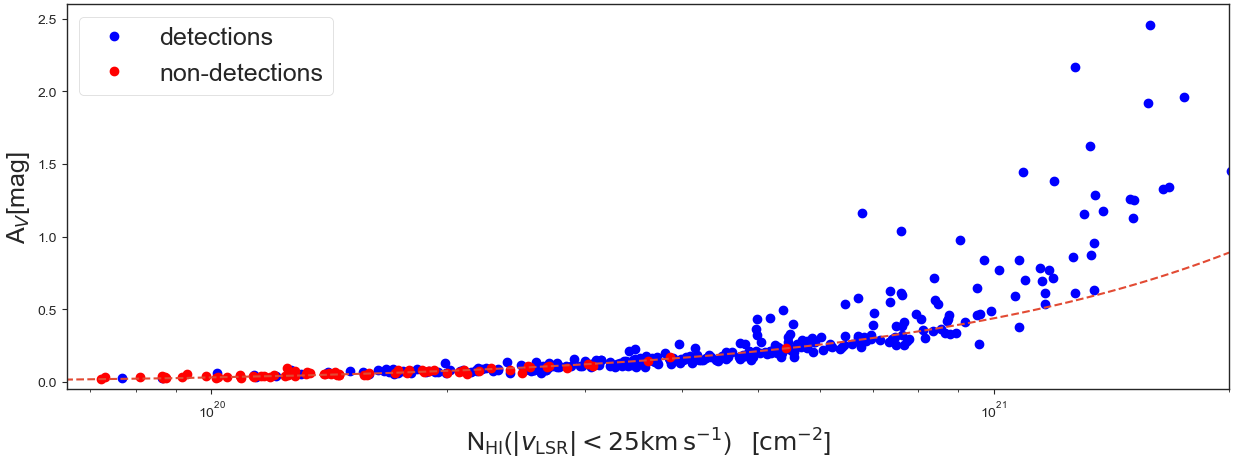}
    }
    \caption{Linear correlation between $A_\mathrm{V}$) and $N_\mathrm{HI}$(HI4PI) toward central lines of sight without absorption detections (red) and with absorption detections (blue).  The dashed lines show the best fit result of the least square approximation given in Eq.\,\ref{Eq:linreg}. Note, sight lines with and without the detection of an \hi\ absorption line populate the lower but largely overlapping column densities.
    }
    \label{Fig:Av_vs_NHI}
\end{figure*}
%
\begin{figure*}[t]
    \centering
    \includegraphics[width=8cm]{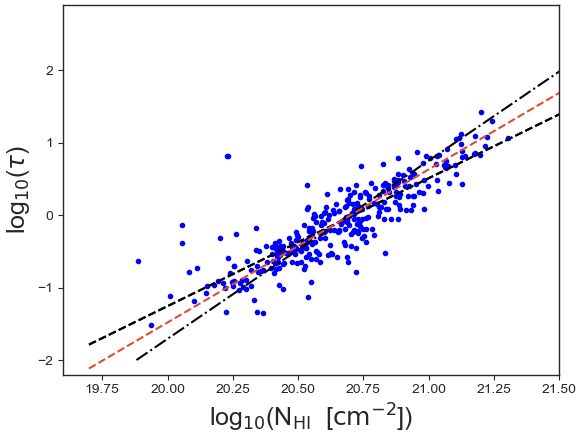}
    \includegraphics[width=8cm]{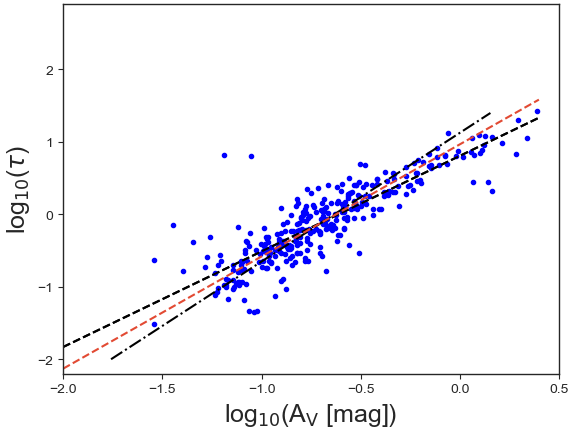}
     \caption{Left:  $\int\tau$dv versus $N_\mathrm{HI}$(HI4PI) is plotted on a double logarithmic scale. The dashed and dash-dotted lines mark the range of uncertainty by fitting a linear function to the data values. The dashed red line represents the average of both least-square approximations. From this we calculated the correlation coefficient of 0.89 and a scaling relation of $\int\tau dv \propto N_\mathrm{HI}^{2.10\pm0.22} $. Right: Same as left but instead of $N_\mathrm{HI}$, the dependence of $\int\tau dv$ from  $A_\mathrm{V}$ is displayed. Here, the correlation coefficient is 0.85 and the scaling relation is $\int\tau dv$ $\propto A_\mathrm{V}^{1.50\pm0.13} $.}
     \label{Fig:Tau_vs_NHI_AV}
\end{figure*}

\end{appendix}

\twocolumn

\end{document}